\documentclass{emulateapj}
\usepackage{amssymb,amsmath,natbib,graphicx}

\begin{document}

\input epsf.tex    
\input epsf.def   
\ifdefined\apj
\else
 \newcommand{\apj}         {Ap. J.}
 \newcommand{\aj}          {A. J.}
 \newcommand{\aap}         {Astron. Astrophys.}
 \newcommand{\araa}        {Annu. Rev. Astron. Astrophys.}
 \newcommand{\mnras}       {MNRAS}
 \newcommand{\pasj}        {PASJ}
 \newcommand{\arcmin}      {\mbox{$\sp{\prime}$}}
 \newcommand{\arcsec}      {\mbox{$\sp{\prime\prime}$}}
 \newcommand{\acknowledgements} {\vspace{0.5in}}

 \jname{ARA\&A}
 \jyear{2012}
 \jvol{}
 \ARinfo{1056-8700/97/0610-00}
\fi
\newcommand{\hi}          {\mbox{\rm H{\small I}}}
\newcommand{\hii}         {\mbox{\rm H{\small II}}}
\newcommand{\htwo}        {\mbox{H$_{2}$}}
\newcommand{\jone}        {\mbox{$J=1\rightarrow0$}}
\newcommand{\jtwo}        {\mbox{$J=2\rightarrow1$}}
\newcommand{\jthree}        {\mbox{$J=3\rightarrow2$}}
\newcommand{\um}          {$\mu$m}
\newcommand{\ha}          {H$\alpha$}
\newcommand{\kmpers}      {\mbox{\rm km~s$^{-1}$}}
\newcommand{\percmcu}     {\mbox{\rm cm$^{-3}$}}
\newcommand{\msun}        {\mbox{\rm M$_\odot$}}
\newcommand{\msunperpcsq} {\mbox{\rm M$_\odot$~pc$^{-2}$}}
\newcommand{\msunperpcsqyr} {\mbox{\rm M$_\odot$~pc$^{-2}$~yr$^{-1}$}}
\newcommand{\msunperyr}   {\mbox{\rm M$_\odot$~yr$^{-1}$}}
\newcommand{\msunperpccu} {\mbox{\rm M$_\odot$~pc$^{-3}$}}
\newcommand{\msunperyrkpcsq} {\mbox{\rm M$_\odot$~yr$^{-1}$~kpc$^{-2}$}}
\newcommand{\xco}         {\mbox{$X_{\rm CO}$}}
\newcommand{\xcot}         {\mbox{$X_{\rm CO,20}$}}
\newcommand{\aco}         {\mbox{$\alpha_{\rm CO}$}}
\newcommand{\xcounits}    {\mbox{\rm cm$^{-2}$(K km s$^{-1}$)$^{-1}$}}
\newcommand{\acounits}  {\mbox{\rm M$_\odot$ (K km s$^{-1}$ pc$^2$)$^{-1}$}}
\newcommand{\Lcounits}  {\mbox{\rm K km s$^{-1}$ pc$^2$}}
\newcommand{\Kkmpers}     {\mbox{\rm K km s$^{-1}$}}
\newcommand{\Kkmperspcsq} {\mbox{\rm K km s$^{-1}$ pc$^2$}}
\newcommand{\co}          {\mbox{$^{12}$CO}}
\newcommand{\cothree}          {\mbox{$^{13}$CO}}
\newcommand{\Ico}         {\mbox{I$_{\rm CO}$}}
\newcommand{\av}          {\mbox{$A_V$}}
\newcommand{\percmsq}     {\mbox{cm$^{-2}$}}
\newcommand{\cii}         {\mbox{\rm [C{\small II}]}}
\newcommand{\nii}         {\mbox{\rm [N{\small II}]}}
\newcommand{\ci}         {\mbox{\rm [C{\small I}]}}
\newcommand{\wco}         {\mbox{\rm W(CO)}}
\newcommand{\fscii}       {($^2$P$_{3/2}\rightarrow^2$P$_{1/2}$)}
\newcommand{\Smol}	  {\mbox{$\Sigma_{\rm mol}$}}
\newcommand{\Mmol}	  {\mbox{${\rm M}_{\rm mol}$}}
\newcommand{\Ssfr}	  {\mbox{$\Sigma_{\rm SFR}$}}
\newcommand{\Sgmc}	  {\mbox{$\Sigma_{\rm GMC}$}}
\newcommand{\Lco}	  {\mbox{$L_{\rm CO}$}}
\newcommand{\ag}{\mbox{ \raisebox{-.4ex}{$\stackrel{\textstyle >}{\sim}$} }}
\newcommand{\al}{\mbox{ \raisebox{-.4ex}{$\stackrel{\textstyle <}{\sim}$} }}
\newcommand{\dgr}         {\mbox{$\delta_{\rm DGR}$}}
\newcommand{\dgrp}         {\mbox{$\delta_{\rm DGR}'$}}
\newcommand{\rco}{R_{\rm CO}}
\newcommand{\rht}{R_{\rm H_2}}
\newcommand{\davdg}{\Delta A_{V}}

\input psfig.sty

\title{The CO-to-H$_2$ Conversion Factor}

\markboth{Bolatto, Wolfire, \& Leroy}{The CO-to-H$_2$ Conversion Factor}

\author{Alberto D. Bolatto, Mark Wolfire,}
\affil{Department of Astronomy, University of Maryland, College Park, MD 20742}
\and
\author{Adam K. Leroy}
\affil{National Radio Astronomy Observatory, Charlottesville, VA 22903}


\begin{abstract}
CO line emission represents the most accessible and widely used tracer
of the molecular interstellar medium. This renders the translation of
observed CO intensity into total \htwo\ gas mass critical to
understand star formation and the interstellar medium in our Galaxy
and beyond. We review the theoretical underpinning, techniques, and
results of efforts to estimate this CO-to-\htwo\ ``conversion
factor,'' \xco, in different environments. In the Milky Way disk, we
recommend a conversion factor $\xco = 2 \times 10^{20}$~\xcounits\
with $\pm 30\%$ uncertainty. Studies of other ``normal galaxies''
return similar values in Milky Way-like disks, but with greater
scatter and systematic uncertainty. Departures from this Galactic
conversion factor are both observed and expected. Dust-based
determinations, theoretical arguments, and scaling relations all
suggest that \xco\ increases with decreasing metallicity, turning up
sharply below metallicity $\approx 1/3$--$1/2$ solar in a manner
consistent with model predictions that identify shielding as a
key parameter.  Based on spectral line modeling and dust
observations, \xco\ appears to drop in the central, bright regions of
some but not all galaxies, often coincident with regions of bright CO
emission and high stellar surface density. This lower \xco\ is also
present in the overwhelmingly molecular interstellar medium of
starburst galaxies, where several lines of evidence point to a lower
CO-to-\htwo\ conversion factor.  At high redshift, direct evidence
regarding the conversion factor remains scarce; we review what is
known based on dynamical modeling and other arguments. 
\end{abstract}

 \maketitle

\section{Introduction}
\label{sec:introduction}

Molecular hydrogen, H$_2$, is the most abundant molecule in the
universe.  With the possible exception of the very first generations
of stars, star formation is fueled by molecular gas. Consequently,
\htwo\ plays a central role in the evolution of galaxies and stellar
systems \citep[see the recent review by][]{KENNICUTT2012}. Unfortunately for astronomers interested in the
study of the molecular interstellar medium (ISM), cold \htwo\ is not
directly observable in emission. \htwo\ is a diatomic molecule with
identical nucleii and therefore possesses no permanent dipole moment
and no corresponding dipolar rotational transitions. The lowest energy
transitions of \htwo\ are its purely rotational quadrupole transitions
in the far infrared at $\lambda=28.22$~$\mu$m and shorter
wavelengths. These are weak owing to their long spontaneous decay
lifetimes $\tau_{\rm decay}\sim100$ years. More importantly, the two
lowest para and ortho transitions have upper level energies $E/k\approx510$~K
and 1015~K above ground \citep{DABROWSKI1984}. They are thus only excited
in gas with $T\gtrsim100$~K. The lowest vibrational transition of
\htwo\ is even more difficult to excite, with a wavelength
$\lambda=2.22$ $\mu$m and a corresponding energy $E/k=6471$~K.
Thus the cold molecular hydrogen that makes up most of the molecular
ISM in galaxies is, for all practical purposes, invisible
in emission.

Fortunately, molecular gas is not pure \htwo. Helium, being
monoatomic, suffers from similar observability problems in cold
clouds, but the molecular ISM also contains heavier elements at the
level of a few $\times 10^{-4}$ per H nucleon. The most abundant of
these are oxygen and carbon, which combine to form CO under the
conditions prevalent in molecular clouds. CO has a weak permanent
dipole moment ($\mu\approx0.11~{\rm D}=0.11\times10^{-18}$~esu~cm) and
a ground rotational transition with a low excitation energy
$h\nu/k\approx5.53$~K. With this low energy and critical density
(further reduced by radiative trapping due to its high optical
depth), CO is easily excited even in cold molecular clouds. At a
wavelength of 2.6~mm, the \jone\ transition of CO falls in a fairly
transparent atmospheric window. It has thus become the workhorse
tracer of the bulk distribution of H$_2$ in our Galaxy and beyond.

As a consequence, astronomers frequently employ CO emission to measure
molecular gas masses. The standard methodology posits a
simple relationship between the observed CO intensity
and the column density of molecular gas, such that

\begin{equation}
{\rm N(\htwo)} = \xco\,{\rm W(^{12}C^{16}O\,\jone)},
\label{eq:xco}
\end{equation}

\noindent where the column density, ${\rm N(\htwo)}$, is in \percmsq\ and
the integrated line intensity, ${\rm W(CO)}$\footnote{Henceforth we
refer to the most common $^{12}$C$^{16}$O isotopologue as simply CO,
and unless otherwise noted to the ground rotational transition
\jone.}, is in traditional radio astronomy observational units of ${\rm
K\,km\,s^{-1}}$. A corollary of this relation arises from integrating
over the emitting area and correcting by the mass contribution of
heavier elements mixed in with the molecular gas,

\begin{equation}
\Mmol = \aco\,\Lco~.
\label{eq:aco}
\end{equation}

\noindent Here \Mmol\ has units of \msun\ and \Lco\ is usually
expressed in \Lcounits. \Lco\ relates to the observed integrated flux
density in galaxies via $\Lco = 2453\,{S_{\rm CO}\Delta v\,D_{\rm
L}^2/(1+z)}$, where $S_{\rm CO}\Delta v$ is the integrated line flux density,
in Jy~\kmpers, $D_{\rm L}$ is the luminosity distance to the source in
Mpc, and $z$ is the redshift
\citep[e.g.,][use Eq. \ref{eq:fluxT} to convert between ${\rm W(CO)}$ and
$S_{\rm CO}\Delta v$]{SOLOMON2005}. Thus \aco\ is simply a mass-to-light
ratio. The correction for the contribution of heavy elements by mass
reflects chiefly helium and amounts to a $\approx36\%$ correction
based on cosmological abundances.

Both \xco\ and \aco\ are referred to as the ``CO-to-\htwo\ conversion
factor.''  For $\xco=2\times10^{20}$~\xcounits\ the corresponding
$\aco$ is $4.3$~M$_\odot\,(\Lcounits)^{-1}$. To translate integrated
flux density directly to molecular mass, Equation \ref{eq:aco} can be
written as

\begin{equation}
\Mmol=1.05\times10^4\,\left( \frac{\xco }{2 \times 10^{20}~\frac{{\rm cm}^{-2}}{{\rm K~km~s}^{-1}} } \right)\,\frac{S_{\rm CO}\Delta v\,D_{\rm L}^2}{(1+z)}~.
\end{equation}

\noindent For convenience we define

\begin{equation}
\xcot \equiv \frac{\xco}{1\times10^{20}\,\xcounits}.
\end{equation}

\noindent We discuss the theoretical underpinnings of these equations in \S\ref{sec:intro_theoretical}.

\begin{figure}[th!]
\centerline{\psfig{figure=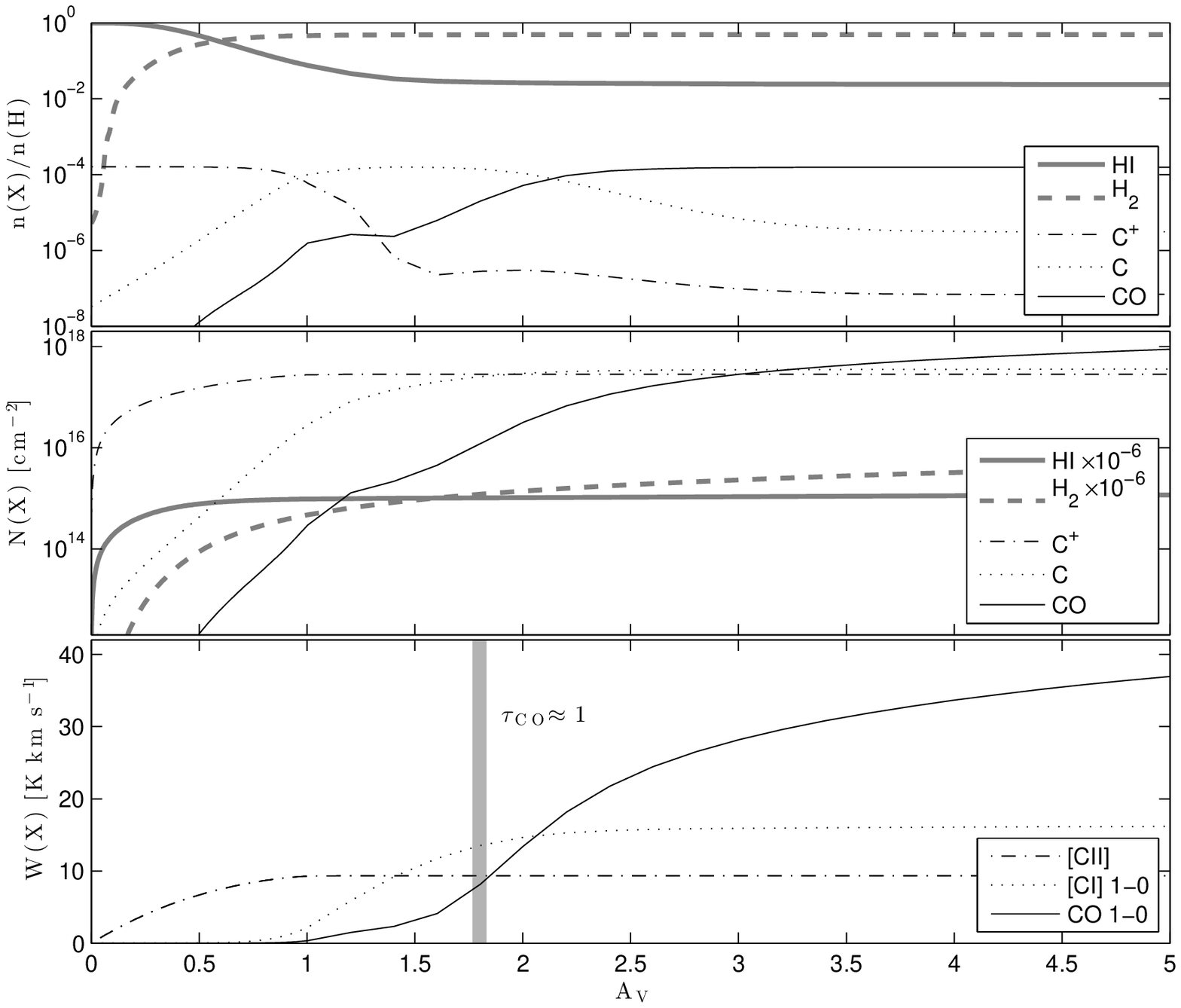,width=\columnwidth}}
\caption {Calculated cloud structure as a function of optical 
depth into the cloud.
Top panel shows the fractional abundance of HI, \htwo, C$^+$, C, and
CO. Middle panel shows their integrated column densities from the
cloud edge. Bottom
panel shows the emergent line intensity in units of K~km~${\rm
s^{-1}}$ for [CII] 158 $\mu$m, [CI] 609 $\mu$m, and CO \jone. The grey
vertical bar shows where CO \jone\ becomes optically thick. At the
outer edge of the cloud gas is mainly
\hi. \htwo\ forms at $\av\sim 0.5$ while the carbon is mainly
C$^+$. The C$^+$ is converted to C at $\av\sim 1$ and CO dominates
at $\av \gtrsim 2$. The model
uses constant H density $n=3\times 10^3$~\percmcu, a radiation field
$\chi = 30$ times the interstellar radiation field of
\cite{DRAINE1978}, a primary cosmic-ray ionization rate of $2\times
10^{-16}$ s$^{-1}$ per hydrogen nucleon, and are based on the PDR
models of \cite{WOLFIRE2010} and \cite{HOLLENBACH2012}. 
\label{fig:COabundances} }
\end{figure}

Note that the emission from CO \jone\ is found to be consistently
optically thick except along very low column density lines-of-sight,
as indicated by ratios of $^{12}$CO to $^{13}$CO intensities much
lower than the isotopic ratio. The reason for this is simple to
illustrate. The optical depth of a CO rotational transition is

\begin{equation}
\tau_J = \frac{8\pi^3}{3h}\mu^2\frac{J}{g_J}\left({e^{h\nu_J/kT_{ex}}-1}\right)\frac{N_J}{\Delta\,v},
\label{eq:tauco}
\end{equation}

\noindent where $J$ and $N_J$ are the rotational quantum number and the 
column density in the upper level of the $J\rightarrow J-1$
transition, $\nu$ is the frequency, $T_{ex}$ is the excitation
temperature (in general a function of $J$, and restricted to be
between the gas kinetic temperature and that of the Cosmic Microwave
Background), $\Delta\,v$ is the velocity width, $\mu$ is the dipole
moment, $g_J=2J+1$ is the statistical weight of level $J$, and $h$ and
$k$ are the Planck and Boltzmann constants respectively.  Under
typical conditions at the molecular boundary, $\tau\approx1$ for the
\jone\ transition requires ${\rm N(\htwo)}\approx2-3\times10^{20}$
\percmsq\ for a Galactic carbon gas-phase abundance
$A_{\rm C}\sim1.6\times10^{-4}$ \citep{SOFIA2004}. At the outer edge
of a cloud the carbon is mainly C$^+$, which then recombines with
electrons to form neutral C (Fig.~\ref{fig:COabundances}). Carbon
is converted to CO by a series of reactions initiated by the
cosmic-ray ionization of H or \htwo\
\citep[e.g.,][]{VANDISHOECK1988} and becomes the dominant carrier of
carbon at $\av\sim 1-2$.  The CO
\jone\ line turns optically thick very quickly after CO becomes a
significant carbon reservoir, over a region of thickness
$\Delta\av\sim0.2-0.3$ for a typical Galactic dust-to-gas ratio (c.f.,
Eq. \ref{eq:bohlin_gdr}).

Equations \ref{eq:xco} and \ref{eq:aco} represent highly idealized,
simplified relations where all the effects of environment, geometry,
excitation, and dynamics are subsumed into the \xco\ or \aco\
coefficients.  
A particular example is the effect that spatial
scales have on the CO-to-\htwo\ conversion factor. Indeed, for
the reasons discussed in the previous paragraph, \xco\ along a
line-of-sight through a dense molecular cloud where $\av\gtrsim10$ is
not expected to be the same as
\xco\ along a diffuse line-of-sight sampling mostly material where
$\av<1$ \citep[see, for example,][]{PINEDA2010,LISZT2012}. 
Thus on small spatial scales we expect to see a large variability in
the CO-to-\htwo\ conversion factor. This variability will 
average out on the large spatial scales, to a typical value
corresponding to the dominant environment. Because of the large
optical depth of the CO
\jone\ transition the velocity dispersion giving rise to the width of
the CO line will also play an important role on ${\rm W(CO)}$, and
indirectly on \xco. Indeed, there is not one value of \xco\ that is
correct and applicable to each and every situation, although there are
values with reasonable uncertainties that are applicable over large
galactic scales.

The plan of this work is as follows: in the remainder of this section
we provide a brief historical introduction. In
\S\ref{sec:intro_theoretical} we present the theoretical background to
the CO-to-\htwo\ conversion factor. In \S\ref{sec:milky_way} we review
the methodology and measurements of \xco\ in the Milky Way, the best
understood environment. We characterize the range of values found and
the underlying physics for each measurement technique. In
\S\ref{sec:normal_galaxies} we review the literature on \xco\
determinations in ``normal'' star-forming galaxies and discuss the
techniques available to estimate \xco\ in extragalactic systems.  In
\S\ref{sec:low_metallicities} we consider the effect of metallicity, a
key local physical parameter. In
\S\ref{sec:starbursts} we review the measurements and the physical
mechanisms affecting the value of the CO-to-\htwo\ conversion factor
in the starburst environments of luminous and ultraluminous
galaxies. In \S\ref{sec:sb_highz} we consider the explicit case of
\xco\ in high redshift systems, where a much more restricted range of
observations exist. In \S\ref{sec:theoretical} we discuss the results
of recent calculations of molecular clouds including the effects of
turbulence and chemistry. Finally, in \S\ref{sec:conclusions} we will
offer some recommendations and caveats as to the best values of \xco\
to use in different environments, as well as some suggestions about
open avenues of research on the topic.

\subsection{Brief Historical Perspective}
\label{sec:intro_historical}

Carbon monoxide was one of the first interstellar medium molecules
observed at millimeter wavelengths. \citet{WILSON1970} reported the
discovery of intense CO emission from the Orion nebula using the 36
foot NRAO antenna at Kitt Peak, Arizona. Surveys of molecular clouds
in the Galaxy
\citep[e.g.,][]{SOLOMON1972,WILSON1974,SCOVILLE1975,BURTON1975}
established molecular gas to be widespread in the inner Milky Way with
a distribution that resembles giant HII regions more closely than that
of atomic hydrogen gas. The combination of CO and $\gamma$-ray
observations demonstrated that \htwo\ dominates over \hi\ by mass in
the inner Galaxy \citep{STECKER1975}. By the end of the following decade,
these studies extended to complete the mapping of the Galactic plane
\citep{DAME1987}.

The first extragalactic detections of CO occurred in parallel with
these early Galactic surveys \citep{RICKARD1975,SOLOMON1975}. They
found CO to be particularly bright in galaxies with nuclear
activity such as M~82 and NGC~253. The number of extragalactic CO
observations grew rapidly to include several hundred galaxies over the
next two decades \citep{YOUNG1991,YOUNG1995}, and CO emission was
employed to determine galaxy molecular masses \citep{YOUNG1982}. By the
late 1980s, the first millimeter interferometers spatially resolved
molecular clouds in other galaxies \citep{VOGEL1987,WILSON1988}.  Such
observations remain challenging, though powerful new interferometric
facilities such as the Atacama Large Millimeter Array (ALMA) will
change that.

The first detection of CO at cosmological redshifts targeted
ultraluminous infrared sources and revealed very large reservoirs of
highly excited molecular gas
\citep{BROWN1991,SOLOMON1992}. Because of the deep integrations required,
the number of high redshift CO detections grew slowly at first
\citep{SOLOMON2005}, but this field is now developing rapidly driven
by recent improvements in telescope sensitivity (see the review by
Carilli \& Walter in this issue). An increased appreciation of the
roles of gas and star formation in the field of galaxy evolution and
the according need to determine accurate gas masses provides one of
the motivations for this review.

\subsection{CO Excitation}
\label{subsec:intro_excitation}

Under average molecular cloud conditions, CO molecules are excited
through a combination of collisions with \htwo\ and radiative
trapping. They de-excite through spontaneous emission and collisions,
except at very high densities where collisions are extremely
frequent. Neglecting the effect of radiative trapping, radiative and
collisional de-excitation will balance for a critical density
$n_{cr,J}\equiv A_J/\gamma_J(T_{kin})$ (neglecting the effects of
stimulated emission), where $T_{kin}$ is the kinetic gas temperature.
Thus, for $n\gg n_{cr,J}$ and excitation temperatures $T_{ex,J}\gg
E_J/k\approx 5.53\, J (J+1)/2$~K the upper level of the $J\rightarrow
J-1$ transition will be populated and the molecule will emit
brightly. In these expressions $A_J$ is the Einstein coefficient for
spontaneous emission (only transitions with $|\Delta J|=1$ are
allowed), $A_J = 64 \pi^4 \nu_J^3 \mu^2 J/(3 h c^3 g_J)$
($A_1\approx7.11\times10^{-8}$~s$^{-1}$).  The parameter $\gamma_J(T)$
is the corresponding collisional coefficient 
(the sum of all collisional rate coefficients for transitions with
upper level $J$), which is a weak function of temperature. For CO,
$\gamma_1\sim3.26\times10^{-11}$~cm$^{3}$~s$^{-1}$ for collisions with
\htwo\ at $T_{kin}\approx30$~K
\citep[][]{YANG2010}. $T_{ex,J}$ refers to the excitation
temperature, defined as the temperature needed to recover the relative
populations of the $J$ and $J-1$ levels from the Boltzmann
distribution. In general, $T_{ex, J}$ will be different for different
transitions.

The critical density for the CO \jone\ transition is
$n(\htwo)_{cr,1}\sim2200$~\percmcu .  Higher transitions require rapidly
increasing densities and temperatures to be excited, as $n_{cr,J}\propto
J^3$ and $E_J\propto J^2$. The high optical depth of the CO emission
relaxes these density requirements, as radiative trapping
reduces the effective density required for excitation by a factor
$\sim 1/\tau_J$ (the precise factor corresponds to an escape probability and
is dependent on geometry).

The Rayleigh-Jeans brightness temperature, $T_J$, measured by a radio
telescope for the $J\rightarrow J-1$ transition will be

\begin{equation}
T_J  \approx 5.53\, J \left({1-e^{-\tau_J}}\right)\left({\frac{1}{e^\frac{5.53\,J}{T_{ex,J}}-1}-\frac{1}{e^\frac{5.53\,J}{2.73(z+1)}-1}}\right)\ {\rm K} ~.
\label{eq:RJ_brightness}
\end{equation}

\noindent The final term accounts for the effect of the Cosmic 
Microwave Background at the redshift, $z$, of interest. Note that
frequently the Rayleigh-Jeans brightness temperature is referred to as
the radiation temperature. Observations with single-dish telescopes
usually yield antenna temperatures corrected by atmospheric
attenuation ($T_A^*$), or main beam temperatures ($T_{MB}$), such that
$T_{MB}=\eta_{MB}\,T_A^*$ where $\eta_{MB}$ is the main beam
efficiency of the telescope at the frequency of the observation
\citep[see][for further discussion]{KUTNER1981}. The Rayleigh-Jeans
brightness temperature $T_J$ is identical to $T_{MB}$ for compact
sources, while extended sources may couple to the antenna with a
slightly different efficiency.

Extragalactic results, and measurements with interferometers, are
frequently reported as flux densities rather than brightness
temperatures. The relations between flux density (in Jy) and
Rayleigh-Jeans brightness temperature (in K) in general, and for CO
lines, are

\begin{equation}
S_{J} \cong 73.5\times10^{-3}\, \lambda^{-2}\,\theta^2\, T_J \approx 10.9\times10^{-3} \,J^2\, \theta^2\, T_J \ ,
\label{eq:fluxT}
\end{equation}

\noindent where $\theta$ is the half-maximum at full width of the
telescope beam (in arcsec), and $\lambda$ is the wavelength of a transition
(in mm).

The Rayleigh-Jeans brightness temperature, $T_{J}$, excitation
temperature, $T_{ex,J}$, and kinetic temperature, $T_{kin}$, are
distinct but related. The $^{12}$CO transitions usually have
$\tau_J\gg1$, making the Rayleigh-Jeans brightness temperature a probe
of the excitation temperature, $T_J\sim T_{ex,J}$ for $T_{ex,J}\gg
5.53 J$~K. In general $T_{ex,J}$ can be shown to be in the range
$T_{cmb}\leq T_{ex,J} \leq T_{kin}$. At densities much higher than
$n_{cr,J}$ the population of the levels $J$ and lower will approach a
Boltzmann distribution, and become ``thermalized'' at the gas kinetic
temperature, $T_{ex,J}\approx T_{kin}$. The corresponding
Rayleigh-Jeans brightness can be computed using
Eq. \ref{eq:RJ_brightness}. When $T_{ex,J}<T_{kin}$, usually
$T_{ex,J}/T_{ex,J-1}<1$ for lines arising in {\em the same} parcel of
gas, and the excitation of the $J$ level is ``subthermal'' (note that
this {\em is not} equivalent to $T_J/T_{J-1}<1$, as is sometimes used
in the literature).



\section{Theoretical Basis} 
\label{sec:intro_theoretical}

At its core the \xco\ factor represents a valiant effort to use the
bright but optically thick transition of a molecular gas impurity to
measure total molecular gas masses. How and why does \xco\ work?

\subsection{Giant Molecular Clouds}
\label{sec:gmcs}

Because the \co\ \jone\ transition is generally optically thick, its
brightness temperature is related to the temperature of the $\tau_{\rm
CO}=1$ surface, not the column density of the gas. {\em Information about
the mass of a self-gravitating entity, such as a molecular cloud, is
conveyed by its line width}, which reflects the velocity dispersion of
the emitting gas.

A simple and exact argument can be laid for virialized molecular
clouds, that is clouds where twice the internal kinetic energy equals
the potential energy.  Following \citet{SOLOMON1987}, the virial mass
$M_{vir}$ of a giant molecular cloud (GMC) in M$_\odot$ is

\begin{equation}
M_{vir}=\frac{3(5-2k)}{G(3-k)}\,R\,\sigma^2,
\label{eq:virial_mass}
\end{equation}

\noindent where $R$ is the projected radius (in pc), $\sigma$ is 
the 1D velocity dispersion (in \kmpers; $\sigma_{3D}=\sqrt{3}\sigma$),
$G$ is the gravitational constant ($G\approx1/232$
M$_\odot^{-1}$~pc~km$^2$~s$^{-2}$), and $k$ is the power-law index of
the spherical volume density distribution, $\rho(r)\propto r^{-k}$
. The coefficient in front of $R\,\sigma^2$ is only weakly dependent
on the density profile of the virialized cloud, and corresponds to
approximately 1160, 1040, and 700 for $k=0$, 1, and 2 respectively
\citep{MACLAREN1988,BERTOLDI1992}.  Unless otherwise specified, we
adopt $k=1$ for the remainder of the discussion. This expression of
the virial mass is fairly robust if other terms in the virial theorem
\citep{MCKEE1992,BALLESTEROS-PAREDES2006}, such as magnetic support,
can be neglected
\citep[for a more general expression applicable to spheroidal clouds
and a general density distribution see ][]{BERTOLDI1992}. As long as
molecular gas is dominating the mass enclosed in the cloud radius and
the cloud is approximately virialized, $M_{vir}$ will be a good
measure of the \htwo\ mass.

Empirically, molecular clouds are observed to follow a size-line width
relation \citep{LARSON1981,HEYER2009} such that approximately

\begin{equation}
\sigma=C\,R^{0.5},
\label{eq:size_linewidth}
\end{equation}

\noindent with $C\approx0.7$ km ${\rm s^{-1}}$ ${\rm pc^{-0.5}}$ 
\citep{SOLOMON1987,SCOVILLE1987,ROMAN-DUVAL2010}. 
This relation is an expression of the equilibrium supersonic
turbulence conditions in a highly compressible medium, and it is
thought to apply under very general conditions \citep[see \S2.1
of][for a discussion]{MCKEE2007}. In fact, to within our current
ability to measure these two quantities such a relation is also
approximately followed by extragalactic GMCs in galaxy disks
\citep[e.g.,][]{RUBIO1993,BOLATTO2008,HUGHES2010}. 

Note that insofar as the size dependence of
Eq. \ref{eq:size_linewidth} is close to a square root, the combination
of Eqs. \ref{eq:virial_mass} and \ref{eq:size_linewidth} yields that
$M_{vir}\propto \sigma^4$, and molecular clouds that fulfill both
relations have a characteristic mean surface density, \Sgmc, at a value related
to the coefficient of Eq. \ref{eq:size_linewidth} so that
$\Sgmc=M_{vir}/\pi R^2\approx331 C^2$ for our chosen density profile
$\rho\propto r^{-1}$. We return to the question of \Sgmc\ in the Milky 
Way in \S\ref{sec:mw_synthesis}.

Since the CO luminosity of a cloud, \Lco, is the product of its area
($\pi\,R^2$) and its integrated surface brightness
($T_B\,\sqrt{2\pi}\,\sigma$), then $\Lco = \sqrt{2\pi^3}
T_B\,\sigma\,R^2$, where $T_B$ is the Rayleigh-Jeans brightness
temperature of the emission (see \S\ref{subsec:intro_excitation}). Using
the size-line width relation (Equation \ref{eq:size_linewidth}) to
substitute for $R$ implies that $\Lco\propto T_B \sigma^5$. Employing
this relation to replace $\sigma$ in $M_{vir}\propto \sigma^4$ we
obtain a relation between $M_{vir}$ and \Lco ,

\begin{equation}
M_{vir}\approx M_{mol}\approx
200\,\left(\frac{C^{1.5}\,\Lco}{T_B}\right)^{0.8}~.
\label{eq:mass_luminosity}
\end{equation}

\noindent That is, for GMCs near virial equilibrium with approximately 
constant brightness temperature, $T_B$, we expect an almost linear relation 
between virial mass and luminosity. The numerical coefficient in
Equation \ref{eq:mass_luminosity} is only a weak function of the density profile of the cloud. Then 
using the relation between $C$ and \Sgmc\ we obtain the following expression for 
the conversion factor

\begin{equation}
\aco\equiv\frac{M_{mol}}{\Lco} \approx 6.1\, \Lco^{-0.2}\,T_B^{-0.8}\,\Sgmc^{0.6}.
\label{eq:virial_aco}
\end{equation}

Equations \ref{eq:mass_luminosity} and \ref{eq:virial_aco} rest on a
number of assumptions. We assume 1) virialized clouds with, 2) masses
dominated by \htwo\ that 3) follow the size-line width relation and 4)
have approximately constant temperature. Equation \ref{eq:virial_aco}
applies to a single, spatially resolved cloud, as \Sgmc\
is the resolved surface density.


%

We defer the discussion of the applicability of the virial theorem to
\S\ref{subsec:virialized}, and the effect
of other mass components to \S\ref{sec:other_sources}.  The assumption
of a size-line width relation relies on our understanding of the
properties of turbulence in the interstellar medium. The result
$\sigma\propto\sqrt{R}$ follows our expectations for a highly
compressible turbulent flow, with a turbulence injection scale at
least comparable to GMC sizes. The existence of a narrow range of
proportionality coefficients, corresponding to a small interval of
GMC average surface densities, is less well understood
\citep[for an alternative view on this point,
see][]{BALLESTEROS-PAREDES2011}. In fact, this narrow range could be
an artifact of the small dynamic range of the samples
\citep{HEYER2009}. Based on observations of the Galactic center
\citep{OKA2001} and starburst galaxies
\citep[e.g.,][]{ROSOLOWSKY2005}, $\Sigma_{\rm GMC}$ likely does vary
with environment. Equation \ref{eq:virial_aco} implies that any such
systematic changes in $\Sigma_{\rm GMC}$ will also lead to systematic
changes in \xco , though in actual starburst environments the picture
is more complex than implied by Equation \ref{eq:virial_aco}. Section
\ref{sec:starbursts} reviews the case of bright, dense starbursts in
detail.

This calculation also implies a dependence of \xco\ on the physical
conditions in the GMC, density and temperature. Combining
Eqs. \ref{eq:virial_mass} and \ref{eq:size_linewidth} with
$M_{vir}\propto \rho\,R^3$, where $\rho$ is the gas density, yields
$\sigma\propto\rho^{-0.5}$.  Meanwhile, because CO emission is
optically thick the observed luminosity depends on the brightness temperature,
$T_B$, as well as the line width, so that $\Lco \propto \sigma T_B$. Substituting
in the relationship between density and line width, 

\begin{equation}
\aco\propto \frac{\rho^{0.5}}{T_B}.
\label{eq:aco_density}
\end{equation}

\noindent The brightness temperature, $T_B$, will depend on the excitation
of the gas (Eq. \ref{eq:RJ_brightness}) and the filling fraction of
emission in the telescope beam, $f_b$. For high density and optical
depth the excitation temperature will approach the kinetic
temperature. Under those conditions, Eq. \ref{eq:aco_density} also
implies $\aco \propto \rho^{0.5} (f_b~T_{kin})^{-1}$. 

Thus, even for virialized GMCs we expect that the CO-to-\htwo\
conversion factor will depend on environmental parameters such as gas
density and temperature. To some degree, these dependencies may offset
each other. If denser clouds have higher star formation activity and
are consequently warmer, the opposite effects of $\rho$ and $T_B$ in
Eq. \ref{eq:aco_density} may partially cancel yielding a conversion
factor that is closer to a constant than we might otherwise expect.

We also note that relation between mass and luminosity expressed 
by Eq. \ref{eq:mass_luminosity} is not exactly linear, which is the reason 
for the weak dependence of
\aco\ on \Lco\ or $M_{mol}$ in Eq. \ref{eq:virial_aco}. As a consequence,
even for GMCs that obey this simple picture, \aco\ will depend (weakly) on the mass of
the cloud considered, varying by a factor of $\sim 4$ over 3 orders
of magnitude in cloud mass.

\subsection{Galaxies}
\label{sec:intro_galaxies}

This simple picture for how the CO luminosity can be used to estimate
masses of individual virialized clouds is not immediately applicable
to entire galaxies. An argument along similar lines, however, can be
laid out to suggest that under certain conditions there should be an
approximate proportionality between the integrated CO luminosity of
entire galaxies and their molecular mass. This is known as the
``mist'' model, for reasons that will become clear in a few
paragraphs.  Following \cite{DICKMAN1986}, the luminosity due to an
ensemble of non-overlapping CO emitting clouds is $\Lco\propto \sum_i
a_i\,T_B(a_i)\,\sigma_i$, where $a_i$ is the area subtended by cloud
$i$, and $T_B(a_i)$ and $\sigma_i$ are its brightness temperature and
velocity dispersion, respectively. Under the assumption that the
brightness temperature is mostly independent of cloud size, and that
there is a well-defined mean, $T_B$, then $T_B(a_i)\approx T_B$.  We
can rewrite the luminosity of the cloud ensemble as
$\Lco\approx \sqrt{2\pi} T_B\,N_{clouds}\,<\pi\,R_i^2\sigma(R_i)>$,
where the brackets indicate expectation value, $N_{clouds}$ is
the number of clouds within the beam, and we have used
$a_i=\pi\,R_i^2$. Similarly, the total mass of gas inside the beam is
$M_{mol}\approx N_{clouds}\,<4/3\pi\,R_i^3\rho(R_i)>$,
where $\rho(R_i)$ is the volume density of a cloud of radius $R_i$.
Using our definition from Eq. \ref{eq:aco} and dropping the $i$
indices, it is then clear that

\begin{equation}
\aco\equiv\frac{M_{mol}}{\Lco}\approx \sqrt{\frac{8}{9\pi}}\,\frac{<R^3\rho(R)>}{T_B\,<R^2\sigma(R)>}.
\label{eq:aco_ensemble}
\end{equation}

If the individual clouds are virialized they will follow Eq. \ref{eq:virial_mass},
or equivalently $\sigma=0.0635\,R\sqrt{\rho}$. Substituting into Eq. \ref{eq:aco_ensemble} we find

\begin{equation}
\aco\propto\frac{<R^3\rho>}{T_B\,<R^3\rho^{0.5}>},
\end{equation}

\noindent which is analogous to Eq. \ref{eq:aco_density} (obtained for 
individual clouds). As \cite{DICKMAN1986} discuss, it is possible to
generalize this result if the clouds in a galaxy follow a
size-line width relation and they have a known distribution of
sizes. Assuming individually virialized clouds, and using 
the size-line width relation (Eq. \ref{eq:size_linewidth}), we
can write down Eq. \ref{eq:aco_ensemble} as

\begin{equation}
\aco = 7.26 \frac{\sqrt{\Sgmc}}{T_B} \frac{<R^2>}{<R^{2.5}>},
\label{eq:aco_galaxy}
\end{equation}

\noindent where we have introduced the explicit dependence of
the coefficient of the size-line width relation on the cloud surface
density, \Sgmc. This equation is the analogue of
Eq. \ref{eq:virial_aco}.

In the context of these calculations, CO works as a molecular mass
tracer in galaxies because its intensity is proportional to the number
of clouds in the beam, and because through virial equilibrium the
contribution from each cloud to the total luminosity is approximately
proportional to its mass, as discussed for individual GMCs. This is
the essence of the ``mist'' model: although each particle (cloud) is
optically thick, the ensemble acts optically thin as long as the
number density of particles is low enough to avoid shadowing each
other in spatial-spectral space.

Besides the critical assumption of non-overlapping clouds, which could
be violated in environments of very high density leading to ``optical
depth'' problems that may render CO underluminous, the other key
assumption in this model is the virialization of individual clouds,
already discussed for GMCs in the Milky Way. The applicability of a
uniform value of \aco\ across galaxies relies on three assumptions
that should be evident in Eq. \ref{eq:aco_galaxy}: a similar value for
\Sgmc, similar brightness temperatures for the CO emitting gas, and a
similar distribution of GMC sizes that determines the ratio of the
expectation values $<R^2>/<R^{2.5}>$. Very little is known
currently on the distribution of GMC sizes outside the Local Group
\citep{BLITZ2007,FUKUI2010}, and although this is a potential source of 
uncertainty in practical terms this ratio is unlikely to be 
the dominant source of galaxy-to-galaxy variation in \aco.

\subsection{Other Sources of Velocity Dispersion}
\label{sec:other_sources}

Because CO is optically thick, a crucial determinant of its luminosity
is the velocity dispersion of the gas, $\sigma$. In our discussion for
individual GMCs and ensembles of GMCs in galaxies we have assumed that
$\sigma$ is ultimately determined by the sizes (through the size-line
width relation) and virial masses of the clouds. It is especially
interesting to explore what happens when the velocity dispersion of
the CO emission is related to an underlying mass distribution that
includes other components besides molecular gas. Following the
reasoning by \citet{DOWNES1993} \citep[see also][]{MALONEY1988,DOWNES1998} and
the discussion in \S\ref{sec:gmcs}, we can write a cloud luminosity \Lco\ 
for a fixed $T_B$ as $L_{\rm CO}^* = \Lco\,\sigma^*/\sigma$,
where the asterisk indicates quantities where the 
velocity dispersion of the gas is increased by other mass components,
such as stars. Assuming that both the molecular gas and the total
velocity dispersion follow the virial velocity dispersion due to a
uniform distribution of mass, $\sigma=\sqrt{GM/5R}$, then
$L_{\rm CO}^* = \Lco \sqrt{M^*/M_{mol}}$, 
where $M^*$ represents the total mass within 
radius $R$. Substituting $\Lco=M_{mol}/\aco$ yields the result
$\aco L_{\rm CO}^* = \sqrt{M_{mol}\,M^*}$.

Therefore, the straightforward application of \aco\ to the observed
luminosity $L_{\rm CO}^*$ will yield an overestimate of the molecular gas
mass, which in this simple reasoning will be the harmonic mean of the
real molecular mass and the total enclosed mass. If the observed
velocity dispersion is more closely related to the circular velocity,
as may be in the center of a galaxy, then $\sigma^*\approx\sqrt{GM^*/R}$
and the result of applying \aco\ to $L_{\rm CO}^*$ will be an even larger
overestimate of $M_{mol}$. The appropriate value of the CO-to-\htwo\
conversion factor to apply under these circumstances in order to
correctly estimate the molecular mass is

\begin{equation}
\alpha_{\rm CO}^* = \frac{M_{mol}}{\Lco^*} = \aco\,\sqrt{\frac{M_{mol}}{{\cal K} M^*}}
\label{eq:aco_star}
\end{equation}

\noindent where ${\cal K}$ is a geometrical correction factor accounting for the
differences in the distributions of the gas and the total mass, so
that ${\cal K}\equiv(\sigma^*/\sigma)^2$. In the extreme case of a
uniform distribution of gas responding to the potential of a rotating
disk of stars in a galaxy center, ${\cal K}\sim 5$. Everything else
being equal, in a case where $M^*\sim 10\, M_{mol}$ the straight
application of a standard \aco\ in a galaxy center may lead to
overestimating $M_{mol}$ by a factor of $\sim 7$. 

Note that for this correction to apply the emission has to be
optically thick throughout the medium. Otherwise any increase in line
width is compensated by a decrease in brightness, keeping the
luminosity constant. Thus this effect is only likely to manifest
itself in regions that are already rich in molecular gas.
Furthermore, it is possible to show that an ensemble of virialized
clouds that experience cloud-cloud shadowing cannot explain a lower
\xco, simply because there is a maximum attainable
luminosity. Therefore we expect \xco\ to drop in regions where the CO
emission is extended throughout the medium, and not confined to
collections of individual self-gravitating molecular clouds. This
situation is likely present in ultra-luminous infrared galaxies
(ULIRGs), where average gas volume densities are higher than the
typical density of a GMC in the Milky Way, suggesting a pervading
molecular ISM \citep[e.g.,][]{SCOVILLE1997}. Indeed, the reduction of \xco\ in
mergers and galaxy centers has been modeled in detail by
\citet{SHETTY2011b} and \citet{NARAYANAN2011,NARAYANAN2012}, and directly observed (\S\ref{sec:normgal_dust},\S\ref{sec:normgal_lines},
\S\ref{sec:sb_LIRGs}, \S\ref{sec:sb_ULIRGs})

\subsection{Optically Thin Limit}
\label{sec:optically_thin}

Although commonly the emission from $^{12}$CO \jone\ transition is
optically thick, under conditions such as highly turbulent gas motions
or otherwise large velocity dispersions (for example stellar outflows
and perhaps also galaxy winds) then emission may turn optically thin.
Thus it is valuable to consider the optically thin limit on the
value of the CO-to-\htwo\ conversion factor. Using Eq. \ref{eq:tauco},
the definition of optically thin emission ($I_J=\tau_J \left[
{B_J(T_{ex})-B_J(T_{cmb})}\right]$, where $B_J$ is the Planck function
at the frequency $\nu_J$ of the $J\rightarrow J-1$ transition,
$T_{ex}$ is the excitation temperature, and $T_{cmb}$ is the
temperature of the Cosmic Microwave Background), and the definition of
antenna temperature $T_J$, $I_J=(2 k \nu_J^2/c^2)\, T_J$, the
integrated intensity of the $J\rightarrow J-1$ transition can be
written as

\begin{equation}
\wco = T_J \Delta v = \frac{8\pi^3\nu_J}{3 k} \mu^2 \frac{J}{g_J} f_{cmb} N_J.
\label{eq:wco}
\end{equation}

\noindent The factor $f_{cmb}$ accounts for the effect of the Cosmic 
Microwave Background on the measured intensity,
$f_{cmb} = 1 - (e^{h \nu_J/k T_{ex}}-1)/(e^{h \nu_J/ k T_{cmb}}-1)$.
Note that $f_{cmb}\sim1$ for $T_{ex} \gg T_{cmb}$.

The column density of \htwo\ associated with this integrated intensity
is simply
$N(\htwo)=\frac{1}{Z_{\rm CO}} \sum_{J=0}^{\infty} N_J$,
where $Z_{\rm CO}$ is the CO abundance relative to molecular 
hydrogen, $Z_{\rm CO}={\rm CO/\htwo}$.
For a Milky Way gas phase carbon abundance, and assuming all gas-phase
carbon is locked in CO molecules, $Z_{\rm CO}\approx3.2\times10^{-4}$
\citep{SOFIA2004}. Note, however, that what matters is the 
integrated $Z_{\rm CO}$ along a line-of-sight, and CO may become
optically thick well before this abundance is reached (for example,
Fig. \ref{fig:COabundances}). Indeed, \citet{SHEFFER2008} analyze $Z_{\rm CO}$
in Milky Way lines-of-sight, finding a steep $Z_{\rm
CO}\approx4.7\times10^{-6} (N(\htwo)/10^{21}~\percmsq)^{2.07}$ for
$N(\htwo)>2.5\times10^{20}$~\percmsq, with an order of
magnitude scatter \citep[see also][]{SONNENTRUCKER2007}.

When observations in only a couple of transitions
are available, it is useful to assume local thermodynamic equilibrium
applies (LTE) and the system is described by a Boltzmann distribution
with a single temperature. In that case the column density will be
$N({\rm CO}) = Q(T_{ex})e^{E_1/k T_{ex}} N_1/g_1$,
where $E_1$ is the energy of the $J=1$ state 
($E_1/k\approx 5.53$~K for CO), and $Q(T_{ex})=\sum_{J=0}^{\infty} g_J
e^{-E_J/kT_{ex}}$ corresponds to the partition function at temperature
$T_{ex}$ which can be approximated as $Q(T_{ex})\sim 2 k T_{ex}/E_1$
for rotational transitions when $T_{ex}\gg 5.5$~K
\citep[note this is accurate to $\sim10\%$ even down 
to $T_{ex}\sim8$~K]{PENZIAS1975}. Using Eq. \ref{eq:wco} we can then write

\begin{equation}
\xco = \frac{N(\htwo)}{\wco} \approx \frac{1}{Z_{\rm CO}} \frac{6 h}{8 \pi^3 \mu^2 f_{cmb} E_1/k} \left(\frac{T_{ex}}{E_1/k}\right) e^\frac{E_1/k}{T_{ex}}.
\label{eq:thinxco}
\end{equation}

\noindent Consequently, adopting $Z_{\rm CO}=10^{-4}$ and using a representative $T_{ex}=30$~K, we obtain

\begin{equation}
\xco\approx1.6\times10^{19}\,\frac{T_{ex}}{30 {\rm K}}\, 
e^{\frac{{\rm 5.53 K}}{T_{ex}} - 0.184} \ \xcounits,
\end{equation}

\noindent or $\aco\approx0.34$~\acounits. 
These are an order of magnitude smaller than the typical values of
\xco\ and
\aco\ in the Milky Way disk, as we will discuss in
\S\ref{sec:milky_way}. Note that they are approximately 
linearly dependent on the assumed $Z_{\rm CO}$ and $T_{ex}$ (for
$T_{ex}\gg5.53$~K).  For a similar calculation that also includes an
expression for non-LTE, see
\citet{PAPADOPOULOS2012}. 


\subsection{Insights from Cloud Models}
\label{sec:Cloud_Models}

A key ingredient in further understanding \xco\ in molecular clouds is
the structure of molecular clouds themselves, which plays an important
role in the radiative transfer. This is important both for the
photodissociating and heating ultraviolet radiation, and for the
emergent intensity of the optically thick CO lines.

The CO \jone\ transition arises well within the photodissociation
region (PDR) in clouds associated with massive star formation, or
even illuminated by the general diffuse interstellar radiation field
\citep{MALONEY1988,WOLFIRE1993}.
At those depths gas heating is dominated by the grain photoelectric
effect whereby stellar far-ultraviolet photons are absorbed by dust
grains and eject a hot electron into the gas. The main parameter
governing grain photoelectric heating is the ratio $\chi
T_{kin}^{0.5}/n_e$, where $\chi$ is a measure of the far-ultraviolet field
strength, and $n_e$ is the electron density. This process will
produce hotter gas and higher excitation in starburst galaxies.
At the high densities of extreme starbursts, the gas temperature and
CO excitation may also be enhanced by collisional coupling between
gas and warm dust grains.

Early efforts to model the CO excitation and luminosity in molecular clouds
using a large velocity gradient model were carried out by
\cite{GOLDREICH1974}. The CO luminosity-gas mass relation was investigated
by \cite{KUTNER1985} using microturbulent models, 
and by \cite{WOLFIRE1993} using both microturbulent and macroturbulent
models.  In microturbulent models the gas has a 
(supersonic) isotropic turbulent velocity field with scales
smaller than the photon mean free path.
In the macroturbulent case the scale size of the turbulence is much
larger than the photon mean free path, and the emission arises from
separate Doppler shifted emitting elements.  Microturbulent models
produce a wide range of CO \jone\ profile shapes, including centrally
peaked, flat topped, and severely centrally self-reversed, while most
observed line profiles are centrally peaked.  Macroturbulent models,
on the other hand, only produce centrally peaked profiles if
there are a sufficient number of ``clumps'' within the beam with
densities $n\gtrsim10^3$~\percmcu\ in order to provide the peak
brightness temperature. 
\cite{FALGARONE1994} demonstrated that a turbulent velocity field can
produce both peaked and smooth line profiles, much closer to
observations than macroturbulent models.

%

\cite{WOLFIRE1993} use
PDR models in which the chemistry and thermal balance was calculated
self-consistently as a function of depth into the cloud.  The
microturbulent PDR models are very successful in matching and
predicting the intensity of low-J CO lines and the emission from many
other atomic and molecular species \citep{HOLLENBACH1997,
HOLLENBACH1999}.  For example, the nearly constant ratio of \cii/CO
\jone\ observed in both Galactic and extragalactic sources
\citep{CRAWFORD1985, STACEY1991} was first explained by PDR models as
arising from high density ($n\ag 10^3$ ${\rm cm^{-3}}$) and high UV
field ($\chi \ag 10^3$) sources in which both the \cii\ and CO are
emitted from the same PDR regions in molecular cloud surfaces.  These
models show that the dependence on CO luminosity with incident
radiation field is weak. This is because as the field
increases the $\tau_{\rm CO}=1$ surface is driven deeper into the
cloud where the dominant heating process, grain photoelectric heating,
is weaker. Thus the dissociation of CO in higher fields regulates the
temperature where $\tau_{\rm CO}=1$.  

%

PDR models have the advantage that they calculate the thermal and
chemical structure in great detail, so that the gas temperature is
determined where the CO line becomes optically thick. We note however,
that although model gas temperatures for the CO \jone\ line are
consistent with observations the model temperatures are typically too
cool to match the observed high-J CO line emission
\citep[e.g.,][]{HABART2010}.  The model density structure is generally
simple (constant density or constant pressure), and the velocity is
generally considered to be a constant based on a single microturbulent
velocity. More recent dynamical models have started to combine
chemical and thermal calculations with full hydrodynamic simulations.

\section{A modern theoretical perspective}
\label{sec:theoretical}

In recent years there has been much progress in updating  PDR models,
and in combining hydrodynamic
simulations, chemical modeling, and radiation transfer codes.
 
\subsection{Photodissociation Regions} 
\label{subsec:PDRs}


 
Since CO production mainly occurs through ion-neutral chemistry
\citep{VANDISHOECK1988} the ionization structure through the PDR 
is important in setting the depth of the CO formation. Recombination
of metal ions on PAHs can modify the abundance of free electrons. For
a typical GMC with $n\sim 10^3$ ${\rm cm^{-3}}$ illuminated by a
radiation field $\sim 10$ times the Galactic ISRF (the interstellar
radiation field in the vicinity of the Sun), the $\tau_{\rm CO}=1$
surface is at a depth of $\av\sim 1$ when PAHs are included. Without
PAHs the electron abundance stays high, and the ion-neutral chemistry
is slowed due to ${\rm H_3^+}$ recombination. Consequently, the
$\tau_{\rm CO}=1$ surface is pushed deeper into the cloud ($\av\sim
2$).  The PAH rates are estimated by
\cite{WOLFIRE2008} by considering the ${\rm C^0/C^+}$ ratio in diffuse
lines of sight, but there is considerable uncertainty both in the rates
\citep{LISZT2011} and in the PAH abundance and their variation with
cloud depth.


\cite{BELL2006} carry out a parameter study of \xco\ using
PDR models with constant microturbulent line width.
The calculated \xco\ versus \av\ plots have a characteristic shape with
high \xco\ at low \av, dropping to a minimum value at $\av\sim 2-4$
and then slowly rising for increasing \av. The \xco\ dependence at low
\av\ arises from molecular gas with low CO abundance. 
After the CO line intensity becomes
optically thick, \xco\ slowly rises again as $N({\rm H_2})$
increases. 
Increasing density up to the critical density
of CO \jone\ ($n(\htwo)_{cr,1}\sim 2200$~\percmcu) 
enhances the CO excitation and causes \xco\ to drop. 
In addition, the  minimum moves closer to the cloud surface
as does the $\tau_{\rm CO}=1$ surface. An
increase in \xco\ by a factor of more than 100 is found when
decreasing dust and metallicity to 1\% of the local Galactic ISM,
reflecting the larger column of \htwo\ at a given \av.  
\cite{BELL2007} suggest that the appropriate \xco\ value to use in
various extragalactic environments can
be estimated from the minimum in the \xco\ versus $A_V$ plot.
 

\cite{WOLFIRE2010} use an updated version of their PDR models, including a self-consistent
calculation for the median density expected from a turbulent density
distribution. The models provide a theoretical basis for predicting
the molecular mass fraction outside of the CO emitting region in terms
of incident radiation field and gas metallicity. They assume that the
median density is given by $\langle n \rangle_{\rm
med}=\bar{n}\exp(\mu)$, where $\bar{n}$ is the volume averaged density
distribution $\bar{n}\propto 1/r$, and $\mu = 0.5\ln (1+0.25{\cal
M}^2)$ \citep{PADOAN1997}.  The sound speed that enters in the Mach
number, ${\cal M}$, is calculated from the PDR model output while the
turbulent velocity is given by the size-linewidth relation (Eq.\
\ref{eq:size_linewidth}). 

\subsection{Numerical Simulations}
\label{subsec:numerical}

As an alternative to PDR models with simple geometries and densities,
hydrodynamical models can be used to calculate the line width, and
density, but with only limited spatial resolution and approximate
chemistry and thermal balance.
\cite{GLOVER2007,GLOVER2007b} carry out simulations of the
formation of molecular clouds using a modified version of the 
magnetohydrodynamical code ZEUS-MP. They include a time-dependent 
chemistry, in particular for ${\rm H_2}$ formation and destruction, and 
thermal balance. \cite{GLOVER2010} enhance the code to account for
CO chemistry in a turbulent GMC. A  turbulent
picture of a GMC is not one with well defined clumps surrounded by
an interclump medium but one with a continuous density distribution and
constant mixing between low and high density regions. The CO abundance
varies within the cloud depending on the gas density and penetration
of the dissociating radiation field.
Calculations of \xco\ are carried out by
\cite{GLOVER2011}, and \cite{SHETTY2011,SHETTY2011b}.
The latter two references include non-LTE CO excitation and line
transfer in the LVG approximation. These authors carry out 3D
turbulent simulations in a box of fixed size (20~pc), and turbulence
generated with uniform power between wavenumbers $1 \le k \le 2$, but
with various initial densities and metallicities. For their standard
cloud model the saturation amplitude of the 1D velocity dispersion is
2.4 km ${\rm s^{-1}}$. Each line of sight has a different CO
intensity, and CO and \htwo\ column density depending on the past and
present physical conditions along it. Thus, for a given $N(\htwo)$
there is a range in
\xco\ values \citep[see Figs.\ 5 and 6 in][]{SHETTY2011}.  The
dispersion generally increases for clouds with lower density and lower
metallicity.

Figure 
7 in \cite{SHETTY2011} shows the calculated mean \xco\ in different
$A_V$ bins for several initial densities and metallicities. The
variation in \xco\ with \av\ is qualitatively similar to that shown in
\cite{BELL2006} for microturbulent models.
Lower densities and metallicities
drive \xco\ to higher values due to lower CO excitation and CO/\htwo\
ratios respectively.  At $\av\gtrsim7$ the models with different initial
densities converge as the CO line becomes optically thick. The minimum in \xco\
occurs at larger \av\ in the hydrodynamic simulations compared to
the one sided PDR models, likely due to the dissociating radiation incident
on all sides of the box and due to its greater penetration along low
density lines of sight. 
A range in density $n=100-1000$ ${\rm cm^{-3}}$ and metallicity
$Z=0.1-1$ ${Z_\odot}$ produces a range of only $\xcot\sim2-10$.  Thus
although there can be large variations in \xco\ along different lines
of sight in the cloud the emission weighted
\xco\ is close to the typical Galactic value (\S\ref{sec:milky_way}). In addition, the low metallicity
case has higher \xco\ at low \av\ but approaches the Galactic value
for higher density (higher \av) lines of sight. Although the models
were run with constant box size, the dependence on \av\ suggests
that clouds with sufficiently small size so that CO does not become
optically thick will have higher \xco. The results
seem to confirm the suggestion by \cite{BELL2006} that the mean \xco\
value should be near the minimum when plotted as \xco\ versus \av.


\cite{SHETTY2011b} investigate the results of varying the temperature,
CO abundance, and turbulent line width. They find only a weak
dependence on temperature with $\xco\propto T_{kin}^{-0.5}$ for
20~K$<T_{kin}<$100~K, and thus over the range of temperatures typically
found in Galactic clouds \xco\ is not expected to vary significantly
due to temperature. At low (fixed) CO abundance ($n_{\rm CO}/n_{\rm
H2}\sim 10^{-6}$) , the line does not become optically thick and thus
${\rm W(CO)}$ follows $N(\htwo)$ with constant \xco\ up to at least
$\log N({\rm H_2}) = 22.5$. Varying the turbulent line width increases
the CO line intensity as expected since decreasing the self-absorption
allows more CO line emission to escape the cloud. They also find,
however, a decreasing brightness temperature and over the range
of velocity dispersion $\sigma = 2-20$ km ${\rm s^{-1}}$, $\wco$ changes
as $\wco \propto \sigma^{0.5}$
instead of changing linearly. The higher turbulent velocities create dense shocks
but also larger voids of low density material.  The competing effects
produce a dependence on $\sigma$ that is slower than
linear. Increasing the column density as well as the line width might
produce a model result closer to $\wco \propto \sigma$.

An interesting result is that \xco\ is not sensitive to the internal
velocity profile within the cloud but only to the total line width,
however that might be generated. A cloud need not obey a power law
size-linewidth relation to have the same \xco\ as one with a Gaussian
distribution and identical dispersion velocity. Thus, clouds need not
be ``virialized'' nor obey a size-linewidth relation to have the same
\xco. 
The modeled \xco\ converges for high \av\ clouds and thus
the \xco\ factor is not expected to vary from cloud to cloud as long
as the there is a large enough column density.
\cite{SHETTY2011b} conclude that a nearly constant \xco\ 
is the result of the limited range in column densities, temperatures,
and linewidths found in Galactic molecular clouds and that applying a
constant \xco\ is approximately correct to within a factor of $\sim2$.

Global  models of \xco\ using hydrodynamic simulations to investigate
 the variation due to  galactic environment are carried out by 
\cite{FELDMANN2012} and \cite{NARAYANAN2011, NARAYANAN2012}.
A significant problem in numerical simulations is how to handle the
physical conditions and/or line emission within regions smaller than
the spatial grid. \cite{FELDMANN2012} use high resolution ($\sim
0.1$~pc) simulations from \cite{GLOVER2011} for ``sub-grid'' solutions
to cosmological simulations of 60~pc resolution.  These models used
constant $T_{kin}=10$~K, LTE excitation for CO, with either constant CO line
width or one proportional to $\Smol^{0.5}$.  By comparing results at
their highest resolutions with those at 1~kpc, and 4~kpc, they asses
the effects of spatial averaging on \xco.  The averaging tends to
reduce the variation of \xco\ on $N(\htwo)$ and UV radiation field
intensity. At greater than kiloparsec scales, and \htwo\ column
densities between $10^{21}$~\percmsq\ and $10^{23}$~\percmsq, \xco\
changes by only a factor of 2. They find essentially no variation in
\xco\ with UV field strength between 0.1 and 100 times the Galactic
ISRF.
\cite{FELDMANN2012} do find a significant variation with metallicity,
with a scaling \xco$\propto Z^{-0.5}$ for virial line widths.  Note
that this relation is much shallower than found by, for example,
\citet[][\S\ref{subsec:HighRedshift}]{GENZEL2012}.  The authors suggest
that low metallicity high-redshift galaxies may not obey the same gas
surface-density to star formation relation observed in local disks.

The results of 
\cite{NARAYANAN2011} and \cite{NARAYANAN2012} will be discussed 
in \S\ref{sec:sb_ULIRGs}. Here we note that their effective
resolution of $\sim 70$~pc requires adopting sub-grid cloud
properties. Although the surfaces of GMCs more massive than $\sim
7\times 10^5$ $M_\odot$ are resolved, their internal structure is
not. The resulting line and continuum transfer, and temperature and
chemical structure can only be approximate. The resolution problem
also enters in simulations of individual clouds.  We showed in
\S\ref{sec:introduction} that CO becomes optically thick
within a column $N(\htwo)\approx2-3\times 10^{20}$~\percmsq.  Thus for
sub parsec resolution ($\sim 0.1$~pc) and density greater than
$n\gtrsim10^3$~\percmcu, the physical conditions (temperature,
density, and abundances) are averaged over the line forming region and
the calculated emitted intensity can be in error.  The resolution
problem is much more severe for \htwo\ dissociation. Self-shielding of
\htwo\ starts within a column of only $N(\htwo)\sim 10^{14}$~\percmsq\
\citep{DEJONG1980}. Thus with a resolution of $\sim0.1$~pc, and
density of $n\gtrsim10$~\percmcu, the optical depth to dissociating
radiation is already $\tau\gtrsim10^4$ in a resolution element. A
complementary approach is to apply a state-of-the-art PDR code with
well resolved
\htwo\ formation to the output of hydrodynamic simulations 
\citep{LEVRIER2012}. With increasing
computing power the issues of resolution will continue to improve.  We
suggest that high priority should be placed on creating large scale
galactic simulations that are well matched to small scale simulations
with resolved cloud structure.  The former provide environmental
conditions and cloud boundary conditions while the latter provide the
chemistry and line emission in a realistic turbulent cloud. Expanding
the library of GMC models with a range of column densities, line
widths, and external heating, and thoroughly checking them against
observations, would be most helpful.

%





\section{\xco\ in the Milky Way}
\label{sec:milky_way}

The Galaxy is the only source where it is possible to determine the 
CO-to-\htwo\ conversion factor in a variety of ways. It thus provides
the prime laboratory to investigate the calibration and the variations
of the proportionality between CO emission and molecular mass.

In the following sections we will discuss three types of \xco\
determinations: 1) employing virial masses, a technique that requires
the ability to spatially resolve molecular clouds to measure their
sizes and kinematics, 2) taking advantage of optically thin tracers of
column density, such as dust or certain molecular and atomic lines, and
3) using the diffuse $\gamma$-ray emission arising from the pion
production process that takes place when cosmic rays interact with
interstellar medium protons. Gamma-ray techniques are severely limited
by sensitivity, and are only applicable
to the Milky Way and the Magellanic Clouds. The good level of
agreement between these approaches in our own galaxy is the foundation
of the use of the CO-to-\htwo\ conversion factor in other galaxies.

\subsection{\xco\ Based on Virial Techniques}
\label{sec:mw_virial}

The application of the virial theorem to molecular clouds has been
discussed by a number of authors, and recently reviewed by
\citet{MCKEE2007}. Here we just briefly summarize the fundamental
points.  The virial theorem can be expressed in the Lagrangian (fixed
mass) or Eulerian (fixed volume) forms, the latter particularly
applicable to turbulent clouds where mass is constantly exchanged with
the surrounding medium. In the somewhat simpler Lagrangian form, the
virial equilibrium equation is

\begin{equation}
2(K-K_s)+B+W=0
\label{eq:virial_full}
\end{equation}

\noindent where $K$ is the volume integral of the thermal plus
kinetic energy, $K_s$ is the surface pressure term, $B$ is the net
magnetic energy including volume and surface terms (which cancel for a
completely uniform magnetic field), and $W$ is the net gravitational
energy which is determined by the self-generated gravitational
potential if the acceleration due to mass external to the cloud can be
neglected. In the simple case of a uniform, unmagnetized sphere virial
equilibrium implies $2K+W=0$.  It is useful to define the virial
parameter, $a_{vir}$, which corresponds to the ratio of total kinetic
energy to gravitational energy \citep{BERTOLDI1992}, so that
$a_{vir} \equiv 5\,R\sigma^2/GM$.

\subsubsection{Are Clouds Virialized?}
\label{subsec:virialized}

In this context, gravitationally bound objects have $a_{vir}\simeq1$.
Whether interstellar clouds are entities in virial equilibrium, even
in a time or ensemble averaged sense \citep{MCKEE1999}, is a matter of
current debate. Observational evidence can be interpreted in terms of
systems out of equilibrium with rapid star formation and subsequent
disruption in a few Myr \citep[e.g.,][]{ELMEGREEN2000}, an
evolutionary progression and a typical lifetime of a few tens of Myr,
long enough for clouds to become virialized
\citep[e.g.,][]{BLITZ1980b,FUKUI2010}, or a lifetime of hundreds of
Myr \citep[e.g.,][]{SCOVILLE1979}. 
\citet{ROMAN-DUVAL2010} find a median
$a_{vir}\approx0.5$ for clouds in the inner Galaxy, suggesting that
they are bound entities where $M_{vir}$ represents a reasonable
measure of the molecular mass, although casting doubt on the
assumption of exact virial equilibrium. \citet[][]{WONG2011} estimate
a very large scatter in $a_{vir}$ in the Large Magellanic Cloud, but
do lack an independent mass tracer so their results rest on the
assumption of a fixed \xco. Observations in the outer Galaxy show
another angle of the situation. \citet{HEYER2001} find that clouds
with $M_{mol}>10^4$ \msun\ are self-gravitating, while small clouds
with masses $M_{mol}<10^3$ \msun\ are overpressured with respect to
their self-gravity, that is, have $a_{vir}\gg1$ and are out of
equilibrium.  Given the observed mass function, however, such clouds
represent a very small fraction of the molecular mass of the Milky
Way.

In any case, observed GMC properties can be understood as a
consequence of approximate energy equipartition, which observationally
is very difficult to distinguish from virial equilibrium
\citep{BALLESTEROS-PAREDES2006}.  Clouds with an excess of kinetic
energy, $a_{vir}\gg1$, perhaps due to ongoing star formation or SNe
would be rapidly dissipated, while clouds with a dearth of kinetic
energy, $a_{vir}\ll1$, would collapse at the free-fall velocity which
is within $40\%$ of the equipartition velocity dispersion and
challenging to distinguish from turbulent motions in
observations. Furthermore, the resulting star formation will inject
energy into the cloud acting to restore the balance. Thus from the
standpoint of determining cloud masses over large samples, the
assumption of virial equilibrium even if not strictly correct, is
unlikely to be very wrong.

\subsubsection{Observational Results}

The most significant study of the relation between virial mass and
\Lco\ (the mass-luminosity relation) in the Milky Way is that 
by \citet{SOLOMON1987}, which encompasses 273 clouds and spans several
orders of magnitude in cloud luminosity and mass. It is dominated by
clouds located in the inner Galaxy, in the region of the so-called
Molecular Ring, a feature in the molecular surface density of the
Milky Way peaking at $R_{GC}\approx4$~kpc galactocentric radius. It
uses kinematic distances with an old value of the distance to the
Galactic Center, $R_\odot=10$~kpc. We
report new fits after a 0.85 scaling in all distances and sizes and
0.72 in luminosities to bring them into agreement with the modern
distance scale ($R_\odot=8.5$~kpc).  The virial mass computations assume a $\rho(r)\propto
r^{-1}$ (see
\S\ref{sec:gmcs}).

\begin{figure}[ht]
\centerline{\psfig{figure=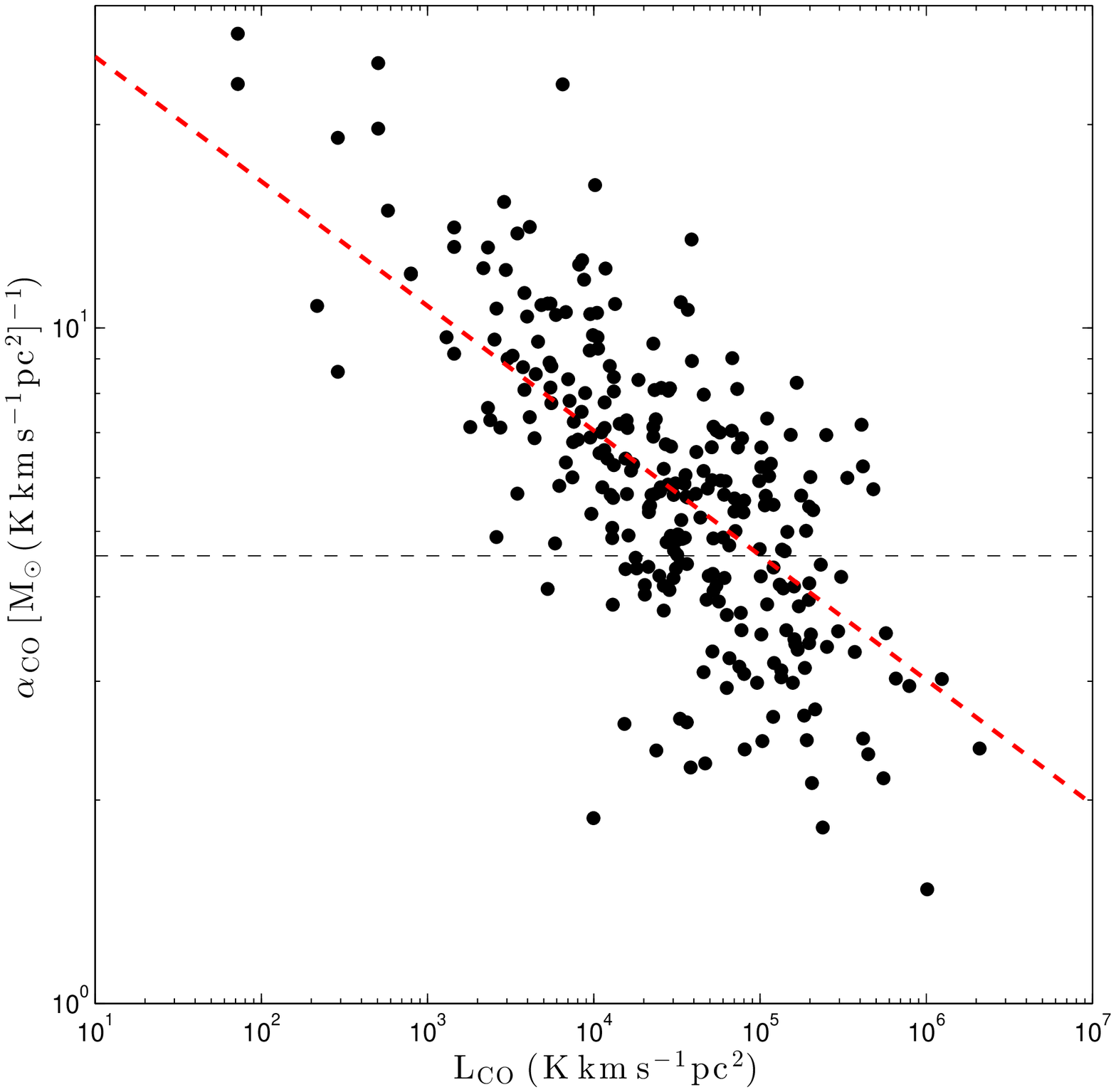,width=\columnwidth}}
\caption{Relation between virial \aco\ and CO luminosity for GMCs in the Milky Way \citep{SOLOMON1987}. We have corrected the numbers in the original table
to reflect the updated distance to the Galactic Center of 8.5~kpc. The
dependence of \aco\ on \Lco\ arises from the fact that the correlation
between $M_{vir}$ and \Lco\ has a nonlinear slope
($M_{vir}\propto\Lco^{0.815\pm0.013}$), following the expectations
from Eq. \ref{eq:mass_luminosity} for approximately constant
brightness temperature. This results in
$\aco\approx4.61\,(\Lco/10^5)^{-0.185}$, denoted by the red thick
dashed line (the dispersion around this relation is $\pm0.15$
dex). The nominal value at $\Lco=10^5$~\Lcounits\ is illustrated by
the thin black dashed line. \label{fig:solomon}}
\end{figure}

\citet{SOLOMON1987} find a very strong correlation between 
$M_{vir}$ and \Lco, such that $M_{vir}=37.9\,\Lco^{0.82}$ with a
typical dispersion of 0.11 dex for $M_{vir}$. Note the excellent
agreement with the expected mass-luminosity relation in
Eq. \ref{eq:mass_luminosity} using a typical CO brightness temperature
$T_B\approx4$~K
\citep{MALONEY1990}. For a cloud at their approximate median
luminosity, $\Lco\approx10^5$~\Lcounits, this yields
$\aco=4.6$~\acounits\ and $\xcot=2.1$. Because the relation is not
strictly linear \aco\ will change by $\sim60\%$ for an order of
magnitude change in luminosity (Fig. \ref{fig:solomon}). Therefore
GMCs with lower luminosities (and masses) will have somewhat larger
mass-to-light ratios and conversion factors than more luminous GMCs.

Independent analysis using the same survey by \citet{SCOVILLE1987}
yields a very similar mass-luminosity relation. After accounting for
the different coefficients used for the calculation of the virial
mass, the relation is $M_{vir}=33.5\,\Lco^{0.85}$. For a
$\Lco\approx10^5$~\Lcounits\ cloud this yields $\aco=6.0$~\acounits\
and $\xcot=2.8$ (this work uses $R_\odot=8.5$~kpc). Interestingly,
there is no substantial difference in the mass-luminosity relation for
GMCs with or without \hii\ regions \citep{SCOVILLE1989}, although the
latter tend to be smaller and lower mass, and have on average half of
the velocity-integrated CO brightness of their strongly star-forming
counterparts. The resulting difference in $T_B$ could have led to a
displacement in the relation, according to the simple reasoning
leading to Eq. \ref{eq:mass_luminosity}, but it appears not to be
significant.

\subsubsection{Considerations and Limitations}

Besides the already discussed applicability of the virial theorem,
there are a number of limitations to virial studies. Some are
practical, while others are fundamental to the virial technique.  On
the practical side, virial studies are sensitive to cloud definitions
and biases induced by signal-to-noise. These will impact both the
values of $R$ and $\sigma$ used to compute the mass. In noise-free
measurements isolated cloud boundaries would be defined using contours
of zero emission, when in reality it is necessary to define them using
a higher contour
\citep[for example,][ use a $T_B\sim4$~K CO brightness
contour]{SOLOMON1987}. \citet{SCOVILLE1987} discuss the impact of this
correction, studying the ``curve of growth'' for $R$ and $\sigma$ as
the definition contour is changed in high signal-to-noise
observations. Moreover, isolated clouds are rare and it is commonly
necessary to disentangle many partially blended features along the line
of sight. To measure a size clouds need to be resolved, and if
appropriate the telescope beam size needs to be deconvolved to
establish the intrinsic cloud size. This is a major concern in
extragalactic studies, but even Galactic datasets are frequently
undersampled which affects the reliability of the $R$ and \Lco\
determinations. Given these considerations, it is encouraging that two
comprehensive studies using independent analysis of the same survey
come to values of \aco\ that differ by only $\sim30\%$ for clouds of
the same luminosity.

A fundamental limitation of the virial technique is that CO needs to
accurately sample the full potential and size of the cloud. For
example, if because of photodissociation or other chemistry CO is
either weak or absent from certain regions, its velocity dispersion
may not accurately reflect the mass of the cloud. This is a particular
concern for virial measurements in low metallicity regions (see
\S\ref{sec:low_metallicities}), although most likely it is not a limitation
in the aforementioned determinations of \xco\ in the inner Galaxy.

\subsection{Column Density Determinations Using Dust and Optically Thin Lines} 
\label{sec:mw_dust}

Perhaps the most direct approach to determining the \htwo\ column
density is to employ an optically thin tracer. This tracer can
be a transition of a rare CO isotopologue or other chemical species
\citep[e.g., CH][]{MAGNANI2003}. It can also be dust, usually optically
thin in emission at far-infrared wavelengths, and used in absorption
through stellar extinction studies.

\subsubsection{CO Isotopologues}

A commonly used isotopologue is $^{13}$CO. Its abundance relative to
$^{12}$CO is down by a factor approaching the
$^{12}$C/$^{13}$C$\approx69$ isotopic ratio at the solar circle
($^{12}$C/$^{13}$C$\approx50$ at $R_{GC}\approx4$~kpc, the
galactocentric radius of the Molecular Ring) as long as chemical
fractionation and selective photodissociation effects can be neglected
\citep{WILSON1999}. Given this abundance ratio and under the
conditions in a dark molecular cloud $^{13}$CO emission may not always
be optically thin, as $\tau_1\sim 1$ requires $\av\sim5$.

The procedure consists of inverting the observed intensity of the
optically thin tracer to obtain its column (or surface) density. In
the case of isotopologues, this column density is converted to the
density of CO using the (approximate) isotopic ratio. Inverting the
observed intensity requires knowing the density and temperature
structure along the line of sight, which is a difficult problem. If
many rotational transitions of the same isotopologue are observed, it
is possible to model the line of sight column density using a number
of density and temperature components.  In practice an approximation
commonly used is local thermodynamic equilibrium (LTE), the assumption
that a single excitation temperature describes the population
distribution among the possible levels along the line of sight. It is
also frequently assumed that $^{12}$CO and $^{13}$CO share the same
$T_{ex}$, which is particularly justifiable if collisions dominate the
excitation ($T_{ex}=T_{kin}$, the kinetic temperature of the
gas). Commonly used expressions for determining $N(^{13}{\rm CO})$
under these assumptions can be found in, for example,
\citet{PINEDA2010}.  Note, however, that if radiative trapping plays
an important role in the excitation of $^{12}$CO, $T_{ex}$ for
$^{13}$CO will generally be lower due to its reduced optical depth
\citep[e.g.,][]{SCOVILLE1987b}.

\citet{DICKMAN1978} characterized the
CO column density in over 100 lines of sight toward 38 dark clouds,
focusing on regions where the LTE assumption is unlikely to introduce
large errors. The combination of LTE column densities with estimates of
\av\ performed using star counts yields
$\av\approx(4.0\pm2.0)\times10^{-16} N(^{13}{\rm
CO})$~cm$^2$~mag. Comparable results were obtained in detailed studies
of Taurus by \citet[][note the nonlinearity in their
expression]{FRERKING1982} and Perseus by \citet{PINEDA2008}, the
latter using a sophisticated extinction determination
\citep{LOMBARDI2001}. Extinction can be converted into molecular column density,
through the assumption of an effective gas-to-dust
ratio. \citet{BOHLIN1978} determined a relation between column density
and reddening (selective extinction) such that
$[N(\hi)+2\,N(\htwo)]/E(B-V)\approx5.8\times10^{21}$~atoms~\percmsq~mag$^{-1}$
in a survey of interstellar Ly$\alpha$ absorption carried out using
the {\em Copernicus} satellite toward 75 lines of sight, mostly
dominated by \hi. For a ``standard'' Galactic interstellar extinction
curve with $R_V\equiv\av/E(B-V)=3.1$, this results in

\begin{equation}
N_H\equiv N(\hi)+2\,N(\htwo)\approx1.9\times10^{21}\,\percmsq\,\av.
\label{eq:bohlin_gdr}
\end{equation}

\noindent A much more recent study using 
{\em Far Ultraviolet Spectroscopic Explorer} observations finds
essentially the same relation \citep{RACHFORD2009}. In high surface
density molecular gas $R_V$ may be closer to $5.5$
\citep{CHAPMAN2009}, and Eq. \ref{eq:bohlin_gdr} may yield a 40\%
overestimate \citep{EVANS2009}.  Using Eq. \ref{eq:bohlin_gdr}, the
approximate relation between $^{13}$CO~\jone\ and molecular column
density is $N(\htwo)\approx 3.8\times10^5\,N(^{13}{\rm
CO})$. \citet{PINEDA2008} find a similar result in a detailed study of
Perseus, with an increased scatter for
$\av\gtrsim5$. \citet{GOLDSMITH2008} use these results together with
an averaging method to increase the dynamic range of their $^{13}$CO
and $^{12}$CO data, a physically motivated variable
$^{12}$CO/$^{13}$CO ratio, and a large velocity gradient excitation
analysis, to determine \htwo\ column densities in Taurus.  They find
that $\xcot\approx1.8$ recovers the molecular mass over the entire
region mapped, while there is a marked increase in the region of low
column density, where \xco\ increases by a factor of 5 where
$N(\htwo)<10^{21}$~cm$^{-2}$. As a cautionary note about the blind use
of \cothree\ LTE estimates, however, \citet{HEIDERMAN2010} find that
this relation between \htwo\ and $^{13}$CO underestimates $N(\htwo)$
by factors of $4-5$ compared with extinction-based results in the
Perseus and Ophiuchus molecular clouds.

\subsubsection{Extinction Mapping}

Extinction mapping by itself can be directly employed to
determine \xco. It fundamentally relies on the assumption of spatially
uniform extinction properties for the bands employed, and on the
applicability of Eq. \ref{eq:bohlin_gdr} to convert extinction into
column density.

\citet{FRERKING1982} determined $\xcot\approx1.8$
in the range $4\lesssim\av\lesssim12$ in $\rho$ Oph, while the same
authors found constant W(CO) for $\av\gtrsim2$ in
Taurus. \citet{LOMBARDI2006} studied the Pipe Nebula and found a best
fit \xco\ in the range $\xcot\approx2.9-4.2$, but only for K-band
extinctions $A_K>0.2$
\citep[equivalent to $\av>1.8$,][]{RIEKE1985}. A simple fit to the
data ignoring this nonlinearity yields
$\xcot\sim2.5$. The
\citet{PINEDA2008} study of Perseus finds
$\xcot\approx0.9-3$ over a number of regions.  The relation between CO and
\htwo, however, is most linear for $\av\lesssim4$, becoming saturated
at larger line-of-sight extinctions.

\begin{figure}[h!]
\centerline{\psfig{figure=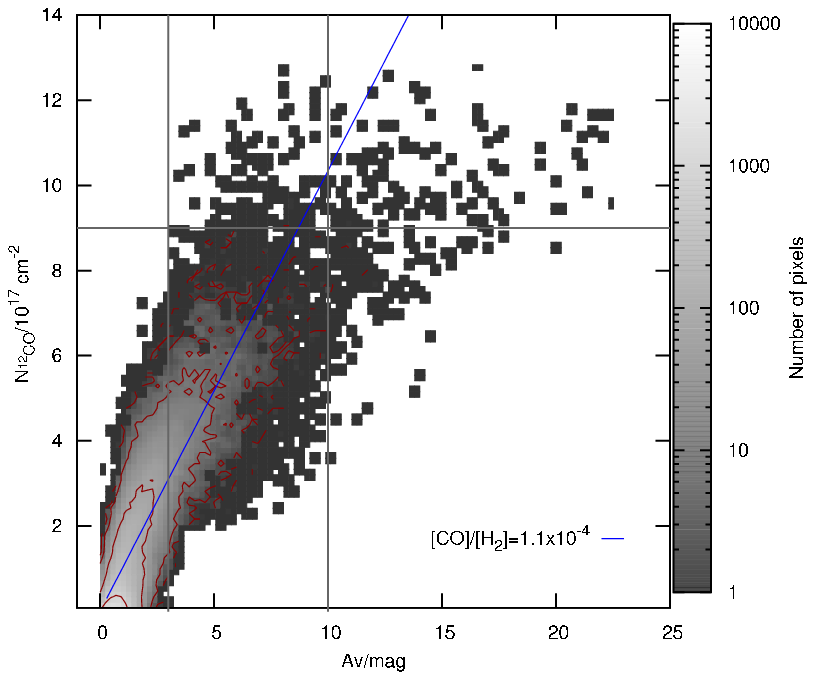,width=\columnwidth}}
\caption{Relation between CO column density and extinction
in the Taurus molecular cloud \citep{PINEDA2010}. The figure shows
the pixel-by-pixel relation between gas-phase CO column density
(obtained from \cothree) and
\av. The blue line illustrates the ``average'' linear relation for $3\lesssim\av\lesssim10$, $N(\co)\approx1.01\times10^{17}\av$~\percmsq\ (implying CO/$\htwo\approx1.1\times10^{-4}$ for the assumed isotopic ratio).
The linearity is clearly broken for $\av\gtrsim10$. \citep{PINEDA2010}
show that linearity is restored to high \av\ after applying a correction
for CO freeze-out into dust grain mantles.
\label{fig:pineda}}
\end{figure}

\citet{PINEDA2010} extend the aforementioned \citet{GOLDSMITH2008} 
study of Taurus by characterizing the relation between reddening (from
the Two Micron All Sky Survey, 2MASS) and CO column density (derived
from $^{13}$CO) to measure $\xcot\approx2.1$. They find that the
relation between \av\ and CO flattens for $\av\gtrsim10$
(Fig. \ref{fig:pineda}), a fact that they attribute to freeze-out of
CO onto dust grains causing the formation of CO and CO$_2$ ice
mantles. Including a correction for this effect results in a linear
relation to $\av\lesssim23$. For $\av\lesssim3$ the column density of
CO falls below the linear relationship, likely due to the effects of
photodissociation and chemical fractionation.
Along similar lines, \citet{HEIDERMAN2010} find that in Ophiuchus and
Perseus CO can underpredict \htwo\ with respect to \av\ for $\Smol>200$
M$_\odot$~pc$^{-2}$ by as much as $\sim30\%$.

\citet{PARADIS2012} recently used a high-latitude extinction map
derived from 2MASS data using an extension of the NICER methodology
\citep{DOBASHI2008,DOBASHI2009} to derive \xco\ in sample of nearby
clouds with $|b|>10^\circ$. They find
$\xcot\approx1.67\pm0.08$ with a somewhat
higher value $\xcot\approx2.28\pm0.11$ for
the inner Galaxy region where $|l|<70^\circ$. They report an excess
in extinction over the linear correlation between total gas and
\av\ at $0.2\lesssim\av\lesssim1.5$, suggestive of a gas phase that
is not well traced by either 21~cm or CO emission. We will return to
this in \S\ref{sec:mw_diffuse}.

\subsubsection{Dust Emission}
 \label{sec:mw_dustemis}
 
The use of extinction mapping to study $N(\htwo)$ is mostly limited to
nearby Galactic clouds, since it needs a background stellar
distribution, minimal foreground confusion, and the ability to resolve
individual stars to determine their reddening.  Most interestingly,
the far-infrared emission from dust can also be employed to map the
gas distribution.  Indeed, dust is an extraordinarily egalitarian
acceptor of UV and optical photons, indiscriminately processing them
and reemitting in the far-infrared.  In principle, the dust spectral
energy distribution can be modeled to obtain its optical depth,
$\tau_d(\lambda)$, which should be proportional to the total gas
column density under the assumption of approximately constant dust
emissivity per gas nucleon, fundamentally the product of the
gas-to-dust ratio and dust optical properties.

How valid is this assumption? An analysis
of the correlation between $\tau_d$ and \hi\ was carried out
at high Galactic latitudes by \citet{BOULANGER1996}, who found 
a typical dust emissivity per H nucleon of 

\begin{equation}
\dgr\equiv\tau_d/N_H\approx1.0\times10^{-25} (\lambda/250\,\mu{\rm m})^{-\beta}\,{\rm cm}^2, 
\label{eq:emissivity_HI}
\end{equation}

\noindent with $\beta=2$, in excellent accord with the recent value for 
high latitude gas derived using {\em Planck} observations
\citep[][who prefer $\beta=1.8$]{PLANCKCOLLABORATION2011XXIV}. They also identified a break in the correlation
for $N(\hi)\gtrsim5\times10^{20}$~\percmsq\ suggestive of an
increasingly important contribution from \htwo\ to $N_H$, in agreement
with results from {\em Copernicus} \citep{SAVAGE1977}. There is
evidence that the coefficient in Eq. \ref{eq:emissivity_HI} changes in
molecular gas. It may increase by factors of $2-3$ at
very high column densities \citep{SCHNEE2008,FLAGEY2009,PLANCKCOLLABORATION2011XXV},
likely due to grain growth or perhaps solid state effects at low
temperatures \citep[e.g.,][]{MENY2007}. Note, however, that recent
work using {\em Planck} in the Galactic plane finds
$\dgr\approx(0.92\pm0.05)\times10^{-25}$~cm$^2$ at 250~$\mu$m, with no
significant variation with Galactic radius \citep{PLANCKCOLLABORATION2011XXI}. This
\dgr\ is almost identical to that observed in dust mixed with mostly
atomic gas at high latitudes, suggesting that the aforementioned
emissivity variations are very localized.

This excellent correlation between $\tau_d$ and $N_H$ is the basis
for a number of studies that use dust emission to determine \htwo\ 
column densities. Most notably, \citet{DAME2001} employed the Columbia
survey of molecular gas in the Galactic plane together with the
Dwingeloo-Leiden \hi\ survey and the IRAS temperature-corrected 
100 $\mu$m spectral density map by \citet{SCHLEGEL1998}. With these data,
$N(\htwo)$ can be obtained using 

\begin{equation}
N(\htwo) = (\tau_d/\dgr - N(\hi))/2,
\label{eq:H2_from_dust}
\end{equation}

\noindent which simply states that the dust optical depth ($\tau_d$) is a
perfect tracer of the total column density of gas when the emissivity
per nucleon (\dgr) is known. As we just discussed, \dgr\ can be
straightforwardly determined on lines of sight dominated by atomic
gas, for example.  The comparison of the molecular column density so
derived with the observed $\wco$ yields $\xcot\approx1.8\pm0.3$, valid
for $|b|>5^\circ$ on large scales across the Galaxy. This study also
finds evidence for a systematic increase in \xco\ by factors $\sim2-3$
at high Galactic latitude ($b>20^\circ$), in regions with typical
$N(\htwo)\lesssim0.5\times10^{20}$~\percmsq\ on $\sim0.5^\circ$
angular scales.

\begin{figure}[t]
\centerline{\psfig{figure=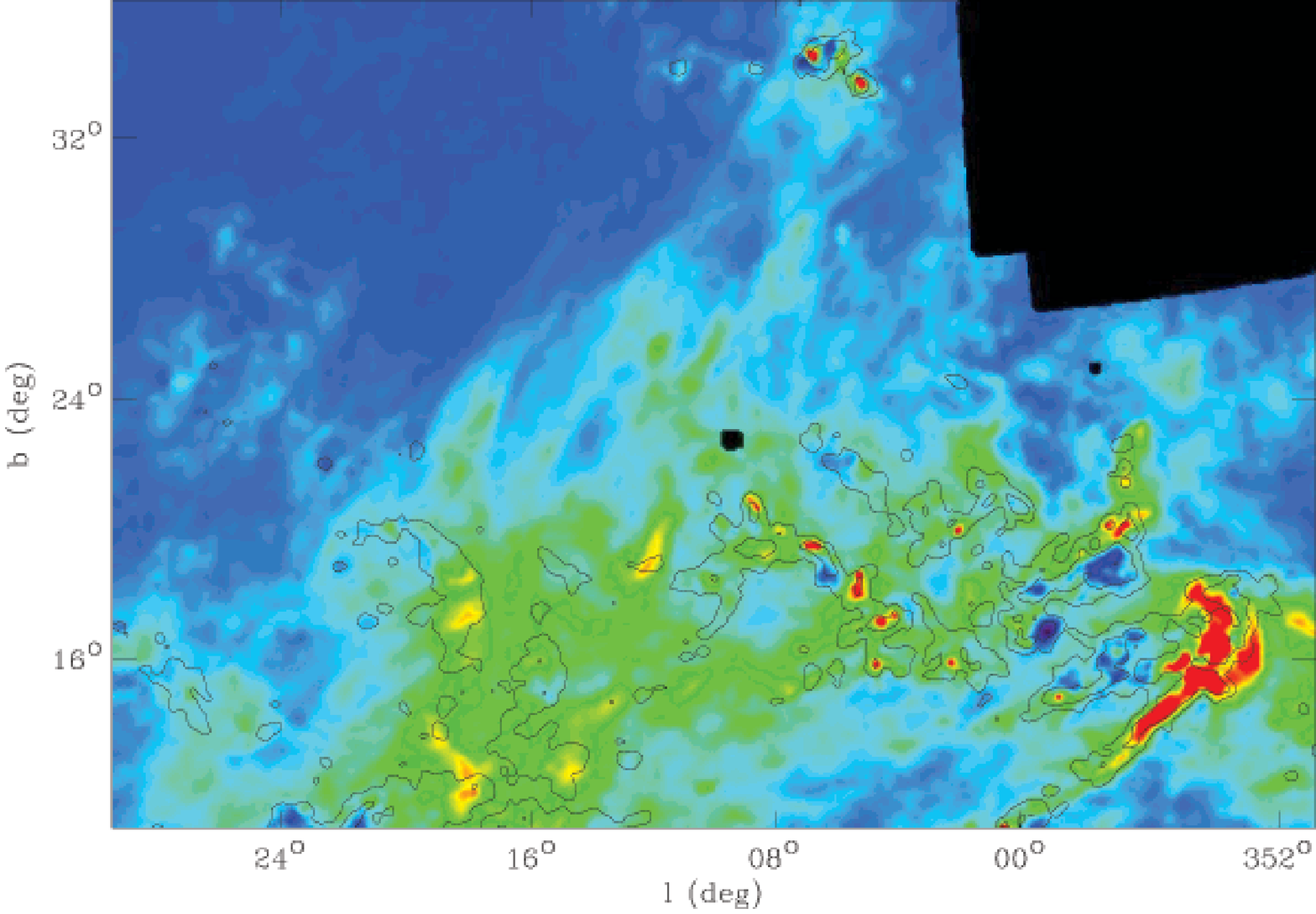,angle=180,width=\columnwidth}}
\caption{{\em Planck} results in the Aquila-Ophiuchus flare 
\citep{PLANCKCOLLABORATION2011XIX}. The figure shows 
molecular gas column density with a color range $N({\rm \htwo})\approx-1.5\times10^{21}$ \percmsq (dark blue) to 
$3.5\times10^{21}$~\percmsq (red). The black contours illustrate the CO emission
from the Columbia survey, with
values $\Ico\approx2, 10, 20$~\Kkmpers. The black region lacks CO information. 
\label{fig:planck_jp}}
\end{figure}

The excellent {\em Planck} dataset has afforded a new view on this
topic (Fig. \ref{fig:planck_jp}). \citet{PLANCKCOLLABORATION2011XIX}
produced a new $\tau_d$ map for the Milky Way and a new determination
of \xco\ using Eq. \ref{eq:H2_from_dust} for lines of sight with
$|b|>10^\circ$. They obtain $\xcot=2.54\pm0.13$, somewhat larger than
the previous study. This difference is likely methodological, for
example the use of different far-infrared wavelengths as well as local
versus global calibrations of \dgr.

\subsubsection{CO-Faint Molecular Gas and Diffuse Lines of Sight}
\label{sec:mw_diffuse}

\begin{figure}[ht!]
\centerline{\psfig{figure=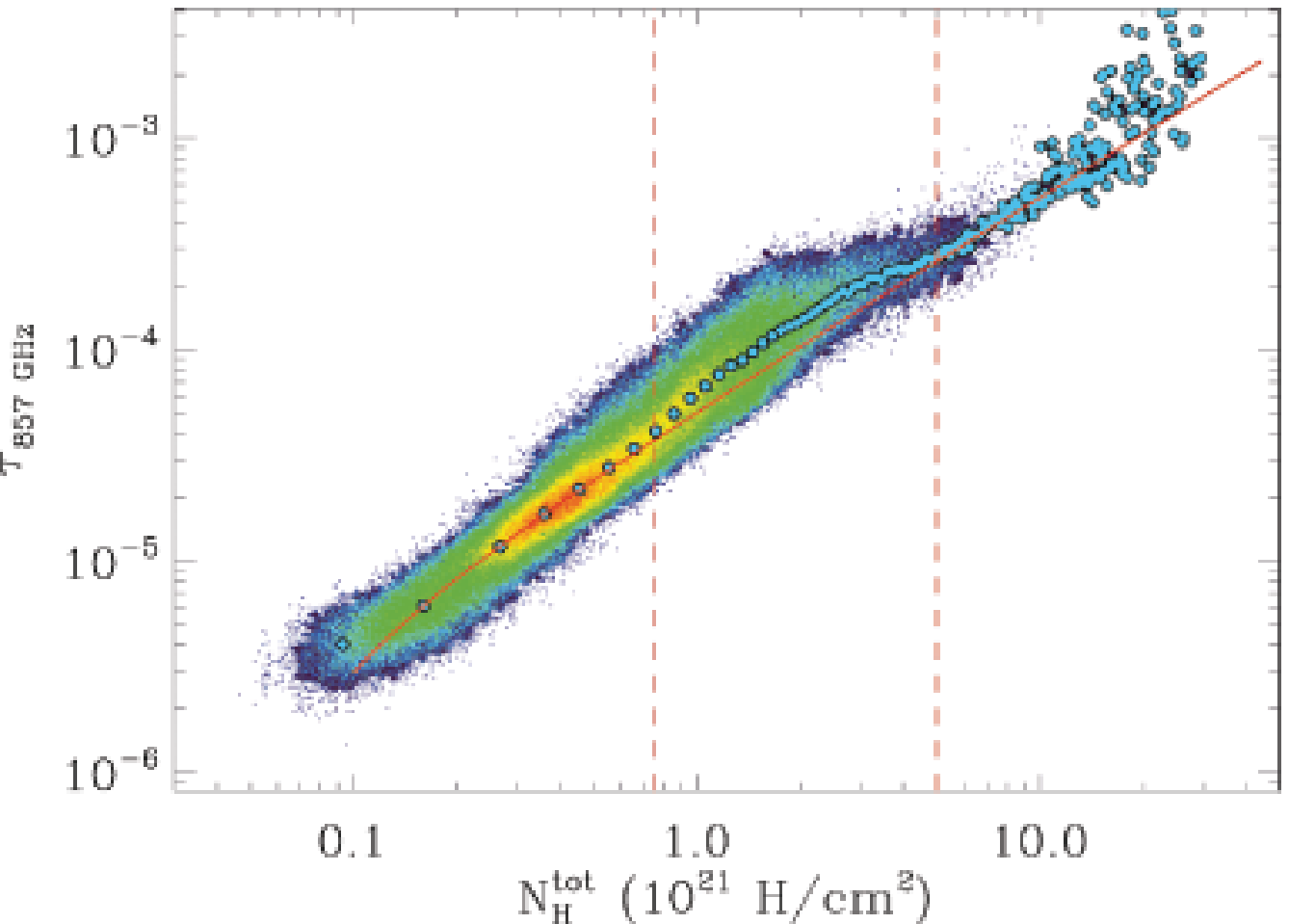,width=\columnwidth}}
\caption{Correlation between $\tau_d$ at 350 $\mu$m and total
hydrogen column density $N_{\rm H}$ for
$\xcot=2.3$ \citep[for
$|b|>10^\circ$,][]{PLANCKCOLLABORATION2011XIX}. The color scale
represents the logarithm of the number of lines-of-sight, and the blue
dots the result of $N_{\rm H}$ binning. The red dashed lines indicate
$\av=0.37$ and $\av=2.5$. The red solid line represents the best
linear fit for low $N_{\rm H}$. Note that it is also a good fit to the
$\av\gtrsim2.5$ points. The excess in the binned correlation over the
red line for $0.37\lesssim\av\lesssim2.5$ is either an indication of
``CO-faint'' molecular gas, or possibly a combination of high
optical-depth ``opaque'' \hi\ with ``CO-faint'' \htwo, or a change in
the dust emissivity over that \av\ regime. \label{fig:dust_excess}}
\end{figure}

Most interestingly,
\citet{PLANCKCOLLABORATION2011XIX} observe a tight linear
correlation between $\tau_d$ and $N(\hi)+2\xco\,{\rm W(CO)}$ for
$\av\lesssim0.4$ and $2.5\lesssim\av\lesssim10$, with an excess in
$\tau_d$ in the intermediate range (Fig. \ref{fig:dust_excess}). This
excess can be understood in terms of a component of \htwo\ (or possibly a
combination of \htwo\ and cold, opaque \hi) that emits weakly in CO
and is prevalent at $0.4\lesssim\av\lesssim2.5$ (the explanation is
not unique, since the methodology cannot distinguish it from a change
in dust emissivity in that narrow \av\ regime, but such possibility appears
unlikely).  This molecular component arises from the region in cloud
surfaces where gas is predominantly \htwo\ but most carbon is not in
CO molecules because the extinction is too low, essentially the PDR surface (see
\S\ref{sec:lm_theory} and Fig. \ref{fig:COabundances}). This component
is frequently referred to as ``CO-dark molecular gas'' or sometimes
simply ``dark gas'' \citep{GRENIER2005,WOLFIRE2010}. In this review we
will refer to it as ``CO-faint,'' which is a more accurately
descriptive name. The existence of molecular gas with low CO abundance
has been noted previously in theoretical models
\cite[e.g.,][]{VANDISHOECK1988}, in observations of diffuse gas and
high-latitude clouds
\citep[e.g.,][]{LADA1988}, and in observations of irregular galaxies
\citep[e.g.,][]{MADDEN1997}. 
In this context, the results by
\citet{PLANCKCOLLABORATION2011XIX} are in qualitative agreement with
already discussed observations that show an increase in \xco\ at low
molecular column densities
\citep[e.g.,][]{GOLDSMITH2008,PARADIS2012}.

The {\em Planck} observations are also in qualitative agreement with
the analysis by \citet{GRENIER2005}, who correlated the diffuse
$\gamma$-ray emission over the entire sky with templates derived from
the \hi, CO, and dust, finding a component of gas not traced by CO
evident in local clouds at high latitudes. Their analysis finds that
this component is as important, by mass, as the ``CO-bright'' \htwo\
component in several of these clouds, and increasingly more important
for smaller cloud masses. The recent analyses based on {\em Fermi}
data by
\citet{ABDO2010d} and
\citet{ACKERMANN2012} are also qualitatively compatible with these results,
finding that the ``CO-faint'' component amounts to $40\%-400\%$ of the
``CO-bright'' mass in the Cepheus, Polaris, Chamaleon, R Cr A, and
Cassiopeia clouds (small local molecular clouds).

The ionized carbon far-infrared fine structure emission provides an
additional probe of molecular gas at low \av. Large scale \cii\
observations of the \fscii\ fine-structure transition in the Milky Way
and external galaxies suggest it is due to a combination of emission
from the Cold Neutral Medium (CNM) and from PDRs located in the
surfaces of GMCs \citep{STACEY1991,SHIBAI1991,BENNETT1994}.  The
contribution from \cii\ in the diffuse ionized gas, however, could
also be important \citep{HEILES1994,MADDEN1993}, particularly along
certain lines-of-sight \citep{VELUSAMY2012}.

In regions where most of the emission arises in PDRs \cii\ has the
potential to trace the ``CO-faint'' molecular regime at low \av.
\citet{LANGER2010} analyze 16 lines-of-sight in the plane of the 
Galaxy and find that in about half of them the observed \cii\
intensity can be entirely explained as due to carbon in atomic gas in
the CNM. The other half, however, exhibits $\cii/N(\hi)$ ratios that
are too large to be due to atomic gas
and may have molecular to atomic ratios as large as
$N(\htwo)/N(\hi)\sim6$. While the very brightest \cii\ components
investigated arise from dense ($n>10^5$~\percmcu) PDRs exposed to
intense radiation fields, most of the \cii\ emission in these
molecular lines-of-sight can be explained as originating in the
surfaces of modestly dense GMCs ($n\sim[3-300]\times10^{3}$~\percmcu)
exposed to at most a few times the local interstellar radiation field
at the solar circle \citep{PINEDA2010b}. Following the reasoning in
\citet{VELUSAMY2010} and \citet{LANGER2010}, a very approximate
relation between \cii\ emission and ``CO-faint'' \htwo\ is
$N(\htwo)\sim 1.46\times10^{20}\,{\rm W(\cii)} - 0.35
N(\hi)$~\percmsq, for ${\rm W(\cii)}$ in \Kkmpers\ (see
Eq. \ref{eq:fluxT} to convert between Jy and K). We caution that the
coefficients correspond to the Milky Way carbon abundance and are very
dependent on the assumed physical conditions, particularly the
densities (we use $n(\hi)\sim200$~\percmcu,
$n(\htwo)\sim300$~\percmcu, and $T_{\rm HI}\sim T_{\rm H2}\sim100$~K).



\citet{LISZT2010} measure \xco\ in diffuse gas by first estimating 
the total hydrogen column from dust continuum emission \citep[using the map
by][]{SCHLEGEL1998}, and subtracting the observed \hi\ column
density (c.f., Eq. \ref{eq:H2_from_dust}). They select lines-of-sight
with ${\rm HCO^+}$ absorption spectra against bright extragalactic
continuum sources. The CO emission, together with the measured ${\rm
H_2}$ column provides a measure of \xco. Surprisingly,
\cite{LISZT2010} and \cite{LISZT2012} find mean values in diffuse gas
similar to those in GMCs. There are large variations, however, about
the mean with low \xco\ (bright CO) produced in warm $T\sim 100$ K
diffuse gas and high \xco\ (faint CO) produced at low ${\rm N(H_2)}$
column densities. The authors argue that these variations mainly
reflect the CO chemistry and its dependence on ultraviolet radiation
field, density, and total column density, rather than the ${\rm H_2}$
column density.  We note that the observed CO column densities cannot be
produced in steady-state PDR models \citep[.e.g.,][]{SONNENTRUCKER2007}. 
Enhanced CO production 
might occur through the ``${\rm CH^+}$ channel'' driven by
non-thermal ion-neutral reactions \citep{FEDERMAN1996, VISSER2009}
or by pockets of warm gas and ion-neutral reactions
in turbulent dissipation regions \citep{GODARD2009}.
Density fluctuations in a turbulent median might also 
increase the CO production \citep{LEVRIER2012}.  
Thus, although the mean \xco\ diffuse cloud value is
similar to GMCs, the CO emission from diffuse gas cannot be easily
interpreted as a measure of the molecular column except perhaps in a
statistical sense.

\subsection{\xco\ Based on Gamma-Ray Observations} 
\label{sec:mw_gammaray}

Diffuse $\gamma$-ray emission in the Galaxy is chiefly due to three
processes: neutral pion production and subsequent decay in collisions
between cosmic-rays and interstellar matter, bremsstrahlung emission
due to scattering of cosmic-ray electrons by interstellar matter, and
inverse Compton scattering of low energy photons by cosmic-ray electrons.
The first of these processes is the dominant production channel for
diffuse $\gamma$-rays with energies above 200 MeV, although at high
Galactic latitude there will be an increasingly important inverse
Compton component
\citep{BLOEMEN1989}. Interestingly, the fact that interactions between
cosmic rays and nucleons give rise to diffuse $\gamma$-ray emission
can be used to count nucleons in the ISM, and indeed the use of \xco\ to
represent the ratio $N(\htwo)/{\rm W(CO)}$ was introduced for the
first time in $\gamma$-ray work using observations from the {\em COS B} 
satellite \citep{LEBRUN1983}.  

Accounting for the pion decay and bremsstrahlung processes, and
neglecting the contribution from ionized gas, the basic idea behind
modeling the emission is to use a relation $I_\gamma =
\sum{\epsilon_{\gamma,{\rm HI}}(R_i)\left[{N(\hi)_i+2\,Y_{\rm
CO}\,\wco_i}\right]}$ \citep{BLOEMEN1989}.  Here $I_\gamma$ is the
diffuse $\gamma$-ray emission along a line-of-sight, $\epsilon_\gamma$
is the \hi\ gas emissivity (a function of Galactocentric radius $R$),
and $Y_{\rm CO}$ is a parameter that takes into account that
emissivity in molecular clouds may be different than in atomic gas,
due to cosmic-ray exclusion or concentration, $\xco=Y_{\rm
CO}\,\epsilon_{\gamma,{\rm HI}}/\epsilon_{\gamma,{\rm H2}}$
\citep{GABICI2007,PADOVANI2009}. This situation is analogous to that presented in
dust emission techniques, where emissivity changes in the molecular
and atomic components will be subsumed in the resulting value of \xco.

Recent analyses use the approach $I_\gamma \approx q_{\rm
HI}N(\hi)+q_{\rm CO}\wco+q_{\rm EBV}E(B-V)_{res}+\ldots,$ where the
first three terms account for emission that is proportional to the
\hi, the ``CO-bright'' \htwo, and a ``CO-faint'' \htwo\ component that is
traced by dust reddening or emission residuals \citep[the additional
terms not included account for an isotropic $\gamma$-ray background
and the contribution from point sources, e.g.,][]{ABDO2010d}.  The
dust residual template consists of a dust map
\citep[e.g.,][]{SCHLEGEL1998} with a linear combination of $N(\hi)$
and $\wco$ fitted and removed.

\subsubsection{Observational Results}

A discussion of the results of older analyses can be found in
\citet{BLOEMEN1989}, here we will refer to a few of the more recent
results using the {\em Compton Gamma Ray Observatory} and {\em Fermi}
satellites.

\citet{STRONG1996} analyzed the EGRET all-sky survey obtaining
$\xcot\approx1.9\pm0.2$. Similar results were obtained by \citet{HUNTER1997}
for the inner Galaxy, and by \citet{GRENIER2005} for clouds in the solar
neighborhood (see also \S\ref{sec:mw_diffuse}).
\citet{STRONG2004} introduce in the analysis a Galactic gradient in \xco,
in an attempt to explain the discrepancy between the derived
$\epsilon_\gamma(R)$ and the distributions of supernova remnants and
pulsars, which trace the likely source of cosmic rays in supernovae
shocks. Matching the emissivity to the pulsar distribution requires a
significant change in
\xco\ between the inner and outer Galaxy ($\xcot\approx0.4$ for
$R\sim2.5$~kpc to $\xcot\approx 10$ for
$R>10$~kpc).


The sensitivity of {\em Fermi} has been a boon for studies of diffuse
$\gamma$-ray emission in our Galaxy. \citet{ABDO2010d} and 
\citet{ACKERMANN2011,ACKERMANN2012b,ACKERMANN2012d} analyze the emission in the solar neighboorhood
and the outer Galaxy. Taken together, they find a similar \xco\ for
the Local arm and the interarm region extending out to
$R\sim12.5$~kpc, $\xcot\sim1.6-2.1$ depending on the assumed
\hi\ spin temperature. The very local high-latitude 
clouds in the Gould Belt have a lower $\xcot\approx0.9$, and appear
not to represent the average properties at the solar circle
\citep{ACKERMANN2012b,ACKERMANN2012c}.

It is important to note, however, that the {\em Fermi\ } studies
follow the convention $\xco\equiv q_{\rm CO}/2q_{\rm HI}$ with the
parameters defined in the previous section. Thus, their
definition of \xco\ does not include the ``CO-faint'' envelope that is
traced by the dust residual template: it only accounts for the \htwo\
that is emitting in CO. This is different from the convention
adopted in the dust studies discussed in the previous section, where
all the \htwo\ along a line-of-sight is associated with the
corresponding CO. \citet{ACKERMANN2012b} report $\xcot\approx0.96,\,0.99,\,
0.63$ for Chameleon, R CrA, and the
Cepheus/Polaris Flare region, respectively using the {\em Fermi}
convention. The ratio of masses associated with the CO and the dust
residual template (reported in their Table 4) suggests that we should
correct these numbers by factors of approximately 5, 2, and 1.4
respectively to compare them with the dust modeling. This correction
results in $\xcot\approx4.8,\,2,\,0.9$, bringing
these clouds in much better (albeit not complete) agreement with the
results of dust modeling and the expectation that they be
underluminous --- not overluminous --- in CO due to the presence of a large
``CO-faint'' molecular component.

\subsubsection{Considerations and Limitations}

We have already mentioned a limitation of $\gamma$-ray studies of
\xco, the degree to which they may be affected by the rejection (or
generation) of cosmic rays in molecular clouds. A major limitation of
$\gamma$-ray determinations is also the poor angular resolution of the
observations. Another source of uncertainty has become increasingly
apparent with {\em Fermi}, which resolves the Magellanic Clouds in
$\gamma$-ray emission
\citep{ABDO2010,ABDO2010b}. In the Clouds, the distribution of
emission does not follow the distribution of gas. Indeed, the emission
is dominated by regions that are not peaks in the gas distribution,
but may correspond to sites of cosmic ray injection
\citep[e.g.,][]{MURPHY2012}, suggesting that better knowledge of the
cosmic ray source distribution and diffusion will have an important
impact on the results of $\gamma$-ray studies.


\citet{ACKERMANN2012} carry out a thorough study of the impact of
systematics on the global $\gamma$-ray analyses that include a
cosmic-ray generation and propagation model, frequently used to infer
Galactic \xco\ gradients. They find that the
\xco\ determination can be very sensitive to assumptions such as
the cosmic ray source distribution and the \hi\ spin temperature, as
well as the selection cuts in the templates (in particular, the dust
template). Indeed, the value in the outer Galaxy in their analysis is
extremely sensitive to the model cosmic ray source distribution.  The
magnitude of a Galactic \xco\ gradient turns out to also be very
sensitive to the underlying assumptions
\citep[see Fig. 25 in][]{ACKERMANN2012}. What appears
a robust result is a uniformly low value of
\xco\ near the Galactic center ($R\sim0-1.5$~kpc). This was already pointed
out by much earlier $\gamma$-ray studies \citep{BLITZ1985}. The authors
also conclude that including the dust information leads to an
improvement in the agreements between models and $\gamma$-ray data,
even on the Galactic plane, suggesting that ``CO-faint'' gas is an
ubiquitous phenomenon.


\begin{table*}[ht]
\def~{\hphantom{0}}
\caption{Representative \xco\ values in the Milky Way disk}\label{tab:xcomw}
\begin{center}
\begin{tabular}{lcl}
\toprule
Method & $\xco/10^{20}$ & \multicolumn{1}{c}{References} \\
       & \xcounits      &           \\
\colrule
Virial          & 2.1  & \cite{SOLOMON1987} \\ 
                & 2.8  & \cite{SCOVILLE1987}\\
Isotopologues   & 1.8  & \cite{GOLDSMITH2008}  \\ 
Extinction      & 1.8  & \cite{FRERKING1982}  \\
                & $2.9-4.2$ & \cite{LOMBARDI2006}  \\
                & $0.9-3.0$   & \cite{PINEDA2008}  \\
                & 2.1  & \cite{PINEDA2010}\\
                & $1.7-2.3$ & \cite{PARADIS2012}\\
Dust Emission   & 1.8  & \cite{DAME2001} \\
                & 2.5  & \cite{PLANCKCOLLABORATION2011XIX} \\
$\gamma$-rays   & 1.9  & \cite{STRONG1996}\\
                & 1.7  & \cite{GRENIER2005} \\
                & $0.9-1.9\ ^*$ & \cite{ABDO2010d}\\
                & $1.9-2.1\ ^*$ & \cite{ACKERMANN2011,ACKERMANN2012d} \\
                & $0.7-1.0\ ^*$ & \cite{ACKERMANN2012b,ACKERMANN2012c} \\
\botrule
\multicolumn{3}{l}{$^*$ Note difference in \xco\ convention (\S\ref{sec:mw_gammaray}).}
\end{tabular}
\end{center}
\end{table*}

\subsection{Synthesis: Value and Systematic Variations of \xco\ in the Milky Way}
\label{sec:mw_synthesis}

There is an assuring degree of uniformity among the values of \xco\
obtained through the variety of methodologies available in the Milky
Way. {\em Representative results from analyses using virial masses, CO
isotopologues, dust extinction, dust emission, and diffuse
$\gamma$-ray radiation hover around a ``typical'' value for the
disk of the Milky Way $\xco\approx2\times10^{20}$~\xcounits\ (Table
\ref{tab:xcomw})}. This fact, combined with the simple theoretical arguments
outlined in \S\ref{sec:intro_theoretical} as to the physics behind the 
\htwo-to-CO conversion factor, as well as the results from elaborate
numerical simulations discussed in \S\ref{subsec:numerical}, strongly
suggests that we know the mass-to-light calibration for GMCs in the
disk of the Milky Way to within $\pm0.3$ dex certainly, and probably
with an accuracy closer to $\pm0.1$ dex (30\%). {\em This is an
average number, valid over large scales}. Individual GMCs will scatter
around this value by a certain amount, and individual lines-of-sight
will vary even more.

There are, however, some systematic departures from this value in
particular regimes, as suggested by the simple theoretical
arguments. As pointed out by several studies, the Galactic center
region appears to have an
\xco\ value $3-10$ times lower than the disk. In addition to the
aforementioned $\gamma$-ray results
\citep{BLITZ1985,STRONG2004,ACKERMANN2012}, a low
\xco\ in the center of the Milky Way was obtained by analysis of
the dust emission \citep{SODROSKI1995}, and the virial mass of its
clouds \citep{OKA1998,OKA2001}. We will see in \S\ref{sec:normal_galaxies}
that this is not uncommonly observed in the centers of other galaxies.
It is likely due to a combination of enhanced excitation (clouds are
hotter) and the dynamical effects discussed in \S\ref{sec:other_sources}.

Departures may occur in some nearby, high-latitude clouds, for example
the Cepheus/Polaris Flare region. A critical step to understand the
magnitude or even existence of such discrepancies will be to derive
\xco\ for diffuse $\gamma$ ray observations using conventions matched
to other fields. We have attempted some estimate of the correction in
the Chameleon, R CrA, and the Cepheus/Polaris Flare.  In
Cepheus/Polaris even after trying to account for the difference in
\xco\ convention, the $\gamma$-ray studies find a low \xco\ value
suggesting that those local clouds are overluminous in CO by a factor
$\sim2$, or maybe that their dust emissivity is different. Low values
are not consistent with results from dust emission techniques
\citep{DAME2001,PLANCKCOLLABORATION2011XIX}.  On the other hand, they
are known to be turbulent to very small ($\sim0.01$ pc) scale
\citep{MIVILLE-DESCHENES2010}, and turbulent dissipation may play a
role in exciting CO \citep{INGALLS2011}.


Finally, observations of \xco\ in the outer galaxy are still sparse,
although some of them suggest high values of \xco\ \citep[for
example,][]{BRAND1995,HEYER2001}. This can be understood in terms of an
increasing dominance of the ``CO-faint'' molecular gas associated
probably with decreasing metallicities in the outer disk (see,
\S\ref{sec:low_metallicities}). In nearby, resolved molecular clouds
there is compelling evidence that a regime of high \xco\ exists at low
extinctions and column densities
\citep[e.g.,][]{GOLDSMITH2008,PINEDA2010}, where most of the carbon
in the gas phase is not locked in CO molecules.

It seems clear that CO becomes a poor tracer of \htwo\ at low column
densities. Most studies agree that this phenomenon occurs for
$\av\lesssim2$ \citep{GOLDSMITH2008,PLANCKCOLLABORATION2011XIX}.  How
much molecular gas exists in this regime is still a subject of study,
but it seems to represent a substantial, perhaps even dominant fraction of
the \htwo\ near the solar circle
\citep{GRENIER2005,WOLFIRE2010,PLANCKCOLLABORATION2011XIX}. 
It is, however, unlikely to be a large fraction of the total
molecular mass of the Galaxy, which is dominated by clouds in the
inner galaxy. Nonetheless, we will see that this ``CO-faint'' phase
almost surely constitutes most of the molecular gas in low metallicity
systems.

Using these conclusions, what is the characteristic surface density of
a GMC, \Sgmc, in the Milky Way? The precise
value, or even whether it is a well-defined quantity, is a matter of
study. Using the survey by
\citet{SOLOMON1987} updated to the new distance scale 
and assuming $\xcot=2$, we find a distribution of surface densities
with $\Sgmc\approx150^{+95}_{-70}$~\msunperpcsq\ ($\pm1\sigma$
interval). Using instead the distribution of $331\sigma^2/R$ to
evaluate surface density (see discussion after
Eq. \ref{eq:size_linewidth}), we obtain
$\Sgmc\approx200^{+130}_{-80}$~\msunperpcsq.
\citet{HEYER2009} use $^{13}$CO observations of these clouds, 
finding $\Sgmc\approx40$~\msunperpcsq\ over the same cloud areas but
concluding that it is likely an underestimate of at least a factor of
2 due to non-LTE and optical depth effects
\citep[$\Sgmc\approx144$~\msunperpcsq\ using a more restrictive $^{13}$CO contour 
instead of the original $^{12}$CO to define surface
area,][]{ROMAN-DUVAL2010}. Real clouds have a range of surface
densities. \citet{HEIDERMAN2010} analyze extinction-based measurements
for 20 nearby clouds, calculating their distribution of
\Smol. It is clear that many of these clouds do not reach
the characteristic \Sgmc\ of clouds in the inner Galaxy; for the most
part the clouds with low mass tend to have low surface
densities. Nonetheless, several of the most massive clouds do have
$\Sgmc\gtrsim100$~\msunperpcsq. The mass-weighted (area-weighted)
\Sgmc\ are 160, 150, 140, and 110 (140, 140, 100, and 90)
\msunperpcsq\ for Serpens-Aquila, Serpens, Ophiuchus, and Perseus
respectively according to their Table 2. By comparison,
$\Sgmc\approx85$ M$_\odot$~pc$^{-2}$ for a sample of nearby galaxies,
many of them dwarf
\citep{BOLATTO2008}. Note that \Sgmc\ is likely to be a function of
the environment. In the Galactic Center, for example,
\citet{OKA1998,OKA2001} report a size-line width relation where the coefficient
$C$ is five times larger than that observed in the Milky Way disk, suggesting
a \Sgmc\ that is 25 times larger \citep[see also][]{ROSOLOWSKY2005}.


\section{\xco\ in Normal Galaxies}
\label{sec:normal_galaxies}

The distance of even the nearest galaxies renders CO the primary
tracer of molecular gas outside the Milky Way, a situation that will
improve but not reverse with ALMA. Other galaxies therefore represent
{\em the} key application of \xco . They also offer a wider range of
environments than the Milky Way and a simpler mapping between local
ISM conditions, especially metallicity, and line of sight. As a
result, for more than two decades observations of the nearest galaxies
have been used to test and extend calibrations of \xco\ as a function
of metallicity and other local ISM properties. Here we review the
techniques available to derive \xco\ in ``normal'' galaxies, meaning
star-forming dwarf, spiral, or elliptical galaxies. We discuss
specific efforts to understand the behavior of \xco\ as a function of
metallicity in \S\ref{sec:low_metallicities} and the special case of
overwhelmingly molecular starburst galaxies, such as the local
luminous and ultraluminous infrared galaxies, in
\S\ref{sec:starbursts}. 
\S\ref{sec:sb_highz} considers \xco\ in galaxies at high redshift.

Only a subset of the techniques used to determine \xco\ in the Milky
Way can be applied to other galaxies. In each case, the limited
sensitivity and resolution of millimeter, submillimeter, and infrared
facilities complicate the calculation. Direct estimates rely almost
exclusively on the use of virial mass measurements
(\S\ref{sec:mw_virial}), dust emission employed as an optically thin
tracer of the total gas reservoir (\S\ref{sec:mw_dust}), or modeling
of multiple CO lines. Ideally such modeling includes optically thin
isotopologues, but studies of high redshift systems must often make
due with a few, or only one, $^{12}$CO line ratios
(\S\ref{sec:sb_highz}).

\subsection{Extragalactic Virial Mass Estimates}
\label{sec:normal_virial}

Since the late 1980s, millimeter telescopes have been able to resolve
CO emission from nearby galaxies into discrete molecular clouds. From
such observations, one can estimate the line width, size, and
luminosity of these objects and proceed as in \S\ref{sec:mw_virial}
\citep[see the recent review by][]{FUKUI2010}.

Resolution and sensitivity have limited virial mass measurements to
the nearest galaxies, those in the Local Group and its immediate
environs. The Magellanic Clouds ($d\sim50$~kpc) are close enough that
single dish telescopes resolve individual clouds
\citep[e.g.,][]{RUBIO1993,ISRAEL2003,MIZUNO2001,FUKUI2008,HUGHES2010,WONG2011}. 
Millimeter-wave interferometers resolve populations of GMCs in other
Local Group galaxies --- M~31, M~33, NGC~6822, and IC~10
\citep[$d\sim1$~Mpc,
e.g.,][]{VOGEL1987,WILSON1990,WILSON1995,ROSOLOWSKY2003,LEROY2006,BLITZ2007,FUKUI2010}.

Early measurements beyond the Local Group focused on very nearby dwarf
galaxies \citep[$d \sim
3$~Mpc,][]{TAYLOR1999,WALTER2001,WALTER2002,BOLATTO2008} or
considered structures more massive than GMCs in more distant galaxies
\citep[Giant Molecular Associations, or GMAs, e.g.,][]{VOGEL1988,RAND1990}. The
nearest spiral galaxies tend to be more distant ($d\sim6$~Mpc), so
that $\sim 1\arcsec$ resolution is required to resolve massive
GMCs. Only recently mm-wave interferometers have begun to achieve this
resolution with the requisite sensitivity, by investing large amounts
of time into dedicated observations of bright regions of the nearest
massive spiral galaxies
\citep{DONOVANMEYER2012,DONOVANMEYER2013,REBOLLEDO2012}. This has allowed the
first cloud-scale virial mass measurements of populations of clouds in
spiral galaxies beyond the Local Group. 

\subsubsection{\xco\ From Extragalactic Virial Mass Analyses}

\begin{figure}[th!]
\centerline{\psfig{figure=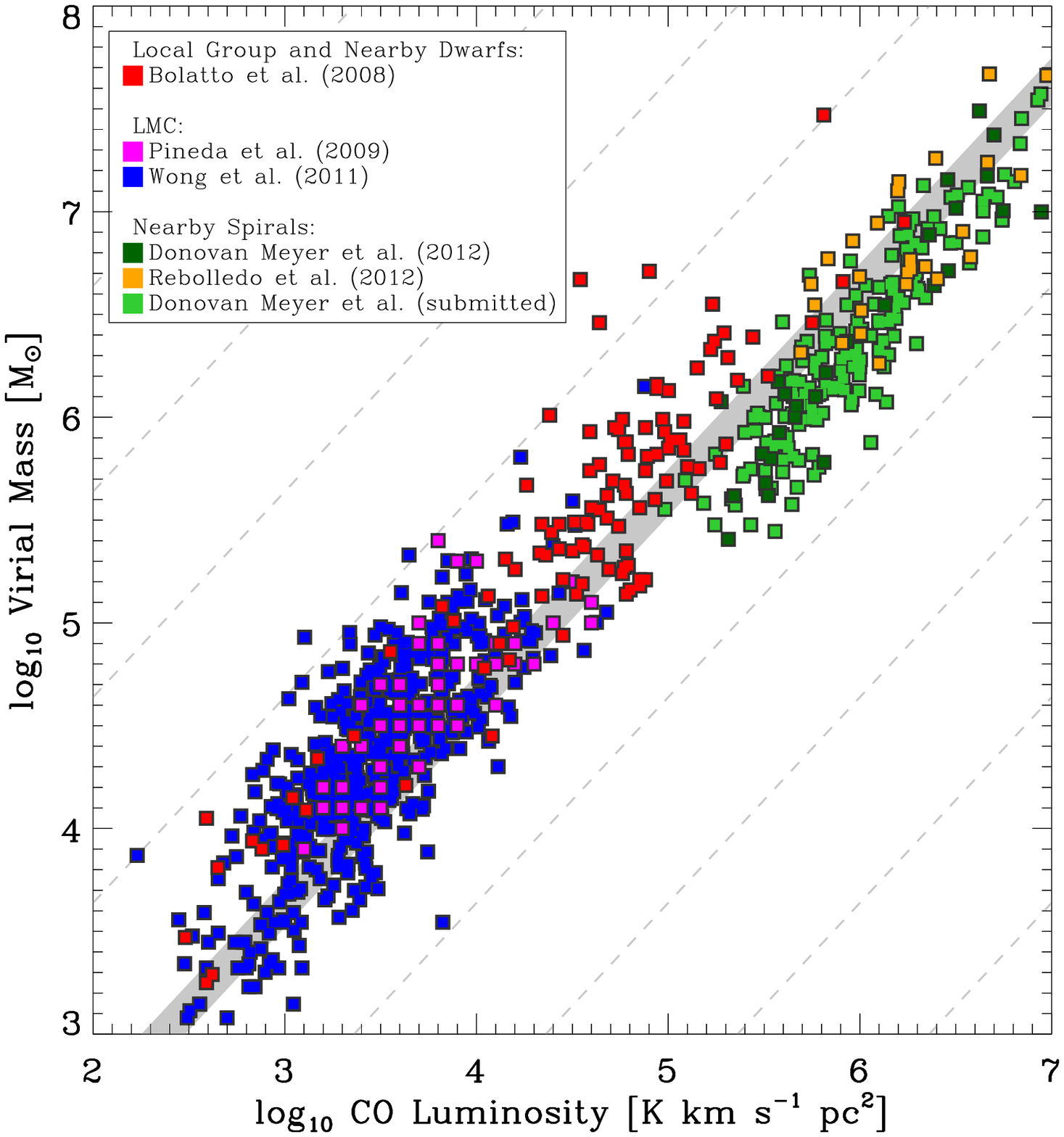,width=\columnwidth}}
\caption{\label{fig:virial_exgal} Relation between virial mass ($y$-axis) and CO luminosity
($x$-axis) for extragalactic GMCs. We show virial masses measured from
selected high spatial resolution CO observations of nearby galaxies:
the compilation of \citet[][including M31, M33, and nine dwarf
galaxies]{BOLATTO2008}, high resolution studies of the LMC by
\citet{PINEDA2009} and \citet{WONG2011}, and high resolution 
studies of the nearby spiral NGC~6946 by \citet{DONOVANMEYER2012} and
\citet{REBOLLEDO2012}, as well as NGC~4826 and NGC~4736 by \citet{DONOVANMEYER2013}. Dashed lines show fixed \xco, with the 
typical Milky Way value $\xcot = 2$ and $\pm 30\%$ indicated by the
gray region. Virial mass correlates with luminosity, albeit with large
scatter, across more than three orders of magnitude in extragalactic
systems. The median across all displayed data is $\xcot = 2.8$ with
$0.4$~dex scatter and the best fit relation has a power law index
$0.90
\pm 0.05$, reflecting that most Local Group clouds show $\xcot \sim 4$
while both studies of the bright spiral NGC 6946 find a lower $\xcot
\approx 1-2$.}
\end{figure}

Broadly, virial mass measurements across a wide range of environments
yield $\xcot\approx1-4$, consistent with Milky Way results and very
similar to \xco\ derived from dust-based techniques applied to high
mass nearby galaxies. \citet{BOLATTO2008} find $\xcot\approx
3.5^{+1.8}_{-1.2}$ in $12$ nearby systems. In the highest resolution
studies of the Large Magellanic Cloud (LMC) to date,
\citet{ISRAEL2003}, \citet{HUGHES2010}, \citet{WONG2011}, 
and \citet{PINEDA2010} all find $\xcot\approx 4$. Considering
NGC~6946, one of the nearest molecule-rich spiral galaxies,
\citet{DONOVANMEYER2012} report $\xcot\approx 1.2$ and
\citet{REBOLLEDO2012} find $\xcot\sim2$. In M~33,
\citet{ROSOLOWSKY2003} show $\xcot\sim2$ independent of radius and
metallicity. In M~31, \citet{ROSOLOWSKY2007} calculate $\xcot\approx
4$ assuming virial equilibrium.

Figure \ref{fig:virial_exgal} shows the relationship between virial
mass and CO luminosity for a subset of these measurements.  A good
correlation extends across three orders magnitude in luminosity and
roughly a dozen systems. As in the Milky way, there may be evidence
for a slightly sub-linear slope (see \S\ref{sec:gmcs} and
\S\ref{sec:mw_virial} and Figure \ref{fig:solomon}), though this may
also reflect methodological or environmental differences among the
galaxies studied. Across the ensemble of points we plot the median
$\xcot = 2.9$ with $0.4$ dex scatter (slightly larger than a factor of
2).

Surprisingly at first, these ``Galactic'' \xco\ values obtained from
virial masses extend to low metallicity, irregular galaxies. In the
low metallicity Small Magellanic Cloud \citep[SMC, with ${\rm
12+log[O/H]\approx8.0}$,][]{DUFOUR1982}, \citet{BOLATTO2003} and
\citet{ISRAEL2003} find $\xcot\approx 4$,
consistent with the Milky Way value. \citet{ROSOLOWSKY2003} find no
dependence of \xco\ on metallicity in M33. \citet{LEROY2006} find an
approximately Galactic \xco\ in the Local Group dwarf IC~10
\citep[${\rm
12+log[O/H]\approx8.17}$,][]{LEQUEUX1979}. \citet{WILSON1994} found
$\xcot\lesssim6.6$ in NGC~6822 \citep[${\rm
12+log[O/H]\approx8.20}$,][]{LEQUEUX1979}. Most of the galaxies
studied by the aforementioned \citet{BOLATTO2008} are dwarf irregulars with subsolar
metallicity.
In each case we highlight the results for the
highest resolution study of the galaxy in question. We discuss the
effects of varying spatial resolution in the next section. Overall,
these studies show that {\em high spatial resolution virial mass
measurements suggest roughly Galactic \xco\ irrespective of
metallicity.}

Virial masses have also been measured for very large structures,
GMAs or superclouds
\citep{VOGEL1988,WILSON2003}, although it is unclear the degree to
which they are virialized or even bound.
\citet{WILSON2003} consider $\sim500$~pc-scale
structures in the Antennae Galaxies and find the \xco\ to be approximately Galactic \citep[see
also][]{UEDA2012,WEI2012}, but this conflicts with results from
spectral line modeling (see \S \ref{sec:starbursts}). Phrasing their 
results largely in terms of
boundedness, \citet{RAND1990} and \citet{ADLER1992} find $\xcot\sim 3$
and $\xcot\sim 1.2$ to be needed for virialized GMAs in M51. Studies
of other galaxies, for example M~83 \citep{RAND1999}, suggest that not
all GMAs are gravitationally bound and that the mass spectrum of GMAs
varies systematically from galaxy to galaxy or between
arm and interarm regions \citep{RAND1990,RAND1999,WILSON2003},
complicating the interpretation of these large-scale
measurements. 

\subsubsection{Caveats on Virial Mass-Based \xco\ Estimates}


As we will see in \S\ref{sec:normgal_dust}, the approximately constant
Galactic \xco\ implied by virial masses on small scales at low
metallicity appears to contradict the finding from dust-based
measurements and other scaling arguments \citep[for
example][]{BLANC2013}, which consistently indicate that
\xco\ increases with decreasing metallicity. 
This discrepancy most likely arises because virial mass measurements
sample the gas that is bright in CO, while dust-based
measurements include all \htwo\ along the line-of-sight.
As discussed below (\S\ref{sec:low_metallicities}), decreasing dust
shielding at low metallicities causes CO to be preferentially
photodissociated relative to \htwo, creating a massive reservoir of
\htwo\ in which C$^+$ and C rather than CO represent the dominant 
forms of gas-phase carbon. Because this reservoir is external to the CO
emitting surface it will not be reflected in the CO size or its line
width (unless the surface pressure term is important). Thus, we expect
that {\em high resolution virial mass measurements preferentially
probe
\xco\ in the CO-bright region.}

The range of values discussed above, $\xcot\sim1-4$, is significant. Does
it indicate real variations in \xco? Measuring cloud properties
involves several methodological choices (\S\ref{sec:mw_virial}).
Different
methods applied to the same Milky Way data shift results by $\sim
30\%$, and biases of $\sim 40\%$ are common in extragalactic data
\citep[e.g.,][]{ROSOLOWSKY2006}. 
As we discuss below, the \xco\ obtained by virial mass analysis
seems to depend on the physical resolution of the observations.
Ideally, results will be compared to ``control'' measurements that
have been extracted and analyzed in an identical way, ideally at
matched spectral and spatial resolution and sensitivity
\citep[e.g.,][]{PINEDA2009}.
Many studies now employ the CPROPS algorithm
\citep{ROSOLOWSKY2006}, which is designed to account for sensitivity
and resolution biases in a systematic way, allowing ready
cross-comparison among data sets. This is not a perfect substitute, however,
for matched analyses.
Our assessment is that in lieu of such careful comparison,
differences of $\pm 50\%$ in \xco\ should still be viewed as
qualitatively similar.

We raised the issue of spatial scale in the discussion of systematics
and implicitly in the discussion of GMAs. Virial mass-based \xco\
exhibit a complex dependence on the spatial scale of the
observations. For dwarf galaxies it seems that even using similar
methodology and accounting for resolution biases, studies with finer
spatial resolution systematically return lower \xco\ than coarser
resolution studies. For example, \citet{HUGHES2010} find
$\xcot\approx4$ in the LMC \citep[as do][]{ISRAEL2003,PINEDA2009,WONG2011}
compared to $\xcot\approx7$ found by
\citet{FUKUI2008} using similar methodology but with $\sim 3$ times
coarser linear resolution. In the SMC, \citet{MIZUNO2001} find
$\xcot\approx14$, while higher resolution studies find $\xco$ as low
as $\xcot\sim2-4$ for the smallest resolved objects
\citep{ISRAEL2003,BOLATTO2003}. Contrasting the interferometer
measurements of \citet{WILSON1994} and the coarser single-dish
observations of \citet{GRATIER2010} reveals a similar discrepancy in
NGC~6822. In one of the first studies to consider this effect,
\citet{RUBIO1993} explicitly fit a dependence for \xco\ in the
SMC as a function of spatial scale, finding $\xco \propto R^{0.7}$
\citep[see also the multiscale analyses
in][]{BOLATTO2003,LEROY2009}.  


This scale dependence may reflect one of several scenarios. First, low
resolution observations can associate physically distinct clouds that 
are not bound,
causing a virial mass analysis to overpredict \xco. Alternatively, the
ensemble of clouds conflated by a coarser beam may indeed be bound. In
the case of a heavily molecular interstellar medium like the Antennae
Galaxies or the arm regions of M~51, most of the material in the
larger bound structure may be molecular and a virial mass measurement
may yield a meaningful, nearly Galactic conversion factor for objects
much bigger than standard GMCs. For a low metallicity
irregular galaxy, the best case for low resolution virial mass
measurements is that large complexes are virialized and that the low
\av\ gas between the bright clouds is \htwo\ associated with \cii. 
We caution, however, that this is only one of many possible
scenarios.

We suggest that the sensitivity of virial mass measurements to extended
CO-free envelopes of \htwo\ is ambiguous at best. In that sense, the
uniformity in \xco\ derived from virial masses probably reflects fairly 
uniform conditions in CO-bright regions of molecular clouds
\citep{BOLATTO2008}.  The spatial scale at which virialized structures emerge
in galaxies is unclear. Given these ambiguities and the assumptions
involved in the calculation of virial masses, we emphasize the need
for careful comparison to matched data to interpret virial results.

\subsection{Extragalactic Dust-Based Estimates of \xco}
\label{sec:normgal_dust}

Dust is expected and observed to be well-mixed with gas, and dust
emission remains optically thin over most regions of normal
galaxies. Following the approach outlined in \S\ref{sec:mw_dust}, dust
emission offers a tool to estimate \xco. Modeling infrared or
millimeter emission yields an estimate of the dust optical depth,
$\tau_{d}$. Using Eq.
\ref{eq:H2_from_dust}, \xco\ can be estimated from $\tau_{d}$, \hi ,
CO, and the dust emissivity per H atom (Eq.
\ref{eq:emissivity_HI}). Conventions in the literature vary, with
$\tau_{\rm d}$ sometimes combined with a dust mass absorption
coefficient and used as a dust mass, and \dgr\ alternately cast as the
emissivity per H atom or a dust-to-gas mass ratio. Regardless of
convention, the critical elements are a linear tracer of the dust
surface density and a calibration of the relation between this tracer
and gas column density. 

\citet{THRONSON1988} and \citet{THRONSON1988b} first suggest and 
apply variations on this technique to nearby galaxies.
\citet{ISRAEL1997} uses data from IRAS to carry out the
first comprehensive dust-based extragalactic \xco\ study. He considers
individual regions in eight (mostly irregular) galaxies and derives
$\tau_{d}$ from a combination of 60 and 100$\mu$m continuum
data. \citeauthor{ISRAEL1997b} estimates \dgr, the dust emissivity per
H atom, from comparison of $\tau_{d}$ and \hi\ in regions within the
galaxy of interest but chosen to lie well away from areas of active
star formation, thus assumed to be atomic-dominated. This internally
derived estimate of \dgr\ represents a key strength of the approach:
\dgr\ is derived self-consistently from comparison of \hi\ and
$\tau_{d}$, and does not rely on assuming a dust-to-gas ratio. This
leads to the cancellation of many systematic errors \citep[e.g.,
see][]{ISRAEL1997b,LEROY2007}, leaving only variations in \dgr\ and
$\tau_{d}$ within the target to affect the determination of \xco\ in
Eq. \ref{eq:H2_from_dust}. 

\subsubsection{Dust-Based Estimates in Normal Disk Galaxies}

\begin{figure}[t]
{\psfig{figure=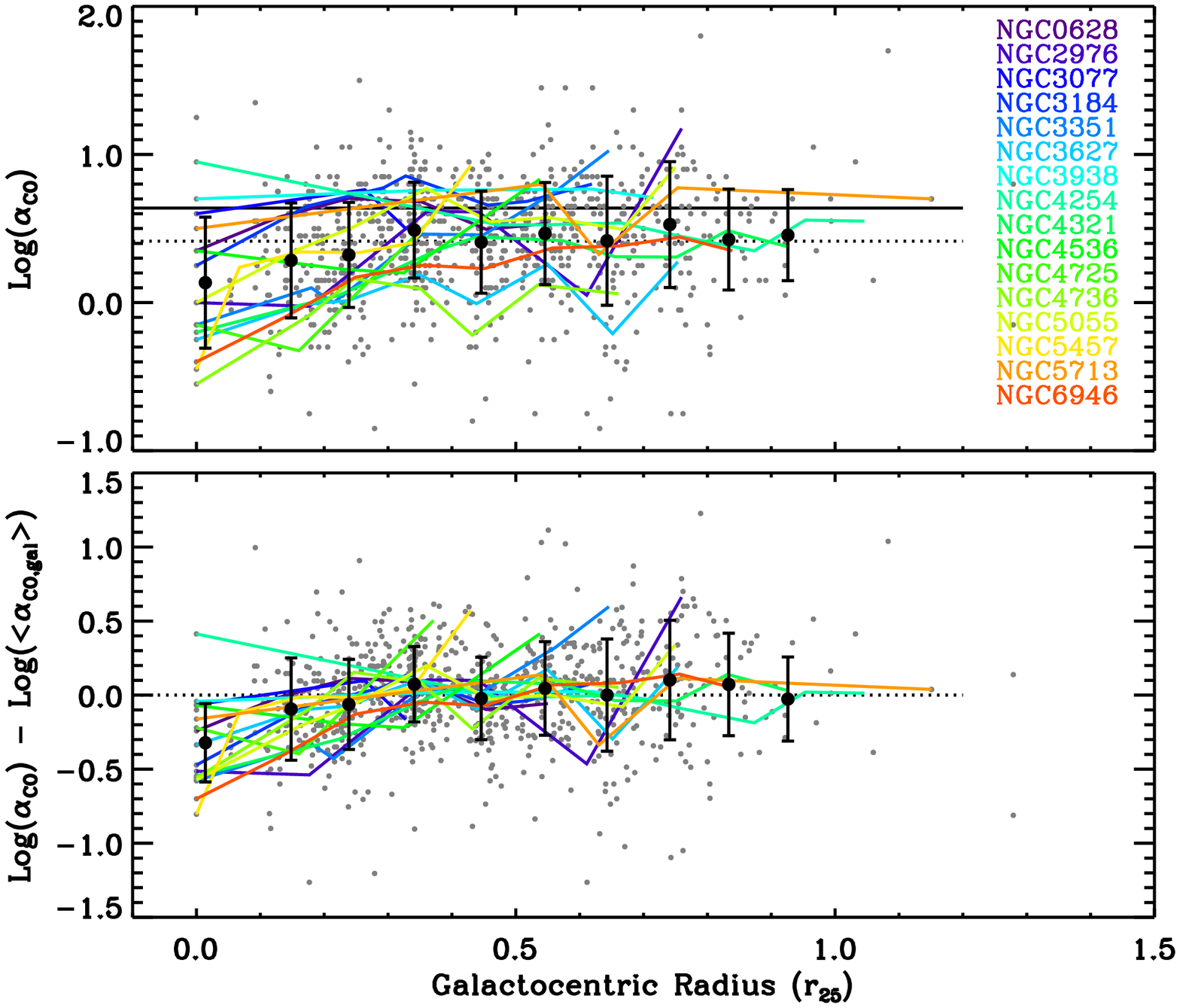,width=\columnwidth}
\caption{Dust-based \aco\ determination across the disks of 22 galaxies 
\citep{SANDSTROM2012}. The top panel shows the results for the
individual regions (gray points), averages in radial bins for each
galaxy (color lines), and the average trend for all data as a function
of radius (black circles with error bars showing the scatter in that
bin). The black solid line in the top panel corresponds to $\xcot=2$,
and the dotted line shows the average, weighting all solution pixels
equally ($\xcot=1.2$). Weighting by galaxy or CO intensity rather than
line of sight, the mean is $\xcot\approx1.4-1.8$ for low inclination
galaxies. The bottom panel shows each galaxy normalized by its mean
\xco\ to highlight the fact that many galaxies exhibit low \xco\ in
their centers.
\label{fig:radial_sandstrom}}
}
\end{figure}

{\em Spitzer} and {\em Herschel} allowed the extension of the dust
approach to more massive, more distant, and more ``normal''
galaxies. \citet{DRAINE2007} compare galaxy-integrated infrared
spectral energy distribution (SED) modeling to CO and \hi\
luminosities for a large sample. They argue that $\xcot\approx4$ (over
entire galaxies) yields the most sensible gas-to-dust ratio results in
their sample. \citet{LEROY2011} perform a self-consistent treatment of
the Local Group galaxies M~31, M~33, LMC, NGC~6822, and SMC. They find
$\xcot\approx 1-4.5$ for regions of M~31, M~33, and the LMC (the
higher metallicity galaxies in the sample). \citet{SMITH2012} use
{\em Herschel} observations to solve for \xco\, finding
$\xcot\approx2$ in M~31.

Other dust tracers such as mm-wave continuum emission and visual
extinction have been used to arrive at \xco\ estimates, although
usually by assuming or scaling a Galactic calibration with the
associated systematic uncertainties. \citet{GUELIN1993} use $1.2$~mm
continuum observations of the edge-on spiral NGC~891 to estimate
$\xcot\sim 1$. Both \citet{NAKAI1995} and
\citet{GUELIN1995} study \xco\ in M~51, the first using extinction
estimates from H$\alpha$/H$\beta$, the second millimeter-wave
continuum emission. \citet{NAKAI1995} arrive at
$\xcot\approx0.9\pm0.1$ with a factor of $\sim 2$ variation with
galactocentric radius, while \citet{GUELIN1995} finds
$\xcot\sim0.6$. \citet{ZHU2009} employs 850~$\mu$m data to check the
\xco\ derived from spectral line modeling in NGC~3310 and NGC~157,
finding good agreement with $\xcot\approx1.3-3.0$ in the disks of
their targets and a much lower \xco\ in the center of NGC~157.

\citet{SANDSTROM2012} carry out the most comprehensive
extragalactic study of \xco\ to date. They combine high-quality CO
$J=2\rightarrow1$ maps with {\em Herschel} and {\em Spitzer} dust
continuum, and high-resolution \hi\ data. Building on the method of
\citet{LEROY2011}, they break apart galaxies into regions
several kpc$^2$ in size and within each region they simultaneously solve for
\dgr\ and $\xco$. This yields resolved, self-consistent $\xco$
measurements across 22 galaxy disks. Their methodology requires good
S/N CO detections and so restricts robust \xco\ measurements to
reasonably CO-bright parts of galaxies, typically half of the optical
disks. There, the authors find $\xcot \approx1.4$--$1.8$
\footnote{The weighting used to derive the average \xco\ affects the mean
value. Weighting all high quality solutions equally,
\citeauthor{SANDSTROM2012} find mean $\xcot \approx 1.4$, median
$\xcot \approx 1.2$. Weighting instead by CO intensity, they find
$\xcot \approx 1.8$.} with a $1\sigma$ scatter among individual
solutions of $0.4$ dex.

The top panel of Fig. \ref{fig:radial_sandstrom} shows \xco\ derived
by \citeauthor{SANDSTROM2012} as a function of galactocentric radius for their
whole sample. The bottom panel shows \xco\ for each region normalized
to the average value for the galaxy, highlighting the internal radial
structure. This study finds clear central \xco\ depressions for a
number galaxies.  The value of \xco\ in galaxy centers relative to
their disks spans a range from no change to a factor of five below the
galaxy average. The authors note that these low central \xco\
values tend to coincide with high stellar masses and enhanced CO
brightness, suggesting that the effects discussed for starbursts in
\S\ref{sec:starbursts} may also be at play in the central parts of
many galaxies, including the Milky Way (\S\ref{sec:milky_way}).

Thus in the disks of normal star-forming galaxies, a dust-based
approach yields $\xcot\approx 1-4$ on kpc scales.  For comparison, in
the Milky Way extinction yields $\xcot\approx1.7-2.3$ and dust
emission yields $\xcot\approx1.8-2.5$ (\S\ref{sec:mw_dust}, Table
\ref{tab:xcomw}). Thus, broadly {\em dust analyses strongly
support a Milky Way conversion factor in the disks of normal, massive
disk galaxies} but methodological differences and, presumably, real
changes in \xco\ with environment produce a factor of 2 spread
among studies and galaxies.

\subsubsection{Dust-Based Estimates in Dwarf Irregular Galaxies}

Dust-based determinations of \xco\ in low metallicity dwarf irregular
galaxies consistently yield values much higher than Galactic, and also
generally higher than virial estimates. \citet{ISRAEL1997} finds in
the SMC, the lowest metallicity target, a notably high
$\xcot\approx120\pm20$. \citet{BOSELLI2002} carry out a similar
calculation and also find systematically higher \xco\ at low
metallicity. Subsequent studies leveraging the sensitivity,
resolution, and wavelength coverage of {\em Spitzer} and {\em
Herschel} qualitatively support this conclusion. \citet{LEROY2007},
\citet{LEROY2009}, and \citet{BOLATTO2011} follow \citet{ISRAEL1997} and
\citet{STANIMIROVIC2000} in analyzing the SMC. They find $\xcot\approx40-120$
using measurements that isolate individual clouds and estimate \dgr\
locally. \citet{GRATIER2010} considers NGC~6822 using a similar
technique, also finding high $\xcot\approx 40$ on large
scales. Attempting to minimize systematics and making very
conservative assumptions, \citet{LEROY2011} still find $\xcot \approx
10$ and $40$ for NGC~6822 and the SMC, significantly higher than they find
for the higher metallicity M~31, M~33, and the LMC.

In the Magellanic Clouds, {\em Herschel} and millimeter-wave bolometer
cameras have mapped individual molecular
clouds. \citet{ROMAN-DUVAL2010} find an increase in \xco\ in the
poorly shielded outer regions of LMC clouds. \citet{RUBIO2004},
\citet{BOT2007}, and \citet{BOT2010} find unexpectedly bright
millimeter-wave emission in SMC clouds. \citet{BOT2010} show that the
emission suggests a factor of four higher masses than returned by a
virial analysis, even on very small scales and even assuming a very 
conservative rescaled Milky Way dust mass emissivity coefficient.

Therefore, {\em dust-based \xco\ estimates indicate high conversion
factors in low metallicity, irregular systems.} This agrees with some
other methods of estimating \xco\ but not
with virial mass results, particularly on small scales, as already
discussed (\S\ref{sec:normal_virial}). We discuss the
discrepancy further below (\S\ref{sec:low_metallicities}). The simplest
explanation is that virial masses based on CO emission do not sample
the full potential well of the cloud in low metallicity systems, where
CO is selectively photodissociated relative to \htwo\ at low
extinctions.  

\subsubsection{Caveats on Dust-Based \xco\ Estimates}

Because dust traces the total gas in the system with little bias
(at least compared to molecules),
this technique represents potentially the most direct way to estimate
\xco\ in other galaxies. We highlight two important caveats:
the possibility of other ``invisible'' gas components and variations
in the emissivity.

As described, the method will trace all gas not accounted for by \hi\ 21~cm
observations and assign it to \htwo. The good qualitative and quantitative
association between \htwo\ derived in this manner and CO emission 
suggests that \htwo\ does represent
the dominant component \citep[e.g.,][]{DAME2001,LEROY2009}, but opaque
\hi\ and ionized gas may still host dust that masquerades as ``\htwo''
using Eq. \ref{eq:H2_from_dust}. Both components could potentially
represent significant mass reservoirs.
Their impact on dust-based \xco\
estimates depends on the methodology. Any smooth component evenly
mixed with the rest of the ISM will ``divide out'' of a
self-consistent analysis. Moreover, the warm ionized medium exhibits a
large scale height compared to the cooler gas discussed here and may
be subject to dust destruction without replenishment \citep[e.g.,
see][]{DRAINE2009}. \citet[][]{PLANCKCOLLABORATION2011XXI} find that
dust associated with their ionized gas template represents a minor
source of emission. On the other hand, absorption line experiments
from our galaxy and others suggest that $\sim 20\%$ \hi\ may be missed
due to opacity effects \citep{HEILES2003} and observations of M~31
suggest that it may lie in a filamentary, dense component
\citep[][]{BRAUN2009}, making it perhaps a more likely contaminant in 
\xco\ determinations.
Nonetheless, the close agreement between \xco\ derived from dust and
other determinations in our Galaxy strongly suggests that
contamination from dust mixed with other ``invisible'' components,
particularly the ionized gas, represents a minor correction.

Concerning the second caveat, the methodology relies on either blindly
assuming or self-consistently determining the dust emissivity per
hydrogen, \dgr. This is a combination of the dust-to-gas mass ratio,
and the mass absorption coefficient of dust at IR wavelengths. In
blind determinations, a linear scaling of Eq. \ref{eq:emissivity_HI}
with metallicity is usually assumed.  In self-consistent \xco\
determinations, \dgr\ is measured locally through the ratio of the
dust tracer to \hi\ somewhere within the galaxy (for example, in an
\hi-dominated region), and ideally close to or inside the region where
\xco\ is estimated. This makes self-consistent determinations very robust.
Even the most robust analyses, however, still assume \dgr\ to be
constant over some region of a galaxy and along a line of sight, and
between the atomic and molecular phases. The \dgr, however, may change
across galaxies and between phases due to metallicity gradients,
varying balance of dust creation and destruction, or changes in dust
grain properties.

As discussed in \S\ref{sec:mw_dust}, recent {\em Planck} results
suggest only mild localized variations in the emissivity per H nucleon
across the Milky Way. Observations of other galaxies do show that
\dgr\ depends on metallicity \citep{DRAINE2007,MUNOZ-MATEOS2009}. 
Large scale ISM density may also
be important; \dgr\ appears depressed in the low-density
SMC Wing and the outer envelopes of dwarf galaxies
\citep[][]{BOT2004,LEROY2007,DRAINE2007}. 
The balance of the dust production and destruction mechanisms is
complex: basic accounting implies that most dust mass buildup occurs
in the ISM \citep[][and references therein]{DWEK1998,DRAINE2009},
suggesting an increase in
\dgr\ in denser regions of the ISM where this accumulation must take
place. Independent of dust mass buildup, if grains become more
efficient emitters for their mass at moderate densities (through the
formation of fluffy aggregates, for example) then the effective \dgr\
will be higher at high densities. Though not strongly favored by the
overall {\em Planck} results, which suggest that emissivity variations
are localized, a number of authors point to evidence for an
environmental dependency of the dust emissivity
\citep[e.g.,][]{CAMBRESY2005,BOT2007}.

In our opinion these caveats concerning emissivity will lead a
dust-based approach to preferentially overpredict the amount of \htwo\
present and consequently \xco, as dust associated with the molecular
ISM will have a higher dust-to-gas ratio or be better at emitting in
the far-infrared. Nonetheless, we still see the dust-based approach as
the most reliable way of producing extragalactic \xco\ estimates.
Placing stronger quantitative constraints on these systematics
requires further work. The {\em Planck} results noted above, and the
close agreement in the Milky Way of dust-based
\xco\ with other techniques, suggests that emissivity variations are
not a major concern.

\subsection{Extragalactic Spectral Line Modeling}
\label{sec:normgal_lines}

Observations of multiple CO lines or a combination of CO and other
chemical species allow one to constrain the physical conditions that give
rise to CO emission. When these observations include optically
thinner tracers like $^{13}$CO, these constraints can be particularly
powerful. Due to sensitivity considerations, most multi-line data sets
have been assembled for bright regions like galaxy centers (or
starbursts, \S\ref{sec:starbursts}). From these data we have
constraints on \xco\ in bright regions with very different systematics
than either virial or dust-based techniques.

Generally, these results indicate that \xco\ lower than $\xcot=2$ is
common but not ubiquitous in the bright, central regions of galaxies.
This is in qualitative agreement with the independent \citet{SANDSTROM2012}
dust-based results already mentioned. For example,
\citet{ISRAEL2006} use large velocity gradient (LVG) modeling of
multiple \co, \cothree, and \ci\ lines in the central region of
M~51. They find $\xcot\approx0.25-0.75$ in the central regions, in
good agreement with earlier work by
\citet{GARCIA-BURILLO1993}. \citet{ISRAEL2009} and \citet{ISRAEL2009b}
extend this work to the centers of ten bright and
starburst galaxies, finding $\xcot\sim0.1-0.3$
\citep[see also][]{ISRAEL2003,ISRAEL2001}, again
significantly lower than Galactic. Based on maps of the
\co/\cothree\ ratio across NGC 3627, \citet{WATANABE2011} argue
for a similar central depression in \xco, with dynamical effects
associated with a galactic bar leading to broader line widths and
optically thinner \co\ emission.  Using optically thin tracers and dust 
emission, \citet{MEIER2001,MEIER2004} find that $\xcot\sim1-0.5$ in the
centers of IC~345 and NGC~6946. Lower conversion factors are not a
universal result, however. Utilizing high resolution data,
\citet{SCHINNERER2010} found \xco\ much closer to Galactic in the arms
of M~51, $\xcot\approx1.3-2.0$. They attribute the difference with
previous studies to the implicit emphasis on GMCs in their high
resolution data, speculating that it removes a diffuse CO-bright
component that would drive \xco\ to lower values.

In addition to degeneracy inherent in the modeling, these line
ratio-based techniques suffer from the fundamental bias of the virial
mass technique. They are only sensitive to regions where CO is bright
and so may miss any component of ``CO-faint'' \htwo. A handful of
observations of low metallicity dwarf galaxies have attempted to
address this directly by combining CO and \cii\ observations, with
\cii\ employed as a tracer of the ``CO-faint'' molecular regime 
\citep{MALONEY1988,STACEY1991}.
\citet{POGLITSCH1995} find that the \cii-to-CO ratio in the 30 Doradus
region of the LMC is $\sim 60,000$, roughly an order of magnitude
higher than is observed in Milky Way star forming regions or other
star-forming spirals \citep[e.g.,][]{STACEY1991}. \citet{ISRAEL1996}
and \citet{ISRAEL2011} also find high values elsewhere in the LMC and
SMC, with large scatter in the ratio. \citet{MADDEN1997} find
that regions in the Local Group dwarf IC~10 also exhibit very high
\cii\ 158$\mu$m emission compared to CO, 2--10 times that found in the
Milky Way. They argue that \cii\ emission cannot be readily explained
by the available ionized or \hi\ gas, and is thus an indicator of
\htwo\ where CO is photodissociated. \citet{HUNTER2001} present ISO
measurements for several more dwarf irregular galaxies, showing that
at least three of these systems also exhibit very high \cii-to-CO
emission ratios. \citet{CORMIER2010} present a {\em Herschel} map of the
158$\mu$m \cii\ line in NGC 4214, again showing very high \cii-to-CO
ratios ($\sim 20,000$--$70,000$) that cannot be explained by emission
from \cii\ associated with ionized gas. Inferring physical conditions
from the \cii\ line requires modeling. When such calculations are
carried out generally imply a massive layer on the photodissociation
region in which the dominant form of hydrogen is \htwo, while carbon
remains ionized as C$^+$ \citep[e.g.,][]{MADDEN1997,PAK1998}.

Thus, {\em detailed spectral line studies of CO-bright sources often,
but not always, suggest lower \xco\ in the bright central regions of
galaxies.} Meanwhile, {\em at low metallicities, \cii\ observations suggest an important reservoir of \htwo\ not traced by
CO}.

%
%

\subsection{Synthesis: \xco\ in Normal Galaxies}

Taken together, the picture offered by extragalactic \xco\
determinations in normal galaxy disks resembles that in the Milky Way
writ large (\S\ref{sec:mw_synthesis}). Virial masses, dust-based
estimates, and spectral line modeling all suggest $\xcot\approx1-4$ in
the disks of normal spiral galaxies. Systematics clearly still affect
each determination at the $50\%$ level, with physical effects likely
adding to produce the factor-of-two dispersion.  Given this, applying
a ``Milky Way'' $\xco=2 \times 10^{20}$~\xcounits\ with an uncertainty
of $\approx 0.3$~dex appears a good first-brush approach for normal
star-forming galaxies. This applies to galaxies where the CO
emission is dominated by self-gravitating clouds or cloud complexes
with masses dominated by \htwo.

Several very strong lines of evidence, as well as simple arguments,
show that \xco\ increases sharply in systems with metallicities below
${\rm 12+log[O/H]\approx8.4}$ \cite[approximately one-half
solar,][]{ASPLUND2009}. High spatial resolution virial estimates find
approximately ``Galactic'' \xco\ in low metallicity clouds, but they
are not sensitive to extended ``CO-faint'' \htwo\ envelopes, only to
CO-bright regions. We expand on the dependence of \xco\ on metallicity
in \S\ref{sec:low_metallicities}. We caution about the usefulness of
virial estimates on large spatial scales, particularly in low
metallicity galaxies. But even in normal galaxies, concerns exist
about whether molecular cloud complexes and associations are bound and
dominated by \htwo\ on large scales.

Finally, in the central parts of galaxies spectral line modeling
suggests that \xco\ is often, but not always, depressed in a manner
similar to that seen in the Galactic center and in molecule-rich
starbursts. Dust observations are consistent with this picture,
revealing central depressions in \xco\ in some galaxies. Broader line
widths, increased excitation, and the emergence of a diffuse molecular
medium likely contribute to more CO emission per unit mass, decreasing
\xco. We expand on this topic when we consider starbursts in
\S\ref{sec:starbursts}.

\section{\xco\ at Low Metallicities}
\label{sec:low_metallicities}

CO emission is frequently very faint or nonexistent in gas-rich,
actively star-forming low metallicity galaxies. This
is true in absolute luminosity terms, with CO emission 
almost completely absent in low mass dwarf irregular galaxies below
metallicity $12+\log {\rm~O/H}\approx8.0$
\citep{ELMEGREEN1980,TACCONI1987,TAYLOR1998}. Even when it is
detected, as in the case of the very nearby Small Magellanic Cloud, CO
is faint \citep{ISRAEL1986}. CO is also faint in a normalized
sense. The CO-to-FIR ratios in blue, vigorously star-forming dwarf
galaxies are clearly lower than in spiral galaxies
\citep{TACCONI1987,YOUNG1996}, and the same applies to other star
formation rate tracers such as H$\alpha$ \citep{YOUNG1996}. The CO per
unit starlight is also depressed in very low mass dwarfs
\citep[e.g.,][]{YOUNG1991}, and the ratio of CO to \hi\ is markedly
decreasing for later morphological type \citep{YOUNG1989}. 


Very importantly for CO emission, the low metal abundance in these
systems implies lower C and O abundances and low dust-to-gas ratios
\citep[e.g.,][]{DRAINE2007,MUNOZ-MATEOS2009}. Both of these factors
will exert a strong effect on the relative distributions of \htwo\ and
CO.  Dust serves as the site of \htwo\ formation and also provides
much of the far-ultraviolet shielding necessary to prevent molecules
that are not strongly self-shielding, such as CO, from
photodissociating. Note that galaxy mass is strongly correlated not
only with metallicity, but also with a number of other parameters that
may affect the equilibrium abundance of \htwo, such as spiral density
waves and interstellar pressure, further complicating the picture of
the relation between CO luminosity and molecular mass. The critical
question is whether low CO luminosities indicate a true deficit of
\htwo, or merely the suppression of CO emission in low metallicity
environments.

Theoretical considerations, described below, lead us to expect a
change in \xco\ at low metallicities. Several strong lines of
observational evidence, described in \S\ref{sec:normal_galaxies}, also
suggest that \xco\ increases sharply in irregular, low-metallicity
systems. Dust-based \xco\ estimates, which are sensitive to the total
gas present, return high values in such targets. The \cii -to-CO ratio
seems dramatically higher in dwarf irregular galaxies than in spiral
galaxies, supporting the hypothesis of a reservoir of \htwo\ at low
\av\ where most carbon is locked in \cii\ rather than CO. The
SFR-to-CO ratio also increases, offering circumstantial evidence for
star-forming gas not traced by CO.  

\subsection{Theoretical Expectations for Low Metallicity Gas}
\label{sec:lm_theory}

\begin{figure}[th!]
\centerline{\psfig{figure=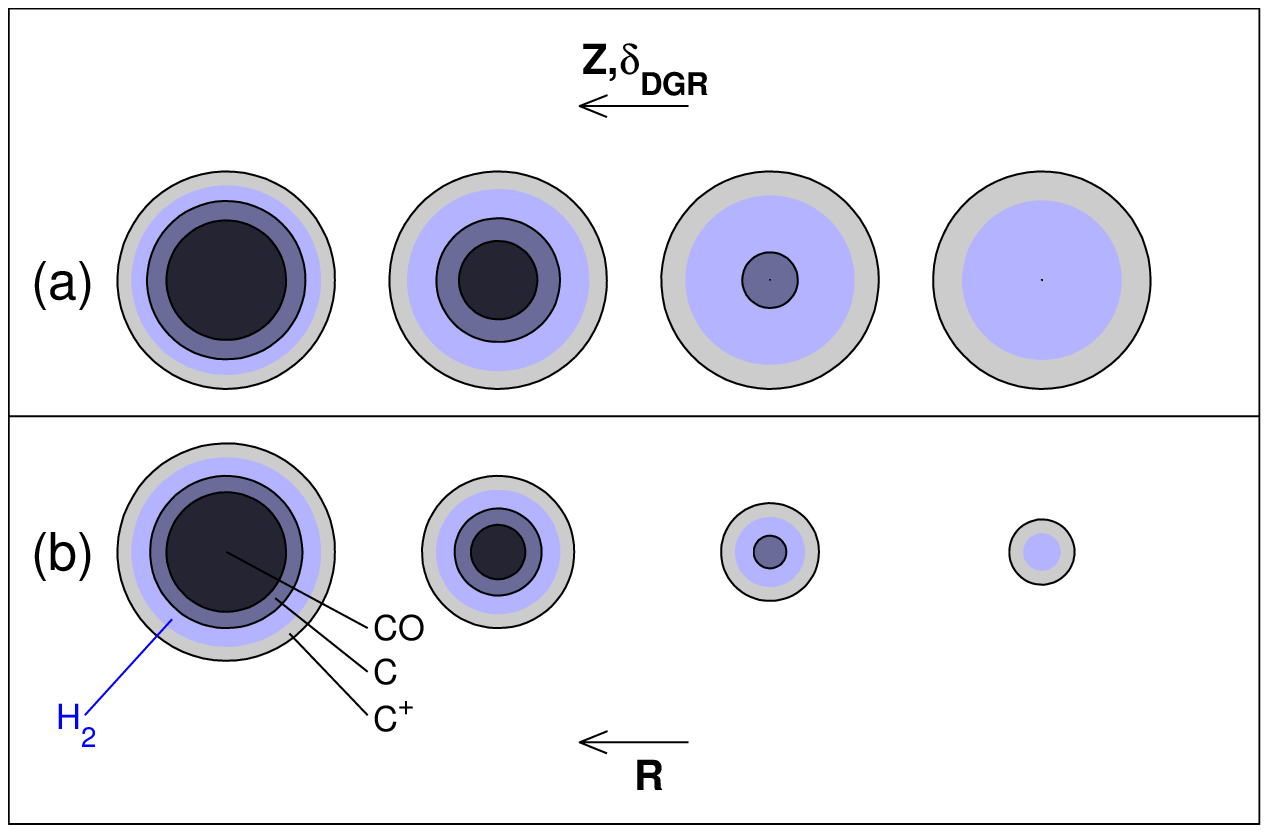,width=\columnwidth}
}
\caption{Effect of metallicity on CO and \htwo\ in a spherical clump
immersed in a uniform radiation field.  Blue shading indicates the
region where the gas is molecular, according to
Eq. \ref{eq:avh2main}. Increasingly darker shading shows the regions
where carbon is found as C$^+$, C, or CO.  The top sequence (a)
illustrates the effect of decreasing metallicity and dust-to-gas ratio
on the distribution of C$^+$, CO, and \htwo. Mostly because of the
increase in $N_{\rm H}$ required to attain a given \av, the CO
emitting region is pushed further into the clump until, for a fixed
cloud size, it disappears at low enough metallicities. The bottom
sequence (b) illustrates the effect of changing the clump size or
column density at a fixed metallicity \citep[adapted
from][]{BOLATTO1999,WOLFIRE2010}.\label{fig:clumpdrawing}}
\end{figure}

Naively, one would expect that a lower C and O would imply
consequently fainter CO emission. Because CO is optically thick,
however, its luminosity is determined by the emitting area and its
brightness temperature and velocity spread. The dependence of \xco\ on
metallicity hinges fundamentally on how the relative sizes of the
\cii\ and CO-emitting regions change with lower heavy element and dust
abundance (Fig. \ref{fig:clumpdrawing}, see the discussion of cloud
structure in \S\ref{sec:Cloud_Models}). \citet{MALONEY1988} carried out
one of the first thorough analyses of the problem. Here we note the
physical drivers for the location of bright CO emission and then
discuss the implications for low metallicity.

The relative distributions of CO and \htwo\ will be a detailed
function of the balance of formation and destruction of CO, processes
discussed for diffuse and dense material by \citet{VANDISHOECK1986,
VANDISHOECK1988} \citep[see also][appendices B and C]{WOLFIRE2010}. CO
formation proceeds mainly through the production of OH via ion-neutral
reactions initiated by cosmic-ray ionization. Once OH is formed the
chemistry proceeds via ion-neutral reactions to form HCO$^+$ followed
by the dissociative recombination of HCO$^+$ to form CO. The formation
rates of CO thus depend on the abundance of C times the abundance of
O. Because the rate of formation depends on the abundance of
OH, the rate of destruction of OH by far-ultraviolet photons
represents an additional limiting factor in setting the CO abundance
\citep[e.g.,][ Appendix C]{WOLFIRE2010}.

At depths where the CO line originates the destruction of CO proceeds
mainly via photodissociation by far-ultraviolet radiation. These
dissociations occur via line transitions between the ground state and
predissociated excited electronic states. That is, the majority of the
transitions to these excited electronic states result in dissociation
rather than CO in bound, excited levels. The longest wavelength
dissociating transition lies at $\lambda\approx 1076$ \AA . Thus the
dissociation of CO in the ISM occurs in the narrow far-ultraviolet
band between $\lambda\approx1076$ \AA\ and the Lyman limit at
$\lambda\approx912$ \AA . These dissociation bands for CO overlap the
Lyman bands of \htwo, and the Lyman lines of \hi. Thus
\hi\ and \htwo\ can shield CO.  Because CO dissociates in line
transitions, similar to \htwo, it can also self-shield at
sufficiently large CO column densities. In addition to gas opacity,
dust absorption and scattering will attenuate the dissociating
radiation field.

A simple expression illustrates the dependence of the CO dissociation
rate on these various process: $R = \chi R_0 \exp(-\gamma \av)\Theta$
s$^{-1}$.  The first two factors reflect the incident radiation
field. $R_0$ is the dissociation rate in free space for a radiation
field with a particular shape and $\chi$ is a scaling factor that
measures the strength of the radiation field. The exponential factor
reflects shielding by dust, with \av\ the line of sight extinction
into the cloud. The factor $\gamma$ includes both cloud geometry
and the translation from \av\ into extinction at the dissociating
wavelengths, accounting for absorption and scattering by grains in the
far-ultraviolet. The parameter $\Theta$ accounts for shielding by \hi\
and \htwo\ gas and CO self-shielding.

For a \citet{DRAINE1978} interstellar radiation field, $R_0 =
2.6\times 10^{-10}$ ${\rm s^{-1}}$ \citep{VISSER2009}; $R_0$ will vary
for radiation fields with other shapes. For typical ISM grains and
penetration into a plane parallel layer, $\gamma = 3.53$
\citep{VANDISHOECK2006}. The parameter $\Theta$ is a monotonically decreasing
function of the \hi, \htwo, and CO column densities and varies
between 1 (for no shielding) and 0. A fit to the line overlap and
self-shielding function, $\Theta$, can be found in \cite{VISSER2009}.


Balancing CO formation and destruction \cite{WOLFIRE2010} find an
expression for the depth into a cloud at which $\tau_{\rm CO}=1$
for the \jone\ transition:
\begin{equation}
A_V(\rco) \simeq 0.102 \ln
 \left[ 3.3\times 10^7 \left( \frac{\chi} {Z'\,n} \right)^2 + 1\right].
\label{eq:avcomain}
\end{equation}
\noindent Here $n$ is the gas density and $\chi$ and $Z'$ are
the far-ultraviolet field and gas phase abundances in units of the local
Galactic values. $A_V(\rco )$ is the depth in units of magnitudes of
visual extinction due to dust at which $\tau_{\rm CO}=1$. In
this expression the explicit dependence on dust-to-gas ratio
cancels out, but the conversion from $A_V$ to column density will
depend on the dust-to-gas ratio.

Variations in \xco\ will depend on the relative extent of the CO and
\htwo\ layers. Therefore, \citet{WOLFIRE2010} present a similar
expression for the depth at which the gas is half molecular
\begin{equation}                                                                
A_V(\rht) \simeq 0.142 \ln \left[ 5.2 \times 10^3 \dgrp 
           \left(\frac{\chi} {\dgrp \,n}\right)^{1.75}+1\right],                          
\label{eq:avh2main}                                                            
\end{equation}
where \dgrp\ is the dust-to-gas ratio relative to the local Galactic value.

The difference between the depth of the \htwo\ and CO layers is thus:
$ \Delta A_V = 0.53 - 0.045\ln \frac {\chi}{n}+0.107 \ln (\dgrp)-0.204 \ln (Z')$.
For a dust-to-gas ratio that scales as the gas phase metallicity we have
\begin{equation}
\Delta A_V = 0.53 - 0.045\ln \frac {\chi}{n}-0.097 \ln (Z')\,.
\label{eq:deltaav}
\end{equation}
\noindent The equations for $A_V(\rco)$, $A_V(\rht)$, and $\Delta A_V$ all 
show weak dependencies on $\chi$, \dgrp, and $Z'$. Thus
they may be expected to change only weakly with changing local
conditions and at lower metallicity, the location of the $\tau_{\rm
CO}=1$ surface, the H$_2$ transition, and the spacing between the two
all remain approximately fixed {\em in units of visual extinction due to
dust.} The ratio of dust to gas also drop with metallicity
\citep{DRAINE2007}. As a consequence of lower
dust abundance and approximately constant $\Delta\av$, the physical
depth of the $\tau_{\rm CO}=1$ surface shifts deeper into the cloud,
producing a larger surface layer of \htwo\ at low \av. A
correspondingly larger layer of \hi\ also exists beyond $A_V(\rht )$
but does not bear directly on this review.  {\em Thus, with lower
metallicity and correspondingly lower dust abundance, CO retreats
further into the cloud than \htwo.} Figure \ref{fig:clumpdrawing}
illustrates the expected interplay of \hi , \htwo , CO, and C$^+$ in a
spherical cloud and the impact of changing metallicity (top row) and
cloud size or column density (bottom row).

If $M(R_{\rm H_2})$ is the molecular mass
within the radius where the molecular fraction is 0.5, and $M({\rm
CO})$ is the mass within the radius of the $\tau_{\rm CO}=1$
surface, then the ``CO-faint'' gas fraction can be defined as $f =
[M(R_{\rm H_2})-M(R_{\rm CO})]/M(R_{\rm H_2})]$.  For a cloud with an
$r^{-1}$ density profile, corresponding to $M(r)\propto r^2 $, the
``CO-faint'' gas mass fraction is given by
\begin{equation}                                                                
f
= 1-\exp\left(\frac{-0.76  \davdg}{\dgrp\overline{N}_{22}}\right)                  
= 1-\exp\left(\frac{-4.0  \davdg}{\overline{A}_V}\right)\, ,                          
\label{eq:fDGAV}                                                                
\end{equation}
where $\davdg$ is the optical depth in the ``CO-faint'' gas layer
(Eq. \ref{eq:deltaav}), 
$\overline{N}_{22}$ is the mean H column density in the CO portion
of the cloud in units of $10^{22}$~\percmsq, and
$\overline{A}_V=5.26\dgrp\overline{N}_{22}$.

The retreat of the CO emitting surface was first noted by early
studies \citep{MALONEY1988,LEQUEUX1994}. This shrinking CO core leads
to lower beam filling factors, and thus a lower observed CO intensity
\citep[e.g.,][]{PAK1998,BOLATTO1999}. The result is that
\xco\ increases with decreasing metal abundance. \citet{MALONEY1988} 
find nearly equal CO and \htwo\ cloud sizes at solar metallicity and
so note that increasing CO abundance above solar will not
substantially change \xco. Recent observational estimates find the
fraction of ``CO-faint'' molecular gas mass to be $\sim 50$\% in the
Milky Way (see \S\ref{sec:mw_diffuse}). This allows slightly more room
for a changing \xco\ moving to supersolar metallicities.  For example,
a change from $\xcot\approx2$ to $\sim1.3$ might be plausible due to
the effect of increasing metallicity.

A secondary consideration is that the brightness temperature of CO may
also be affected by systematic excitation changes with
metallicity. The temperature of the gas represents a balance between
heating and cooling. If grain photoelectric heating dominates at the
$\tau_{\rm CO}=1$ surface, then to first order both the heating rate
(proportional to the dust-to-gas ratio for photoelectric heating) and
the cooling rate (proportional to metal abundance for line cooling)
will scale similarly with metallicity, yielding no change in gas
temperature with metal abundance. It remains unclear, however, whether
the photoelectric heating efficiency increases, decreases, or stays
the same at low metallicities \citep[e.g.,][]{ROLLIG2006,ISRAEL2011}.


In summary, models predict an increase in \xco\ for large regions of
low metallicity galaxies due to the contraction of the CO-emitting
surface relative to the area where the gas is \htwo\ for a fixed cloud
size.  This effect may be offset (or compounded) by a mild increase
(or decrease) in brightness temperature brought about by changes in
heating sources, chiefly the photoelectric heating effect. Note that
the usefulness of CO as a tracer of total
\htwo\ mass will ultimately break down at low enough metallicity, where
it will be found only at the highest column densities and in
well-shielded environments.

\subsection{Metallicity-Dependent Calibrations of \xco}
\label{sec:lm_calibrations}

\begin{figure*}[t]
\centerline{
\psfig{figure=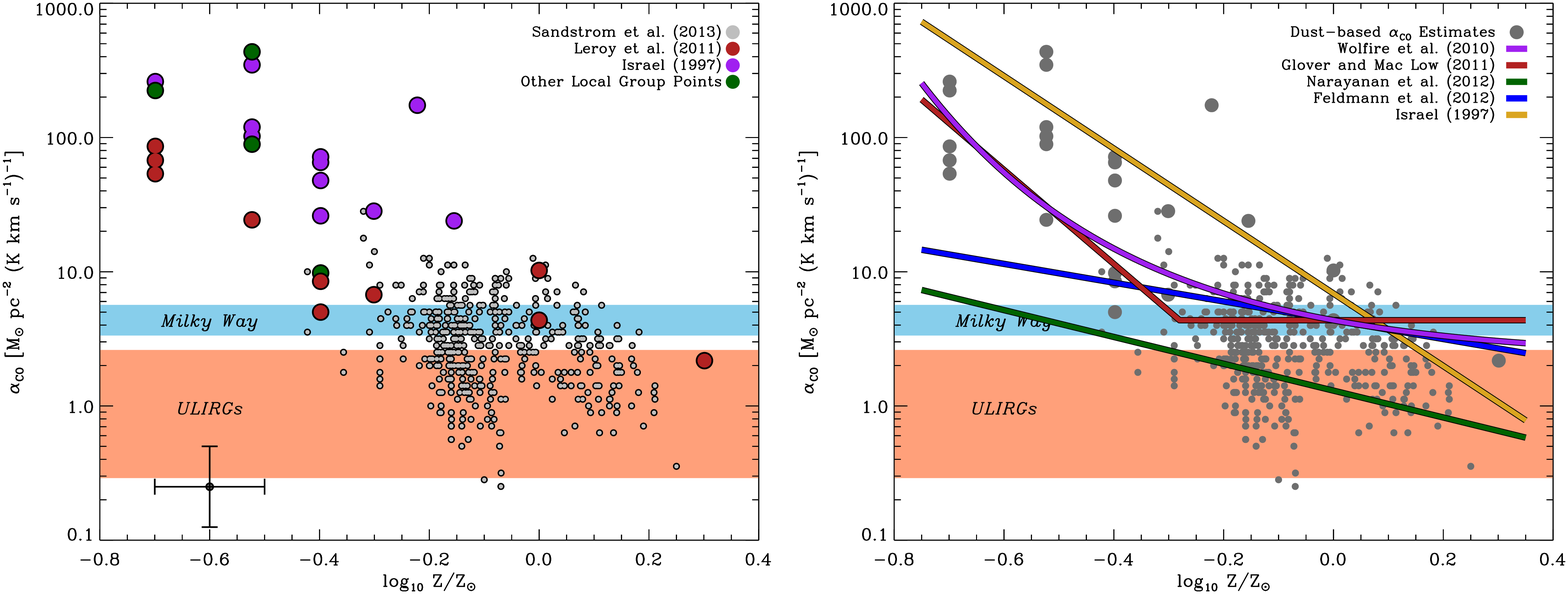,width=\textwidth}
}
\caption{Conversion factor, estimated from dust-based approaches,
as a function of gas-phase abundance. {\bf (Left)} Color points show estimates for
very nearby galaxies \citet{ISRAEL1997}, \citet[][based on \cii\
]{MADDEN1997}, \citet{LEROY2007}, \citet{GRATIER2010},
\citet{ROMAN-DUVAL2010}, \citet{LEROY2011}, \citet{BOLATTO2011}, and
\citet{SMITH2012}. Gray points show high quality solutions from
analysis of 22 nearby disk galaxies by
\citep{SANDSTROM2012}, with typical uncertainties illustrated by the error bars
near the bottom left corner. Metallicities are from
\citet{ISRAEL1997}, \citet{BOLATTO2008}, and \citet{MOUSTAKAS2010} and
quoted relative to solar in the relevant system \citep[$12+\log{\rm
[O/H]}=8.7$ for the first two, $12+\log{\rm [O/H]}=8.5$ for the latter
which uses the metallicity calibration by][]{PILYUGIN2005}. Note that
significant systematic uncertainty is associated with the
$x$-axis. The color bands illustrate our recommended ranges in \aco\
for the Milky Way and ULIRGs.  {\bf (Right)} Colored lines indicate
predictions for
\xco\ as a function of metallicity from the references indicated, 
normalized to $\xcot = 2$ at solar metallicity where necessary. For
these predictions, we assume that GMCs have $\left<
\Sgmc \right> = 100$~M$_\odot$~pc$^{-2}$, which we translate to a
mean extinction through the cloud using
Eq. \ref{eq:bohlin_gdr}. Dust-based determinations find a sharp
increase in \xco\ with decreasing metallicity below $Z \sim
1/3$--$1/2~Z_\odot$.}
\label{fig:lowz}
\end{figure*}

Calibrating \xco\ as a function of metallicity has represented a key
goal of extragalactic CO studies for two decades. The issue remains
complicated for several reasons: some practical, some theoretical. 
We have already discussed the biases and caveats of the different
\xco\ extragalactic estimators.
On practical grounds, the faintness of CO in low luminosity, low
metallicity systems renders observations very difficult. The abscissa
in any calibration, metallicity, also remains one of the hardest quantities
in extragalactic astronomy to measure with precision. The basic
cloud structure arguments presented in
\S\ref{sec:lm_theory} suggest that the sub- and supersolar regimes
should be treated differently --- or at least that a single power law
represents a poor choice across all regimes. We will discuss here
several existing calibrations of \xco\ with metallicity and contrast
them against the data, presenting a simple tentative formula that
includes metallicity effects in \S\ref{sec:conclusions}.

Some of the most comprehensive and widely used work comparing \xco\ to
metallicity relies mostly on virial mass-based \xco\ estimates 
\citep[e.g.,][]{WILSON1995,ARIMOTO1996,BOSELLI2002,BOLATTO2008}. 
These calibrations should best be viewed as calibrations of {\em \xco\
within the CO-emitting region,} and not be used to predict the total
\htwo\ content from CO on large scales. We suggest that the lack of an
\xco\ vs. metallicity trend in \citet{BOLATTO2008} and the weak trend in
\citet{WILSON1995} reflect the uniformity of opaque, bright
CO-emitting structures across many types of galaxies. Similarly, the
calibrations by \citet{ARIMOTO1996} and \citet{BOSELLI2002} rely on
virial masses to derive \xco\ as a function of metallicity
\citep[and so do][]{OBRESCHKOW2009}.


What are the appropriate calibrations to consider for \xco\ as a
function of metallicity? Dust-based approaches such as
\citet{ISRAEL1997} or the ``alternative approach'' of
\citet{BOSELLI2002} or combined CO, \cii , and dust modeling
\citep{MADDEN1997,PAK1998} appear to offer the best, or at least the 
most accessible, extragalactic approach. These have the potential to
capture the whole \htwo\ distribution. 
In the following pagraphs we employ these observations to test recent
theoretical work on the subject
\citep[e.g.,][]{WOLFIRE2010,GLOVER2010, NARAYANAN2012,FELDMANN2012}.
Note that, as discussed in
\S\ref{sec:lm_theory}, we do not expect \xco\ to change significantly
due to metallicity in the supersolar metallicity regime. Instead,
variations due to optical depth, line width, or excitation temperature
are likely to dominate.

Figure \ref{fig:lowz} presents an attempt at such a comparison. We
show an ensemble of dust and FIR determinations
\citep[][]{ISRAEL1997,MADDEN1997,LEROY2007,GRATIER2010,LEROY2011,BOLATTO2011,SMITH2012}
targeting local galaxies along with \xco\ estimates determined
\citet{SANDSTROM2012} for a sample of 22 nearby disk galaxies.  In the
right panel we plot theoretical predictions for \xco\ as a function of
metal abundance, assuming dust-to-gas ratio is linearly dependent on
metallicity. For \citet{WOLFIRE2010} we plot the equation

\begin{equation}
\label{eq:wolfire_xco}
\frac{\xco \left(Z' \right)}{\xco \left(Z^\prime = 1\right)} = {\exp{\frac{+4.0~\Delta \av}{{Z^\prime \overline{A}_{V,MW}}}}}{\exp{\frac{-4.0~\Delta \av}{\overline{A}_{V,MW}}}}\ ,
\end{equation}

\noindent which is obtained from Eq. \ref{eq:fDGAV} assuming that
dust-to-gas ratio tracks metallicity. Here 
$Z^\prime$ is the abundance of heavy elements and dust relative to solar,
$\xco \left(Z^\prime = 1\right)$ is the CO-to-H$_2$ conversion factor
at solar metallicity, and
$\overline{A}_{V,MW}$
is the mean extinction through a GMC at Milky
Way metallicity ($\overline{A}_{V,MW}\approx5$ for $\Sigma_{\rm
GMC}\approx100$~M$_\odot$pc$^{-2}$). 
The prediction for \citet{GLOVER2011} simply adapts their Equation 16, which gives \xco\
as a function of $A_V$. We assume that the mean extinction through a
cloud scales as the metal abundance, $A_V = Z^\prime A_{V,MW}$

\begin{equation}
\frac{\xco \left(Z^\prime \right)}{\xco \left(Z^\prime = 1\right)} = \left\{ \begin{array}{rl} 1 &\mbox{for~}Z^\prime \overline{A}_{V,MW} > 3.5 \\
\left(Z^\prime \overline{A}_{V,MW} \right)^{-3.5} & \mbox{for~}Z^\prime \overline{A}_{V,MW} < 3.5 \end{array} \right.
\end{equation}

\noindent In the full prediction of \citet{NARAYANAN2012}, \xco\ depends on a
combination of $I_{\rm CO}$ and $Z$, rather than metallicity alone. We
cannot readily place their predictions in this plot. Instead we plot
their Equation 6, in which \xco\ depends on $Z^\prime$ and
$\Sigma_{\rm H2}$. For each of these predictions, we assume
$\Sigma_{\rm H2} = 100$~M$_\odot$~pc$^{-2}$, translating this to
$A_{V,0}$ assuming $R_V=3.1$ and the dust-to-gas ratio of
\citet{BOHLIN1978}. We also plot the metallicity-only fit by
\citet[][note that his fit includes the radiation field, with a 
definition that renders it heavily degenerate with the
dust-to-gas ratio]{ISRAEL1997}.

Figure \ref{fig:lowz}
shows large scatter, even among determinations using similar
techniques for the same galaxy, but provides reasonable evidence for
an upturn in \xco\ with decreasing metallicity below $Z^\prime \sim
1/3$--$1/2$. The fit by \citet{ISRAEL1997} skirts the upper envelope
of measured values, while the prediction by \citet{FELDMANN2012}
predicts weaker-than-observed variations at very low $Z^\prime$,
though data remain scarce (those points represent only $\sim 4$
galaxies). The sharp nonlinear increases predicted by
shielding-oriented models like \citet{WOLFIRE2010} or the simulations
of \citet{GLOVER2011} provide the best fits to the existing data.

Less direct approaches to constrain \xco\ also exist. A recently
popular technique is to assume an underlying relationship between star
formation and molecular gas and to use this relationship and an
estimate of the recent star formation rate to arrive at the \htwo\
present. \citet{SCHRUBA2012} applied this technique to estimate \xco\
as function of metallicity in a sample of nearby
galaxies. \citet{GENZEL2012} combined a wide set of low and high
redshift measurements to estimate the dependence of \xco\ on
metallicity. Both studies find a significant dependence of \xco\ on
metallicity but with a wide range of possible power law exponents,
$\xco \propto Z^1$ to $Z^3$. The strength of this approach is
that the observations needed to make such estimates are widely
accessible. The weakness, of course, is that it requires assuming an
underlying relationship between H$_2$ and star formation. Any true
dependence of the star formation efficiency on metallicity, or any
other quantity covariant with metallicity, will be recast as
additional variations in \xco .

If \xco\ does increase rapidly moving to low metallicity, our
knowledge of the distribution function of molecular column density will present a practical limit to the
usefulness of CO to trace \htwo. 
At metallicities perhaps as high one half solar, half of the
H$_2$ mass will exist outside the CO-emitting surface, and that fraction
will rapidly increase for decreasing metallicity. Thus,
application of a CO-to-H$_2$ conversion factor at very low metallicity
ultimately involves extrapolating the total mass of a cloud from only
a small inner part; by $Z^\prime \sim 0.1$ this may already be
analogous to measuring the total H$_2$ mass of a Milky Way cloud from
HCN emission or some other high density tracer. In the Milky Way,
the fraction of cloud mass in high column density lines of sight can vary
dramatically from cloud to cloud
\citep[e.g.,][]{KAINULAINEN2009,LADA2010}.  In the LMC,
\citet{ISRAEL1996} found large region-to-region variations in the 
\cii-to-CO ratio.  As one moves to increasingly low metallicities, 
the use of CO emission to quantify the \htwo\ reservoir becomes more
and more extrapolative. While difficult to quantify, this effect
should add significant scatter that increases with decreasing
metallicity, and will eventually present a practical floor past which
CO is not a useful tracer of total \htwo\ mass.


\section{\xco\ in Starbursts and other Luminous Galaxies}
\label{sec:starbursts}

Molecular gas in starbursts exists under conditions very different
from those found in most normal galaxies. Observations of starbursts
suggest widespread gas volume and column densities much higher than
those typical of normal disks
\citep[e.g.,][]{JACKSON1995,IONO2007}. Molecular gas in 
starbursts is also warmer, exciting higher rotational transitions than
those found in less active objects
\citep[e.g.,][]{BRADFORD2003,WARD2003,RANGWALA2011}.  In fact, a negative
correlation is observed between molecular gas depletion time (a
parameter that characterizes how long a galaxy can maintain its
current star formation rate), and excitation
\citep[e.g.,][]{PAGLIONE1997}. Similarly, a positive correlation
exists between gas density and star formation rate 
\citep[e.g.,][]{GAO2004}. These observations show that there is a
fundamental relation between the density and temperature of the
molecular gas and the existence of starburst activity, such that the
gas present in starbursting galaxies or regions has higher densities
and temperatures than those prevalent in quiescent systems.

What are the effects of higher temperatures, densities, and column
densities on \xco? To first order, higher gas temperatures yield
brighter CO emission, decreasing \xco.
Note, however, that while
increasing the temperature decreases \xco, increasing the density and
surface density of the self-gravitating clouds of gas has the opposite
effect, increasing \xco\ (see Eqs. \ref{eq:virial_aco},
\ref{eq:aco_density}, and \ref{eq:aco_galaxy}). Therefore we expect
a certain level of compensation to occur, lessening the impact of
environment on the conversion factor as long as most of the CO
emission arises from GMCs. In regions where the average gas density is
comparable to that of a GMC, however, the entire medium will turn
molecular and CO emission will originate from an extended warm phase.
These conditions are thought to be prevalent throughout the active
regions of the brightest starbursts, such as ULIRGs. Adding to this
fact, many of these luminous galaxies are mergers, or the starbursts
occur in regions such as galaxy centers. The gas correspondingly
experiences motions in excess of the velocity dispersion due to its
self-gravity. The large column densities conspire to make the medium
globally optically thick, thus setting up the conditions discussed in
\S\ref{sec:other_sources}.  In this situation the CO emission will be
disproportionally luminous, driving \xco\ to lower values. 

We can quantitatively explore this scenario with PDR model calculations
\citep[adapted from][]{WOLFIRE2010}, which incorporate self-consistently the
chemistry, heating, and radiative transfer. Setting up a ``typical''
Milky Way GMC (a virialized structure of size $\sim30$~pc with
$\Smol\approx170$~M$_\odot$~pc$^{-2}$ and
$M\approx1\times10^5$~M$_\odot~$) we reproduce a Galactic \xco. In a
$\sim160$~pc virialized cloud with $\Smol\approx10^4$~M$_\odot$~pc$^{-2}$
and M$\approx2\times10^8$~M$_\odot$ (representing the molecular
structures observed in existing high-resolution observations of
ULIRGs) under the same Galactic conditions, we obtain $\xcot\sim80$
in rough agreement with Eq. \ref{eq:virial_aco}. Increasing the 
gas velocity dispersion to include $2\times10^9$~M$_\odot$ of stars
(see \S\ref{sec:other_sources}), and increasing the UV and cosmic
ray fluxes by $10^3$ to account for the larger SFR, decreases the
conversion factor to $\xcot\sim0.6$. Most of this effect is due to
the velocity dispersion: increasing the UV and cosmic ray fluxes by
only $10$ with respect to the Galactic case yields $\xcot\sim2$, still
much lower than the starting value of $80$.

\citet{NARAYANAN2011} 
explore these effects in detail using a series of computational models
of disks and merging galaxies. They find that \xco\ drops throughout
the actively star-forming area in merger-driven starbursts due to
increased gas temperatures (caused in part by collisional thermal
coupling between dust and gas, which occurs at high densities), and
the very large velocity dispersion in the gas (in excess of
self-gravity) which persists for at least a dynamical time after the
burst.  The magnitude of the drop in
\xco\ depends on the parameters of the merger, with large \xco\
corresponding to low peak SFR. Thus, the large drop in \xco\ occurs in
massive mergers during the starburst phase, and \xco\ settles to
normal values when the star formation activity and the conditions that
caused it subside. The simulated normal disk galaxies experience less
extreme conditions of density and turbulence, and accordingly possess
higher values of \xco\ except in their centers. The mean values of the
CO-to-\htwo\ factor for the simulated mergers and disks are
$\xcot\approx0.6$ and $4$ respectively, with a very broad distribution
for the mergers.  In a follow-up study \citet{NARAYANAN2012} introduce
a calibration of \xco\ where it becomes a function of ${\rm W(CO)}$,
as well as metallicity, $Z$. This calibration captures the fact that
the factors that cause a drop in \xco\ occur increasingly at higher
surface densities. Note, however, that because \xco\ is a non-linear
function of ${\rm W(CO)}$ obtained from luminosity-weighted
simulations, the calibration must be applied carefully to
observations. The observed CO intensity corresponds to the intrinsic
intensity multiplied by a filling factor $f_{beam}<1$, while the
luminosity-weighted ${\rm W(CO)}$ employed by the model-derived
calibration is much closer to the intrinsic CO intensity, which is not
directly observed because observations do not completely resolve the
source. 

In the following sections we will discuss the observational findings
in starburst galaxies in the local universe.

\subsection{Luminous Infrared Galaxies}
\label{sec:sb_LIRGs}

One of the earliest comprehensive studies of the state of the
molecular gas and the value of \xco\ in a luminous infrared galaxy
(LIRG) was performed on the prototypical starburst
M~82. \citet{WILD1992} use multi-transition observations and detailed
excitation calculations to determine the proportionality between the
optically thin C$^{18}$O emission and the \htwo\ column
density.  Bootstrapping from this result, they find that \xco\ varies
across the disk of M~82, and is in the range
$\xcot\approx0.5-1$.  They attribute this low
\xco\ mostly to high temperatures, as they estimate
$T_{kin}>40$~K for the bulk of the gas from their modeling. Using a
similar technique, \citet{PAPADOPOULOS1999} analyze the inner region
of NGC~1068, a well known starburst with a Seyfert 2 nucleus. They
conclude that a diffuse, warm, molecular phase dominates the $^{12}$CO
emission, while the mass is dominated by a denser phase that is better
observed in the C$^{18}$O isotopologue. The $^{12}$CO \jone\ emission
from the diffuse phase has low optical depth ($\tau_{1}\sim1-2$) and
is not virialized, and is thus overluminous with respect to its
mass. They argue that the CO-to-\htwo\ conversion factor in the
nuclear region is $\xcot\sim0.2-0.4$, although the precise value
depends critically on the assumed CO abundance.

Along similar lines, \citet{ZHU2003} perform a multi-transition
excitation study of molecular gas in the Antennae pair of interacting
galaxies (NGC~4038/9). They find that the \htwo\ gas mass in their
analysis depends critically on the $^{12}$CO/\htwo\ abundance ratio,
which is an input to their model. They conclude that
$^{12}$CO/\htwo$\sim0.5-1\times10^{-4}$, and consequently
$\xcot\sim0.2-0.4$ for the center of NGC~4038 and $\xcot\sim0.5-1$ for
the ``overlap'' region between both galaxies. They argue that the
adoption of this smaller-than-standard \xco\ is consistent with the
gas distribution, including \hi, in the interacting pair, while the
very large mass for the ``overlap'' region that would result from
adopting a standard
\xco\ would be very difficult to reconcile with the gas
dynamics. The authors find that their excitation analysis points
to gas with high velocity dispersion, large filling fraction, and low
optical depth as the reason why CO is overluminous in the Antennae.
Recently, \citet{SLIWA2012} find also a low \xco\ in their excitation
and dynamical analysis of Arp~299 using interferometric data,
$\xcot\approx0.2-0.6$.

The broad conclusions of these in-depth studies are in agreement with
findings from studies carried over large samples. \citet{YAO2003}
survey 60 local infrared-luminous (starburst) galaxies spanning FIR
luminosities $L_{\rm FIR}\sim10^9 - 10^{12}$ L$_\odot$, analyzing
their \co\ \jone\ and \jthree\ emission. Using the dust temperature as
a proxy for the gas temperature, and assuming coextensive emission and
a CO/\htwo\ abundance ratio similar to what would be expected if all
the gas phase carbon was locked in CO molecules for Milky Way
abundances, they conclude that $\xcot\sim0.3-0.8$ is consistent with
the available data.

\citet{PAPADOPOULOS2012} perform a detailed analysis on 
another large sample of luminous infrared galaxies, with $L_{\rm
FIR}\gtrsim10^{11}$~L$_\odot$. They find that one-phase radiative
transfer models generally match the observations for the lower $J$
transitions of CO with a typical $\xcot\sim0.3$
($\aco\approx0.6\pm0.2$~\acounits), a value compatible with previous
studies \citep[e.g.,][]{DOWNES1998}. The authors conclude that
although the gas temperature is partially responsible for the lower
\xco, the most important factor is the gas velocity dispersion. They
also point to a caveat with this result, which has to do with the
possible existence of a dense, bound phase with lower velocity
dispersion and a much higher \xco. This phase is associated with dense
gas tracers such as high-J CO or heavy rotor molecules (HCN, CS,
HCO$^+$).  It has the potential to dominate the molecular mass of the
system, and raise the conversion factor to $\xcot\sim2-6$. Because the
uniqueness of this explanation for the observed excitation is unclear,
and the derived masses conflict in some cases with dynamical mass
estimates, we consider the results of multi-component models an
interesting topic for further research.
\citeauthor{PAPADOPOULOS2012} also point out that although the global CO \jone\
luminosity is dominated by the warm, low \xco\ component, it is
possible to hide a cold, normal \xco\ component that could add a
significant contribution to the molecular mass of the system. This
component would most likely be spatially distinct, for example an
extended molecular disk, and thus could be separated in high spatial
resolution studies with interferometers.


\subsection{Ultraluminous Infrared Galaxies}
\label{sec:sb_ULIRGs}

Ultra-luminous infrared galaxies (ULIRGs) are extreme cases of
dust-enshrouded stabursts and active galactic nuclei, with
far-infrared luminosities in excess of L$_{\rm
FIR}\sim10^{12}$~L$_\odot$.  These objects are the products of
gas-rich major mergers, and possess very large CO luminosities
\citep{MIRABEL1988,SANDERS1996}. Despite that, their ${\rm L_{FIR}/L_{CO}}$
ratios frequently exceed those found in spiral galaxies, including
interacting pairs, and Milky Way GMCs. The fundamental question is how
those CO luminosities relate to their molecular gas masses. Understanding
\xco\ in ULIRGs is particularly interesting because they likely
provide the best local templates for the most luminous high-redshift
submillimeter galaxies, which have very high star formation rates compared
to other galaxies of similar stellar mass at their redshift 
\citep[i.e., are off the main sequence,][]{TACCONI2006,NARAYANAN2010}.

Studies presented in a series of papers
\citep{DOWNES1993,SOLOMON1997,DOWNES1998,BRYANT1996,BRYANT1999} show that the molecular gas masses
obtained for ULIRGs using the Galactic disk CO-to-\htwo\ conversion
factor are uncomfortably close to (or exceed) their dynamical masses,
suggesting that the Galactic \xco\ overestimates their total molecular
mass. These authors develop a consistent one-component model that
explains the high-resolution observations as rotating, highly
turbulent remnants of the merging process. The CO emission is
dominated by low density gas, and although it is optically thick it is
only moderately so \citep[for example, see also][]{IONO2007}. This
lower molecular mass is consistent with the observed (optically thin)
millimeter dust continuum emission for Galactic dust-to-gas ratio. The
``typical'' \xco\ they derive is $\xcot\sim0.4$, approximately a
factor of 5 lower than in the Milky Way disk
\citep[$\aco\sim0.8$~\acounits, with individual results ranging between 0.3
and 1.3 in Table 9;][]{DOWNES1998}.  

Detailed high resolution studies of individual ULIRGs find similar
results. \citet{BRYANT1996,BRYANT1999} study the kinematics of seven
LIRGs and ULIRGs, determining interferometric sizes and dynamical
masses.  They conclude that the use of a Milky Way conversion factor
results in molecular masses larger than the dynamical mass of the
system in the cores of all seven objects, although in some cases that
can be explained away assuming a face on orientation. They find
$\xcot<0.7$ and $\xcot<1.5$ for Mrk~231 and NGC~6240
respectively.
Modeling the CO kinematics of Arp~220, including the effects of the
stellar components in the dynamics of the nuclear disk,
\citet{SCOVILLE1997} find $\xcot\sim1$. These estimates are comparable with those from
one-component radiative models for the same galaxies, and typically
much lower than the two-component results
\citep{DOWNES1998,PAPADOPOULOS2012}.



\subsection{Synthesis: \xco\ in Starbursts}

There has been much recent progress, through both observations and
modeling, on the determination of \xco\ in starbursts. Studies of CO
excitation in large samples, as well as detailed studies of individual
cases, strongly point to lower than Galactic values of \xco,
particularly in extreme starbursts such as ULIRGs. These low values of 
\xco\ are driven by a globally molecular medium coupled with high gas 
temperatures and, more importantly, very high velocity dispersions in
the CO emitting gas due to a combination of merger activity and the
stellar potential
\citep{NARAYANAN2011,PAPADOPOULOS2012}. The standard practice has been
to adopt $\aco\approx0.8$~\acounits\ \citep{DOWNES1998}, a value
similar to the $\aco\approx0.6\pm0.2$~\acounits\ resulting from recent
one-component modeling in a large sample \citep{PAPADOPOULOS2012}.
{\em Note, however, that this is an average value and there are likely
large galaxy-to-galaxy variations} \citep{NARAYANAN2011}. To first
order these variations should be mostly related to the total surface
density of the galaxy, if the gas is bound and experiencing the global
gravitational potential (c.f., Eq. \ref{eq:aco_star}). We will use
this to suggest a tentative \xco\ correction in
\S\ref{sec:conclusions}.

Large uncertainties exist 
in the estimates from observations, stemming from assumptions about
\co/\htwo\ and \co/\cothree\ ratios, coextensive emission in the
different transitions, and the general need for simplifications in
excitation models. Note also, as a persistent caveat, the possibility
of hiding significant molecular mass in ULIRGs in a low velocity
dispersion, dense component. This posibility, however, seems
disfavored in some cases where high quality dynamical estimates
are available.


\section{\xco\ at High Redshifts}
\label{sec:sb_highz}

The improved sensitivity of millimeter and centimeter-wave instruments
has allowed detection of CO emission from an increasingly diverse
range of systems out to higher and higher redshifts. Most of the
objects observed in CO to date represent the bright, rare end of the
luminosity distribution: so-called submillimeter galaxies (SMGs) and
QSO hosts. Some recent studies, however, target rotating disks that
lie at the high-mass end of the star-forming blue sequence at their
redshift (i.e., have typical star formation rates given their stellar
mass --- we will refer to them as ``main sequence'' galaxies). CO
observations thus now begin to sample the regime of ``main sequence''
galaxies \citep{TACCONI2010,DADDI2010}, and will expand to lower
luminosity systems over the next decade.

Estimating molecular gas masses is frequently the main goal of CO
observations at high redshifts. Unfortunately, direct determination of
\xco\ in high redshift objects remains tremendously challenging. An
additional complicating factor is that most high-redshift
observations do not measure the \jone\ transition of CO, but higher
rotational transitions. Thus translating these measurements into
molecular masses requires, at least in principle, understanding CO
excitation in addition to the other physics governing \xco .

At this stage, the modeling of optically thin isotopologues
\citep[e.g.,][]{PAPADOPOULOS2012} may offer the best opportunity for
direct \xco\ measurements at high redshift.  Such observations have
been too costly to undertake, and even with ALMA may only be practical
for the brightest objects. Dust continuum observations offer another
route to estimate molecular masses, although their application at high
redshift requires understanding dust-to-gas ratios in relation to
metallicities --- a complex problem that ultimately requires knowledge
of the balance of ISM enrichment, dust production, and destruction
processes in galaxies. In lieu of direct measurements, the best route
is to understand the physical drivers of
\xco\ and to apply knowledge acquired from local galaxies to systems
at high redshift. 

\subsection{Observed CO Line Ratios}

The ratio of the $J=3\rightarrow2$ to $J=1\rightarrow0$ transition,
$r_{31} \equiv T_3/T_1$, exhibits a wide range of values across
galaxies and has been observed at both low and high
redshift. \citet{MAUERSBERGER1999} and \citet{YAO2003} consider
$r_{31}$ for large samples of nearby galaxies with infrared
luminosities $9\lesssim \log(L_{\rm FIR})\lesssim 12$. They find a
mean $r_{31}\approx 0.63$--$0.66$ (median $r_{31} \sim 0.5$) and a
broad distribution of values.  No strong correlations between $r_{31}$
and dust temperature or luminosity are evident, but they note that
$r_{31}$ increases with increasing concentration and star formation
efficiency (measured as $L_{\rm FIR}/M(\htwo)$, though note the CO
intensity in the denominator).  \citet{IONO2009} examine 15 luminous
LIRGs and ULIRGs with $\log(L_{\rm FIR})\gtrsim11.5$ and find mean
$r_{31}\approx 0.48$ (median $0.4$) integrated over whole galaxies,
with a higher mean $r_{31}\approx 0.96$ at the location of peak CO
emission. 
\citet{MAO2010} discuss $r_{31}$ in a sample of over 60
galaxies, finding that barred galaxies and starbursts have the higher
averaged $r_{31}$ values ($r_{31}\sim0.89\pm0.11$ for starbursts),
followed by AGNs (see their Table 4).  More recently,
\citet{PAPADOPOULOS2011} report the results of a large sample of LIRGs
and ULIRGs composed of new observations and literature compilation,
where they find mean $r_{31}\approx0.67$ and $r_{21}\approx0.91$.  By
contrast with these samples of active galaxies, the inner portion of
the Milky Way has a ratio $r_{31}\approx 0.28\pm0.17$
\citep{FIXSEN1999}, denoting a correspondingly lower excitation in a
$z=0$ ``normal'' star forming galaxy. 


To first order $r_{31}$ offers a tool to distinguish excited,
star-forming gas similar to that seen in LIRGS and ULIRGS
from the gas found in more normal systems like the Milky Way. As
emphasized by a number of authors, however, these line ratios are
hardly unambiguous indicators of local conditions. For example,
\citet{PAPADOPOULOS2011} show that for their sample of LIRGs and
ULIRGs possible densities and temperatures are often poorly
constrained,
with data allowing high density ($n\sim10^4$~\percmcu), normal
temperature ($T_{kin}\sim20$~K) or low density ($n\sim300$~\percmcu)
high temperature ($T_{kin}\sim150$~K) solutions.
Note as a limitation
the assumption of similar filling fractions for galaxy-wide averages
in all transitions, while high-resolution maps frequently find
excitation gradients. It is thus key to appreciate that $r_{31}$,
although useful as a first order excitation indicator, combines the
effects of both excitation and beam filling.

With the improving sensitivity of the available instrumentation,
measurements of line ratios including CO \jone\ at high redshift are
rapidly increasing. For SMGs, the presumed high redshift simile of
local ULIRGs, the measured line ratios resemble those found for local
LIRGs and ULIRGs. \citet{HARRIS2010}, \citet{SWINBANK2010}, and
\citet{IVISON2011} report mean $r_{31}\approx0.68$, 0.66, and 0.55
respectively for a handful of SMGs. \citet{CARILLI2010} and
\citet{BOTHWELL2012} note that fitting observed SMG spectral line
energy distributions requires a cool and a warm component, with most
of the mass in the former. \citet{RIECHERS2011b} find low integrated
$r_{41}\approx0.25-0.6$ in two SMGs, noting the existence of
excitation gradients, with CO \jone\ in a more extended distribution
than the higher transition. At the brightest end of the galaxy
distribution \citet{RIECHERS2011} report $r_{31}\sim0.96-1.06$ for four
high-redshift SMGs hosting quasars, suggesting that the excitation in
these sources is such that the emission is thermalized to higher
rotational transitions \citep[see also][for a CO excitation discussion
of a number of bright SMG and QSO-host high-redshift
galaxies]{WEISS2007}.

The lowest galaxy luminosities currently probed by high redshift CO
observations correspond to the high luminosity end of the star-forming
galaxy main sequence. Unlike SMGs or local ULIRGS, these systems are
not major mergers. Excitation data remain scarce for these ``main sequence''
disk galaxies. \citet{DANNERBAUER2009} observed a $BzK$ galaxy at
$z\approx1.5$ report $r_{32}\sim0.5$ and $r_{21}\sim0.6$. \citet{ARAVENA2010}
report $r_{31}\approx0.61$ and $r_{21}\sim0.4-1.2$ in a few $BzK$ galaxies.
These observations have large uncertainties, but suggest that the
conditions in high-z disk galaxies are not as extreme at those in
local ULIRGs.
As a counterpoint, \citet{RIECHERS2010} discuss the CO
\jone\ detection of two highly magnified ($\mu\sim30$) 
low mass galaxies where they find $r_{31}\approx0.72$ and
$0.78$ with errors of $\sim0.2$, suggesting higher excitation than the
previous examples if there is no differential lensing. For further
discussion see the review by Carilli \& Walter in this same issue.

\subsection{Estimates of \xco\ in High Redshift Systems}
\label{subsec:HighRedshift}

A number of observational studies have considered the problem of \xco\
in high-redshift sources, combining line ratio measurements,
consistency arguments, and scaling relations to argue for plausible
values.

\begin{figure}[th!]
\centerline{\psfig{figure=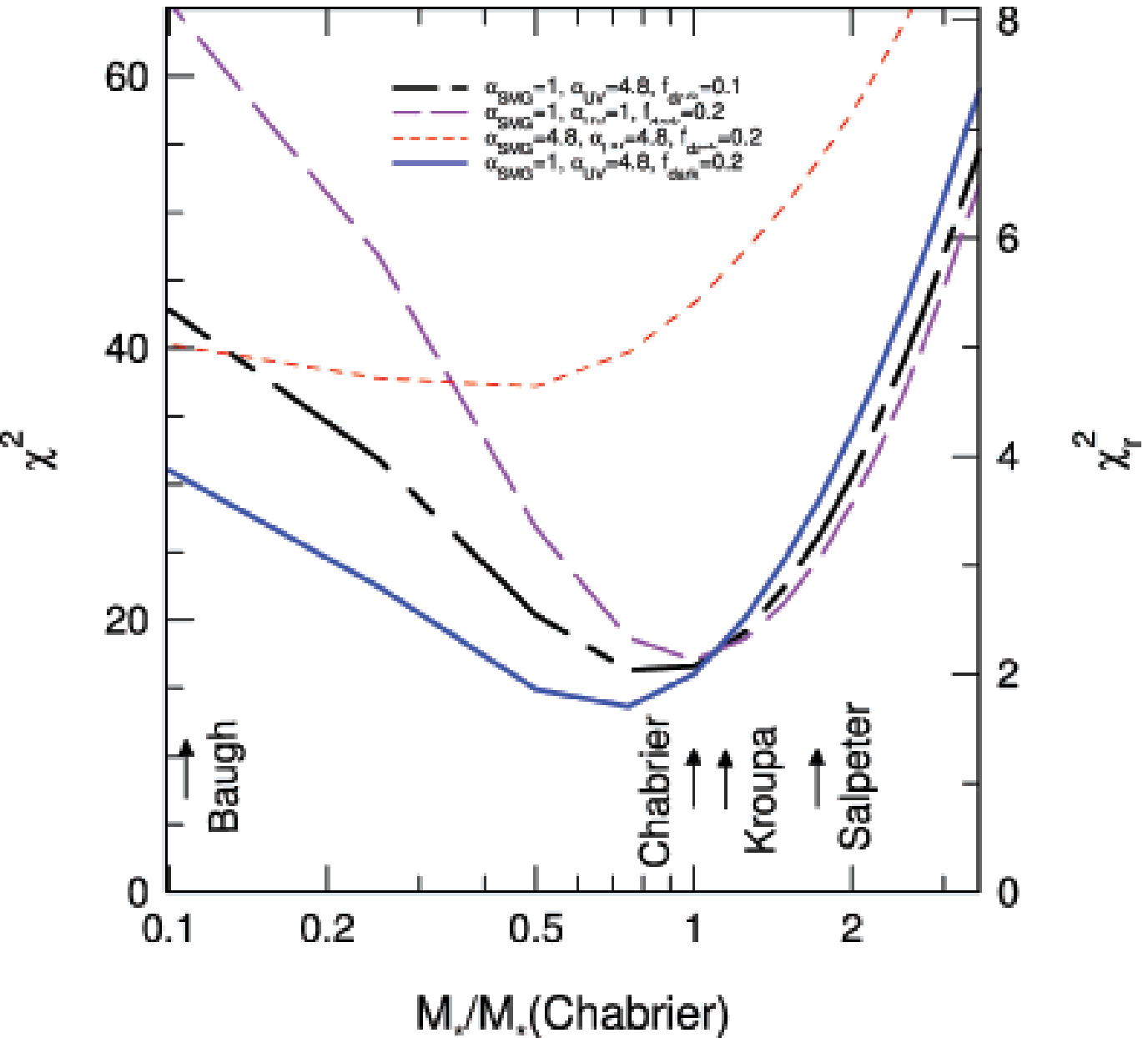,width=\columnwidth}}
\caption{Minimization of the difference between dynamical
mass and the sum of the gaseous, stellar, and dark matter components,
for a sample of 9 high-z galaxies with high quality data, dominated by
SMGs \citep{TACCONI2008}. $\chi^2$ and reduced $\chi^2$ (left and
right side axes, respectively) are shown here as a function of chosen
Initial Mass Function, for different gas ($\alpha$) and dark matter
($f_{dark}$) parameters. The analysis distinguishes between $\alpha$
in the SMG subsample ($\alpha_{\rm SMG}$) and in the UV selected subsample
($\alpha_{\rm UV}$) of lower mass galaxies.
\label{fig:Tacconi2008}}
\end{figure}

\citet{TACCONI2008} discuss the values of \xco\ applicable to a
sample of high-z star-forming galaxies dominated by SMGs but also
including lower mass galaxies.  Allowing for a dark matter
contribution of $10\%-20\%$ within a 10~kpc radius, they minimize the
$\chi^2$ of the difference between the dynamical mass and the sum of
the stellar, gas, and dark matter. They find a Galactic \xco\ to be
strongly disfavored, while a smaller, ULIRG-like $\xcot\approx0.5$
produces much more satisfactory results for
SMGs (Fig. \ref{fig:Tacconi2008}).

In their SMG sample, \citet{IVISON2011} find 
$\xcot\sim0.9-2.3$ to be compatible with the observed star
formation histories and dynamical and stellar masses. They argue for
lower $\xcot\lesssim 0.5$ in most cases, however, based on
star formation efficiency considerations. They decompose the molecular
gas into a warm, star-forming phase (with intrinsic $r_{31}\sim0.9$)
and a cold, quiescent phase ($r_{31}\sim0.3$), and invoke an
approximately maximal starburst for the star-forming phase. The
maximal starburst, $L_{\rm
IR}/M(\htwo)\approx500$~L$_\odot$~M$_\odot^{-1}$, represents the
largest allowed ratio of star formation to \htwo\ before radiation
pressure disperses a starburst \citep[e.g.,][]{THOMPSON2005}.

While \citet{TACCONI2008} and \citet{IVISON2011}
point to similarities between ULIRGs and SMGs, note that
\citet{BOTHWELL2010} find that several SMGs have gas distributions
(imaged in $J\geq3$ transitions) more extended than those of local
ULIRGs. This may suggest that local ULIRGs may not be the best
analogues of SMGs, and that some SMGs may not arise from major
mergers. The excitation gradients seen by \citet[][]{IVISON2011} and
\citet{RIECHERS2011} also point to extended quiescent gas reservoirs
in SMGs that are not frequently observed in local ULIRGs.

\citet{DADDI2010b} and \citet{GENZEL2010} study high-redshift galaxies 
selected from optically identified objects at $z\sim1.2-2.2$. These
are still massive systems, but drawn from the main sequence at their
redshifts rather than major merger-driven starbursts. Kinematically,
most of these systems are extended rotating disks, although with
larger velocity dispersions than local disks. \citet{DADDI2010},
\citet{GENZEL2010}, and
\citet{TACCONI2010} argue that, because their CO emission likely
arises from collections of self-gravitating GMCs \citep[as suggested
by local estimates of Toomre Q,][]{GENZEL2011} with modest
temperatures $T\sim20-35$~K, these systems have Galactic
\xco. Additional constraints come from dynamical measurements that,
together with simulations and modeling of the stellar populations,
yield a gas mass resulting in $\xcot\approx1.7\pm0.4$, a value
considerably higher than observed in ULIRGs or estimated for SMGs
\citep{DADDI2010b}.  Note that these authors also provide a
calibration for the dynamical mass estimation in this type of galaxy.

A different approach to quantify the amount of gas in a galaxy is to
try to use the continuum dust emission. \citet{MAGDIS2011} use this
approach in two very well studied high-z galaxies, a massive SMG at
$z\sim4$ and a ``main sequence'' disk star-forming galaxy at $z\sim1.5$. They
take advantage of the well characterized SEDs of these objects to
calculate their dust masses employing different approaches to dust
modeling \citep[traditional modified blackbody as well as the type of
models used by ][]{DRAINE2007}.  This dust mass is converted into a
gas mass (assumed to be dominated by molecular gas) using a
metallicity derived through the corresponding SFR-mass-metallicity
relation \citep[e.g.,][]{MANNUCCI2010} and assuming a Galactic
dust-to-gas ratio corrected (approximately linearly) by
metallicity. The result is $\xcot\sim0.4\pm0.2$ for the SMG, and
$\xcot\sim1.9\pm1.4$ for the disk galaxy. \citet{MAGNELLI2012} follow
a similar procedure in a mixed sample of 17 high-z galaxies, including
``merger-like'' and ``main sequence'' disks, finding values of \xco\
consistent with Galactic for the ``main sequence'' disks and and a factor of
several lower for the ``merger-like'' galaxies. They also find
$\xco\propto T_{dust}^{-0.8}$, similar to the behavior expected from
Eq. \ref{eq:aco_density}.

Although this new use of dust observations at high-redshift is
promising, we caution that going from dust SEDs to gas masses involves
a large chain of assumptions that is fraught with potential
problems. At a fundamental level, we are far from understanding the
dust creation-destruction balance which sets the dust-to-gas ratio in
galaxies \citep[e.g.,][and references
therein]{DRAINE2009,DWEK2011}. Moreover, the translation of a SED into
a dust mass relies {\em almost entirely} on dust grain properties
(composition, mass emissivity, size distribution) that have been
derived for the Milky Way \citep{DRAINE2001,DRAINE2007b}, and are
extremely uncertain even for local galaxies
\citep[e.g.,][]{GALLIANO2011}. Note that the self-consistent
dust-based \xco\ estimates discussed \S\ref{sec:normgal_dust} are
emphasized precisely because they are free to a large degree from
these problems. Nonetheless, this is an interesting approach that
reinforces what seems to be the outstanding trend discussed above.
ULIRG-like conversion factors for SMG (merger) galaxies, and Milky
Way-like conversion factors for disks at high redshift.

\begin{figure}[th!]
\centerline{\psfig{figure=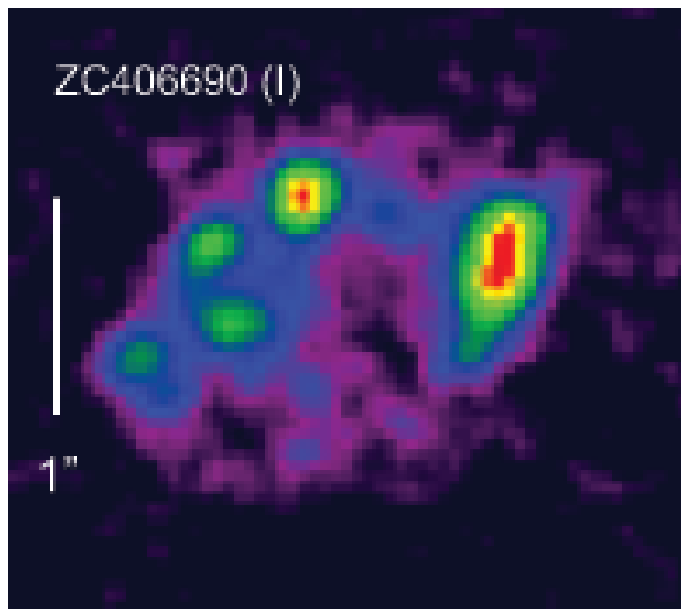,angle=0,width=10pc}
\psfig{figure=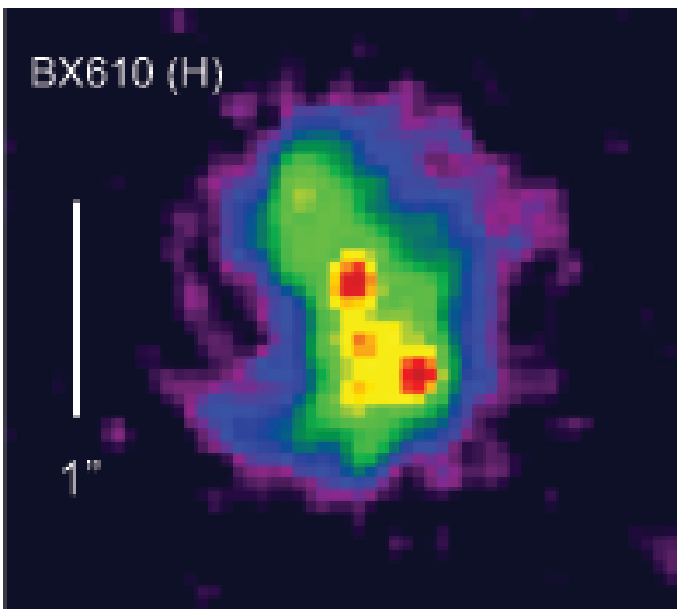,angle=0,width=9.8pc}}
\centerline{\psfig{figure=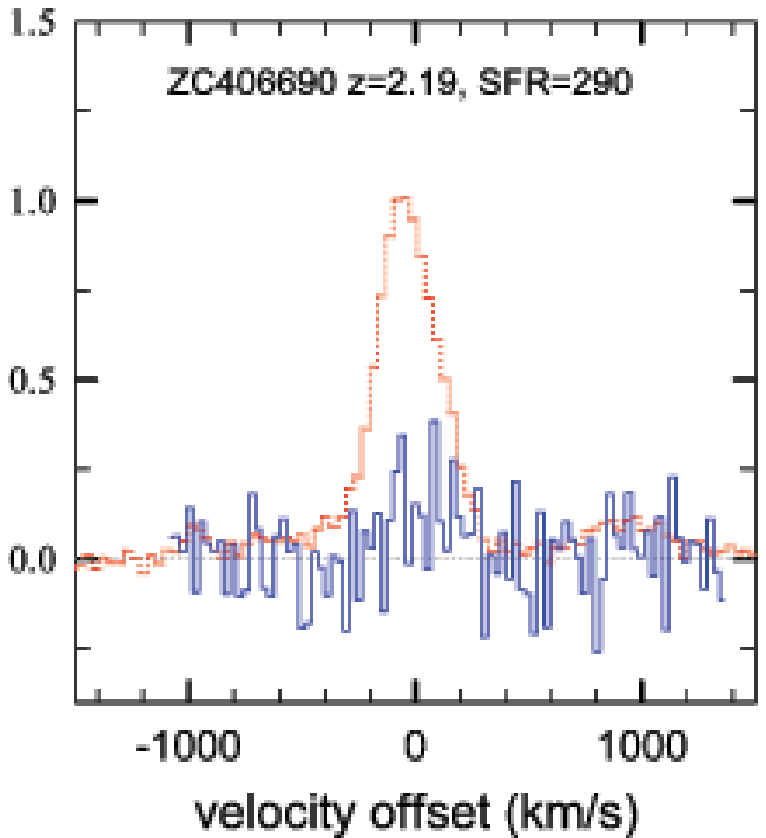,angle=0,width=10pc}
\psfig{figure=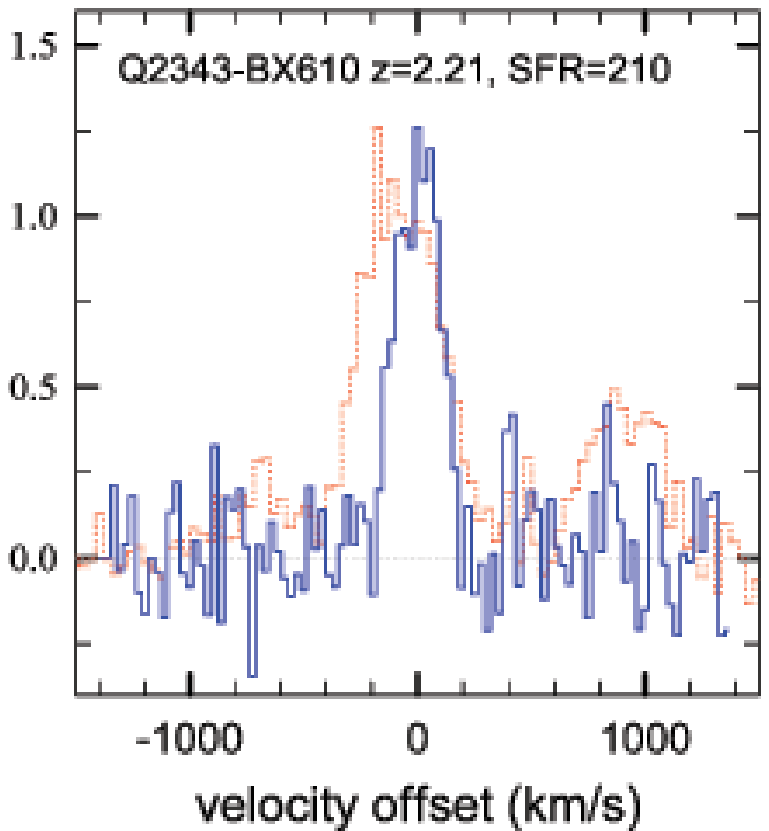,angle=0,width=10.5pc}}
\caption{Metallicity effects in high redshift ``main sequence''
galaxies \citep{GENZEL2012}. The top panels show rest-frame H$\alpha$
images of zC406690 and Q2343-BX610, two $z\sim2$ galaxies with similar
rotational velocities ($v_{rot}\approx224$ and $216$ \kmpers\
respectively), star formation rates (SFR$\approx480$ and $212$
M$_\odot$~yr$^{-1}$, including extinction correction), and stellar
masses (M$_*\approx4.3$ and $17\times10^{10}$ M$_\odot$). The bottom
panels show the corresponding rest frame optical spectrum containing the
H$\alpha$ and \nii\ transitions at 6563 and 6585 \AA\ (red), and the 
CO \jthree\ emission (blue). The galaxy zC406690 has a low metallicity,
indicated by a low \nii/H$\alpha$ ratio, and a correspondingly low
CO emission, despite its large H$\alpha$ flux and star formation activity.
\label{fig:genzel2012}}
\end{figure}

\citet{GENZEL2012} present 44 low-mass high-redshift galaxies, showing that
observations are starting to probe the low metallicity regime at high
redshift. The authors find the effect already discussed for local low
metallicity galaxies in
\S\ref{sec:low_metallicities}, where CO is disproportionally faint
for the star formation activity (Fig. \ref{fig:genzel2012}). By
assuming that these galaxies obey the same gas surface-density to star
formation relation observed in local disks \citep[e.g.,][]{LEROY2008},
\citeauthor{GENZEL2012} estimate \htwo\ masses and derive a
metallicity-dependent calibration for \xco\ at high redshift. They
find $\xco\propto Z^{-1.3}-Z^{-1.8}$, in approximate agreement with local
measurements \citep{LEROY2011,SCHRUBA2012}.

\subsection{Synthesis: \xco\ at High Redshift}

Estimates of the appropriate \xco\ for high redshift galaxies rely
largely on scaling and consistency arguments. Because they probe the
physical state of the molecular gas, line ratio measurements represent
a particularly powerful tool for such arguments. Given the
difficulties in observing the \jone\ line, they also represent a
critical measurement to understand how to translate observations of
high-$J$ transitions into CO \jone. Studies of local LIRGs and ULIRGs
show $r_{31}\gtrsim0.5$ globally in these sources, and $r_{31}\sim1$
when focusing on the active regions. Similar ratios are observed in
very massive galaxies at high redshifts (SMGs), where
$r_{31}\sim0.5-0.7$ for integrated fluxes, and $r_{31}\sim1$ is
observed in the very compact starbursts surrounding QSOs. CO \jone\
data for main sequence star-forming galaxies at high-z remain scarce,
making it is difficult to assert whether they possess large low
excitation reservoirs of molecular gas. As the data quality,
resolution, and availability improves, we expect that the simple
single-component models used today will evolve into more realistic
multi-component models for the molecular ISM of these sources.

A number of authors have explored the value of \xco\ applicable to
samples of high-redshift galaxies. The consensus, validated to first
order by numerical modeling \citep[e.g.,][]{NARAYANAN2011}, is that massive
merger-driven starbursts such as SMGs are most consistent with a low
\xco\ similar to local ULIRGs, while blue-sequence galaxy disks most
likely have higher \xco , similar to local disks. This is reasonable
in terms of the physics that drive the value of \xco. High density
environments with an extended warm molecular phase not contained in
self-gravitating clouds will result in low \xco, while molecular gas
contained in collections of self-gravitating GMCs will have \xco\
close to the Galactic disk value. We expect the most heavily
star-forming of these disks to have higher \htwo\ temperatures, and
consequently somewhat lower \xco\ values. Because of the opposing
effects of density and temperature in self-gravitating GMCs (for
example, Eq. \ref{eq:aco_density}), however, it appears that values as
low as those observed in ULIRGs will only occur when the CO emission
is dominated by an extended warm component that is not self
gravitating
\citep[e.g.,][]{PAPADOPOULOS2012}.

Multi-line studies including both low- and high-$J$ transitions allow
more rigorous constraints on the density and kinetic temperature of
the gas \citep{PAPADOPOULOS2012}.  
In the near future, multi-line studies of CO
isotopologues or paired observations of CO and optically thin sub-mm
dust emission will offer additional, more direct constraints on
\xco, though with their caveats and shortcomings.

An exciting development at high redshift is the emergence of main
sequence ``normal'' galaxy surveys that begin to sample the lower
metallicity regime prevalent at early times.  In these samples we see
\xco\ effects that are consistent with those 
already discussed for local galaxies (\S\ref{sec:low_metallicities}):
metal-poor galaxies are underluminous in CO for their star formation
activity \citep{GENZEL2012}. This situation will be increasingly
common, and more extreme, as CO surveys sample lower galaxy masses and
earlier times: by $z\sim3-4$ galaxy metallicities drop by $0.5-0.7$
dex for a fixed stellar mass, and the characteristic galaxy mass also
becomes lower \citep{MANNUCCI2009}. Ultimately the best probe of the
molecular gas content of these low metallicity young galaxies may be
the fine structure line of \cii, or perhaps the dust continuum.



\section{Conclusions and Open Problems}
\label{sec:conclusions}

We have summarized the efforts to measure \xco\ in the Milky Way
and other galaxies, as well as the theoretical arguments and studies
that show that CO can be used as a tracer of molecular mass, under
certain conditions, through the adoption of a CO-to-\htwo\ conversion
factor.
 
In the following paragraphs, we would like to offer concise answers to a few
key questions that have been developed elsewhere in this review:

\begin{itemize}
\item What is the ``best'' value for the Milky Way \xco?

\item Under what circumstances is the assumption of ``Galactic \xco'' likely
to be good or bad?

\item How well can we calibrate \xco\ as a function of metallicity?

\item What is the most appropriate value of \xco\ in ULIRGs?

\item What can we say about \xco\ in the distant universe?

\item Can we offer a practical prescription for \xco\ based on observables?

\item What observations and calculations are necessary to move forward?
\end{itemize}

Measurements in the Milky Way have achieved a very good level of
sophistication and consistency, beyond what is possible in external
galaxies.  We have shown that there is an assuring degree of
uniformity among the large scale values of \xco\ obtained through
different techniques, particularly in the inner disk ($1$~kpc$\lesssim R
\lesssim 9$~kpc). {\em We recommend adoption of a
constant $\xco=2\times10^{20}$~\xcounits\ ($\aco=4.3$~\acounits) with
an uncertainty of $\pm0.1$ dex (a factor of $1.3$) for the inner disk
of the Milky Way.} The evidence for a large scale Galactic \xco\
gradient and its magnitude are, presently, at best unclear, and the
simplicity of a constant conversion factor is preferable. Nonetheless,
there is convincing evidence that \xco\ in the Milky Way center region
is smaller than in the disk by factors of $3-10$, and this is
reaffirmed by extragalactic observations that find that a low \xco\ is
not uncommon in other galaxy centers.  Following the results obtained
by a number of studies, we recommend using
$\xco\approx0.5\times10^{20}$~\xcounits\ for $R\lesssim500$~pc in the
Milky Way. The uncertainty is difficult to quantify, but taking the
different measurements obtained as well as the results discussed for
galaxy centers and starbursts, we recommend $\pm0.3$ dex (a factor of
$2$). We expect that further CO multitransition modeling,
$\gamma$-ray, and dust continuum studies will help constrain better
this value as well as that of a possible large scale Galactic
gradient.  For the outer Milky Way we expect \xco\ to increase, in
principle, following the same physics underlying its increase in low
metallicity environments, as we discuss below.

It is important to recognize that these are average numbers, strictly
valid for GMCs on scales of tens of pc. The validity of invoking
anything like a constant \xco\ on a line-of-sight by line-of-sight basis
is considerably less defined. Both theoretically and
observationally, we have shown that considerable variation can exist
on small scales, reflecting local chemistry and physical conditions.

The most mature techniques in ``normal'' galaxies remain virial mass
measurements and the use of dust as an optically thin tracer. Both
techniques have their drawbacks, and we particularly emphasize the
ambiguous (at best) sensitivity of virial mass measurements to any
extended envelope of H$_2$ mixed primarily with ${\rm C^+}$ rather than
CO. At solar metallicities, a wide range of measurements yield $\xcot
\approx 1$--$4$, but with large (still $\gtrsim$ factor of 2) scatter
and uncertainties related to the dynamical state of clouds and the
environmental dependence of dust properties. In the absence of further
characterization or studies, {\em we recommend the conservative
approach of adopting $\xco = 2\times10^{20}$~\xcounits\ with an
uncertainty of $\pm0.3$ dex (a factor of $2$) in the disks of normal,
solar metallicity galaxies.} This applies to galaxies where the CO
emission is dominated by self-gravitating \htwo\ clouds or cloud
complexes.  This value can approximately be applied down to
metallicities $\sim0.5 Z_\odot$, and in regions where the total gas
plus stars surface density is $\lesssim300$~\msunperpcsq. ALMA will
greatly expand the application of both dust and virial mass
techniques. Other future prospects will include modeling of resolved
\cii\ emission, now widely available thanks to {\em Herschel}, the
extension of spectral line modeling beyond bright galaxy centers,
and further exploitation of galaxy scaling relations.

\begin{figure}[t!]
\centerline{
\psfig{figure=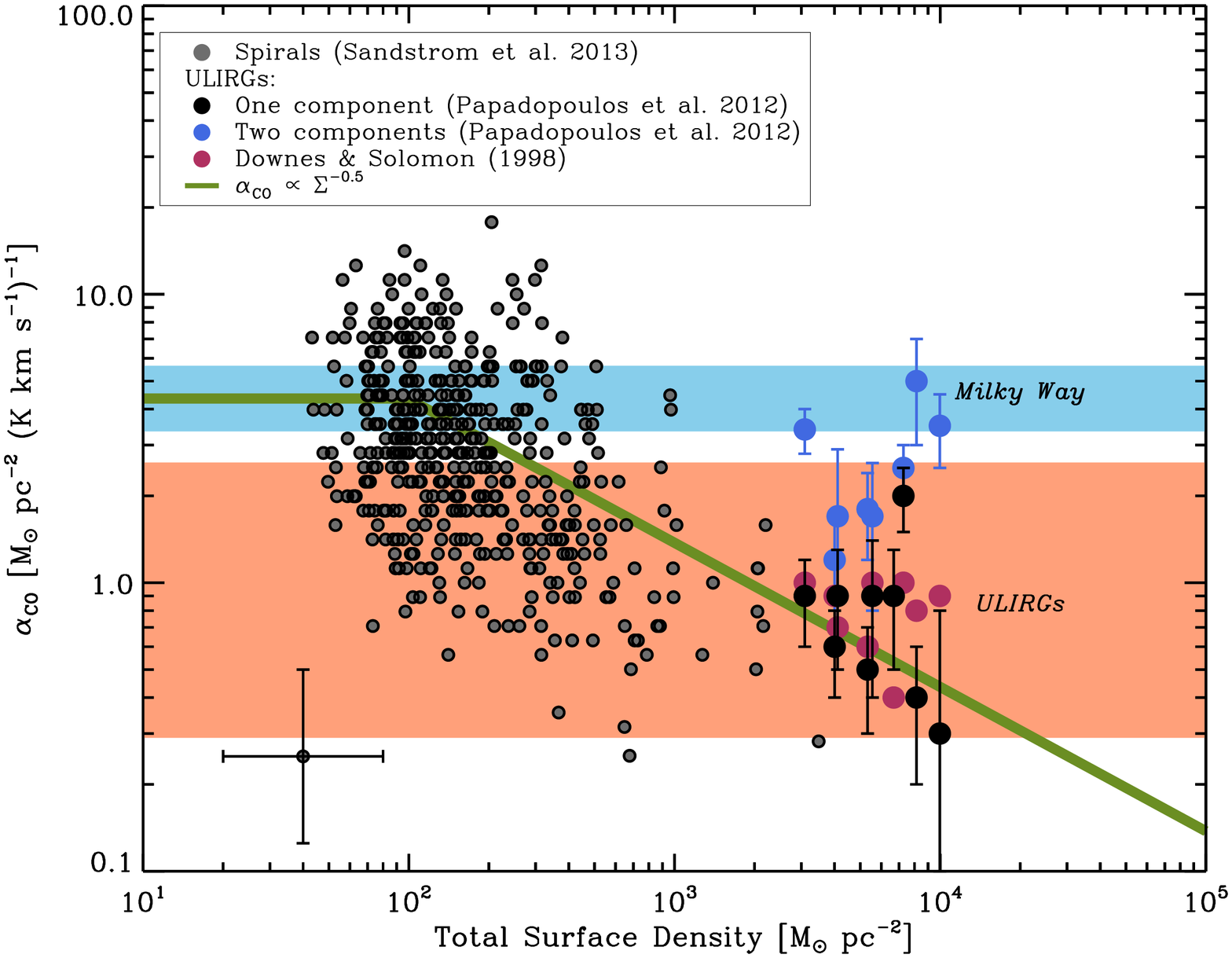,width=\columnwidth}
}
\caption{Conversion factor as a function of total surface density
for nearby disk galaxies and ULIRGs. The gray points illustrate the
high S/N solutions for \aco\ based on dust emission on kpc scales in a
sample of nearby disks, with typical errors illustrated in the lower
left corner \citep{SANDSTROM2012}. The corresponding surface densities are
dominated by the stellar component. For the ULIRGs we plot the 
\aco\ determinations by \citet[magenta]{DOWNES1998} and \citet[][multitransition one-component
fits in black and two-component fits in blue, error bars represent
possible range]{PAPADOPOULOS2012}. The abcisa is from the dynamical
mass measurements by \citet{DOWNES1998}, thus we plot only the overlap
of both samples. The color bands illustrate the recommended ranges for
Milky Way and ULIRG conversion factors.  The $\Sigma^{-0.5}$ line for
$\Sigma>100$~\msunperpcsq\ is a reasonable representation of the
overall trends when considering the one-component fits.
\label{fig:sd}}
\end{figure}

Several regimes exhibit clear departures from a Galactic \xco . Some,
but not all galaxy centers share a value of \xco\ lower than disks,
qualitatively similar to that observed in overwhelmingly molecular
massive starburst galaxies such as ULIRGs.  Dust and spectral line
modeling of these central regions show depressed \xco\ compared to
galaxy disks, with the depression spanning a wide range of
\xco\ up to an order of magnitude below Galactic. These central \xco\
depressions are not universal, although they seem correlated with the
stellar surface density $\Sigma_*$ \citep{SANDSTROM2012}, and
additional information appears necessary to predict the appropriate
\xco\ for use in any specific galaxy center.

A wide variety of evidence points to high \xco\ in low metallicity
regions: the absolute and normalized faintness of CO, high \cii-to-CO
ratios, high SFR-to-CO ratios, large dust-based \xco\ determinations,
and theoretical calculations of cloud structure. Virial masses
represent a significant exception to this body of evidence. When
derived at high spatial resolution, these tend to show little or no
enhancement in \xco\ above the Galactic value, even in low metallicity
systems. We review the theoretical expectations for the shrinking of
the opaque CO emitting surface relative to \htwo\ as metallicity
decreases. We favor a self-consistent use of dust, an optically thin
tracer of gas, as the currently most mature methodology to robustly
estimate molecular mass at low metallicity. We also highlight the
problems with the blind use of a dust-to-gas to metallicity
calibration. The present self-consistent dust-based
\xco\ estimates offer support for the picture of rapidly increasing
\xco\ at low metallicity, but still
yield a wide range of \xco\ even for similar techniques applied to the
same galaxy. {\em We recommend adopting a shielding-based prescription
such as that presented by \citet{WOLFIRE2010} or \citet{GLOVER2011} to
account for the effects of metallicity, with the normalization chosen
to match a ``Galactic'' \xco\ at solar metallicity}.

The uncertainties involved in any metallicity-dependent correction
remain substantial. As a first order picture, we expect \xco\ will
change slowly for metallicities larger than $12+\log[{\rm O/H}]\sim8.4$
(approximately $Z_\odot/2$), and considerably faster at lower
metallicities.  By the time $\xco$ reaches
10 times the Galactic value the CO emitting surface encompasses only
$\sim 5$--$10\%$ of the \htwo\ cloud, suggesting that the utility of
CO as a global tracer of \htwo\ will become more and more marginal as
one moves to progressively metal-poorer environments. Rather, CO will
in fact be a tracer of high column density peaks and well-shielded
regions.

There is general agreement that the processes operating in
overwhelmingly molecular and turbulent starbursts, where high gas
temperatures are also present, drive \xco\ to values that can be
substantially lower than in solar metallicity galactic disks.  Because
of the role of velocity dispersion in setting \xco, in the absence of
a self-regulating mechanism it is almost certain that there exists a
continuum of values rather than a unique \xco\ value that is
applicable in all cases. The typical result of the one-component
modeling is $\xco\sim0.4\times10^{20}$~\xcounits\ for massive,
turbulent, ultraluminous starbursts. The range around this value is
large, at least $\pm0.5$ dex, most of it likely representing real
physical variation among sources. 


The picture of \xco\ at high-redshift is still emerging, and
instruments like ALMA will make a crucial contribution to better
understand it. Lacking direct measurements, the best approach is to
use knowledge of the physical drivers of \xco\ developed in local
galaxies, as well as scaling and consistency arguments.  The simplest
approach is to identify the brightest, off-main sequence massive SMGs
likely due to starbursting mergers with local ULIRGs, while disky,
rotation-dominated ``main sequence'' galaxies are to first order more
likely similar to local disks dominated by self-gravitating or
virialized molecular clouds. This is an area of active research. The
picture will become more nuanced as new observations revealing the
resolved kinematics of the molecular gas and its excitation are
obtained. In particular, observations of ``main sequence'' galaxies at
$z\sim1-2$ suggests that metallicity effects will become an
increasingly important consideration at high-z, as observations push
to lower galaxy masses at higher redshifts and consequently more
metal-poor environments \citep{GENZEL2012}.

\subsection{Toward a Single Prescription}

Ultimately, we desire a prediction for \xco\ based on observable
properties, for objects ranging from low-metallicity dwarf galaxies to
high surface-density ULIRGs.  In the following paragraphs we present
some steps in that direction, referring to \aco\ since that is the
quantity most often useful for distant galaxies.  

Based on the discussions in \S\ref{sec:lm_theory} and
\S\ref{sec:starbursts}, \aco\ can be thought of as having two
primary dependencies; one related to the temperature and velocity
dispersion effects driving a low value in ULIRGs, the other related to
the dominance of CO-faint molecular gas driving a high value at low
metallicities. Treating the two effects as separable, $\aco =
\alpha_{\rm CO,MW} f_{\rm COF} f_{\rm SB}$, where $\alpha_{\rm CO,MW}$
represents an overall normalization under Milky Way disk
conditions. The factor $f_{\rm COF}$ corresponds to a correction that
accounts for the fraction of H$_2$ mass associated with the outer
layers of clouds where most CO is photodissociated.  The factor
$f_{\rm SB}$ accounts for changes in
\aco\ due to temperature and velocity dispersion.  

Drawing from \S\ref{sec:lm_theory}, $f_{\rm COF}$ may be approximated
by considering Eq. \ref{eq:fDGAV} applied to a population of
identical, fixed surface density clouds,

\begin{equation}
f_{\rm COF} \approx 0.67~\exp \left( \frac{+0.4}{Z^\prime~\Sigma_{\rm GMC}^{100}} \right)~.
\end{equation}

\noindent Here we assume that dust-to-gas ratio tracks 
metallicity, $Z^\prime$ is the metallicity normalized to the solar
value, and $\Sigma_{\rm GMC}^{100}$ is the average surface density of
molecular clouds in units of $100$~M$_\odot$~pc$^{-2}$.

The factor $f_{\rm SB}$ is considerably more tentative. The simple
theoretical arguments we outline in \S\ref{sec:intro_theoretical}, as
well as simulations \citep[e.g.,][]{SHETTY2011}, suggest that both the
gas velocity dispersion and temperature are key
parameters. Nonetheless, keep in mind that the fundamental driver of
\xco\ is what fraction of the CO luminosity arises from gas in
self-gravitating clouds, versus an extended not self-gravitating
component bound by the total mass of the system.  Given current
observational constraints and our desire to parametrize in terms of
measurable quantities, we suggest that the variations between normal
disks, galaxy centers, and ULIRGs are mostly captured by a surface
density-dependent factor of the form $f_{\rm SB}
\propto\Sigma_{\rm total}^{-\gamma}$, where $\Sigma_{\rm total}$ refers to
the combined gas plus stellar surface density on kpc scales.

Present constraints remain scarce, but we make an effort to present
them in Fig. \ref{fig:sd}. The data corresponds to the kpc-scale
dust-based measurements in nearby disks by
\citet{SANDSTROM2012}, as well as the overlap between the
ULIRG samples by \citet[][from which we take dynamical masses and
\aco]{DOWNES1998} and \citet[][from which we take \aco\
estimates]{PAPADOPOULOS2012}. In this latter case, \aco\ is derived
from one-component (similar to the results by \citeauthor{DOWNES1998})
or two-component multi-transition fits (which include contributions
from a dense phase). The dynamical surface density is dominated by the
stellar component, even in ULIRGs \citep{DOWNES1998}. Informed by the
theoretical arguments leading to Eq. \ref{eq:aco_star}, and by the
results of detailed modeling \citep{SHETTY2011}, we plot
$\aco\propto\Sigma_{\rm total}^{-0.5}$ normalizing to our recommended
Galactic \aco\ value at $\Sigma_{\rm
total}=100$~\msunperpcsq. Obviously this correction should not extend
to surface densities below those of resolved self-gravitating GMCs.

Given the large uncertainties and the small dynamic range of the \aco\
measurements this simple prescription seems to reproduce the trends
present in the data reasonably well, particularly for the results of
one-component models for the ULIRGs (which we consider most
mature). The observations may be fit with a smaller $\gamma$ although
with considerable uncertainty
\citep[e.g.,][]{SANDSTROM2012}, which leads us to prefer the
theoretically motivated $\gamma\approx0.5$.  Density increases in the
self-gravitating molecular material with respect to the Milky Way
average GMC properties will drive the \aco\ points up, while increases
in temperature will drive them down. The sample spans a factor of
$\sim 2$ in $T_{dust}$, which should be a reasonable proxy for gas
temperature in the ULIRGs. Although we have searched for the signature
of temperature effects in the data, we see no discernible correlation
with $T_{dust}$ \citep[e.g.,][]{MAGNELLI2012}. Likely, the sample
lacks the necessary dynamic range to pull those effects out of the
data. Possibly, as previously discussed, the lack of a temperature
correlation could be in part due to cancelations between the opposite
effects the density of self-gravitating clouds and their temperature
have on \xco. Thus, as a tentative first step for a simple conversion
factor prescription, we suggest using

\begin{equation}
\aco \approx 2.9\exp \left(\frac{+0.4}{Z^\prime~\Sigma_{\rm GMC}^{100}} \right) 
\left(\frac{\Sigma_{\rm total}}{100~\msunperpcsq}\right)^{-\gamma}
\end{equation}

\noindent in \acounits, with $\gamma\approx0.5$ for $\Sigma_{\rm total}>100$~\msunperpcsq\ 
and $\gamma=0$ otherwise. Note that we still expect a fair dispersion
around this average prescription, representing the variation in local
parameters such as temperature or \Sgmc.

There has been an exciting range of theoretical and numerical
developments on calculations of \xco\ in the last few years. The
coupling of high resolution hydrodynamical simulations including
chemistry and radiative transfer, with increasingly sophisticated
theoretical modeling of photodissociation regions and molecular
clouds, and galaxy scale simulations offers an exciting avenue of
progress. Numerically derived calibrations, such as those obtained on
small scales by \citet{GLOVER2011} or \citet{SHETTY2011} and on large
scales by
\citet{NARAYANAN2012}, show much promise. Such simulations are likely
to become increasingly reliable as the modeling is able to better
incorporate and couple the physics, kinematics, and radiative transfer
on the small and large scales. Grounded on observations, simulations may
offer the ultimate way to calibrate the CO-to-\htwo\ conversion
factor in a variety of environments.

\acknowledgements
We especially thank the following people for providing extensive
comments on earlier versions of this manuscript: Leo Blitz, Ewine van
Dishoeck, Neal Evans, Reinhard Genzel, Erik Rosolowsky, and Nick
Scoville.  We also thank the following people for providing figures,
comments, advise, and/or for enduring one of the partial or complete
drafts of this manuscript: Jean-Philippe Bernard, Chris Carilli,
Thomas Dame, Jennifer Donovan Meyer, Isabelle Grenier, Andrew Harris,
Remy Indebetouw, Frank Israel, Gu\"olaugur J\'ohannesson, Douglas
Marshall, Desika Narayanan, Eve Ostriker, Padelis Papadopoulos, Jorge
Pineda, Karin Sandstrom, Rahul Shetty, Andrew Strong, Linda Tacconi,
Stuart Vogel, Fabian Walter, and Zhi-Yu Zhang.  A.D.B. wishes to
acknowledge partial support from a CAREER grant NSF-AST0955836,
NSF-AST1139998, and from a Research Corporation for Science
Advancement Cottrell Scholar award, as well as full support from his
wife, Liliana.

\bibliographystyle{Astronomy}

\bibliography{XCO_ARAA} 

\begin{thebibliography}{}
\expandafter\ifx\csname natexlab\endcsname\relax\def\natexlab#1{#1}\fi

\bibitem[{Abdo et~al.(2010)Abdo, Ackermann, Ajello, Atwood, Baldini
  et~al.}]{ABDO2010}
Abdo AA, Ackermann M, Ajello M, Atwood WB, Baldini L, et~al. 2010.
\newblock \textit{\aap} 512:7

\bibitem[{Abdo et~al.(2010b)Abdo, Ackermann, Ajello, Baldini, Ballet
  et~al.}]{ABDO2010b}
Abdo AA, Ackermann M, Ajello M, Baldini L, Ballet J, et~al. 2010b.
\newblock \textit{\aap} 523:46

\bibitem[{Abdo et~al.(2010d)Abdo, Ackermann, Ajello, Baldini, Ballet
  et~al.}]{ABDO2010d}
Abdo AA, Ackermann M, Ajello M, Baldini L, Ballet J, et~al. 2010d.
\newblock \textit{\apj} 710:133--149

\bibitem[{Ackermann et~al.(2012c)Ackermann, Ajello, Allafort, Antolini, Baldini
  et~al.}]{ACKERMANN2012c}
Ackermann M, Ajello M, Allafort A, Antolini E, Baldini L, et~al. 2012c.
\newblock \textit{\apj} 756:4

\bibitem[{Ackermann et~al.(2012b)Ackermann, Ajello, Allafort, Baldini, Ballet
  et~al.}]{ACKERMANN2012b}
Ackermann M, Ajello M, Allafort A, Baldini L, Ballet J, et~al. 2012b.
\newblock \textit{\apj} 755:22

\bibitem[{Ackermann et~al.(2012d)Ackermann, Ajello, Allafort, Baldini, Ballet
  et~al.}]{ACKERMANN2012d}
Ackermann M, Ajello M, Allafort A, Baldini L, Ballet J, et~al. 2012d.
\newblock \textit{\aap} 538:71

\bibitem[{Ackermann et~al.(2012)Ackermann, Ajello, Atwood, Baldini, Ballet
  et~al.}]{ACKERMANN2012}
Ackermann M, Ajello M, Atwood WB, Baldini L, Ballet J, et~al. 2012.
\newblock \textit{\apj} 750:3

\bibitem[{Ackermann et~al.(2011)Ackermann, Ajello, Baldini, Ballet, Barbiellini
  et~al.}]{ACKERMANN2011}
Ackermann M, Ajello M, Baldini L, Ballet J, Barbiellini G, et~al. 2011.
\newblock \textit{\apj} 726:81

\bibitem[{Adler et~al.(1992)Adler, Lo, Wright, Rydbeck, Plante \&
  Allen}]{ADLER1992}
Adler DS, Lo KY, Wright MCH, Rydbeck G, Plante RL, Allen RJ. 1992.
\newblock \textit{\apj} 392:497--508

\bibitem[{Aravena et~al.(2010)Aravena, Carilli, Daddi, Wagg, Walter
  et~al.}]{ARAVENA2010}
Aravena M, Carilli C, Daddi E, Wagg J, Walter F, et~al. 2010.
\newblock \textit{\apj} 718:177--183

\bibitem[{Arimoto, Sofue \& Tsujimoto(1996)}]{ARIMOTO1996}
Arimoto N, Sofue Y, Tsujimoto T. 1996.
\newblock \textit{\pasj} 48:275--284

\bibitem[{Asplund et~al.(2009)Asplund, Grevesse, Sauval \& Scott}]{ASPLUND2009}
Asplund M, Grevesse N, Sauval AJ, Scott P. 2009.
\newblock \textit{\araa} 47:481--522

\bibitem[{Ballesteros-Paredes(2006)}]{BALLESTEROS-PAREDES2006}
Ballesteros-Paredes J. 2006.
\newblock \textit{\mnras} 372:443--449

\bibitem[{Ballesteros-Paredes et~al.(2011)Ballesteros-Paredes, Hartmann,
  V{\'a}zquez-Semadeni, Heitsch \& Zamora-Avil{\'e}s}]{BALLESTEROS-PAREDES2011}
Ballesteros-Paredes J, Hartmann LW, V{\'a}zquez-Semadeni E, Heitsch F,
  Zamora-Avil{\'e}s MA. 2011.
\newblock \textit{\mnras} 411:65--70

\bibitem[{Bell et~al.(2006)Bell, Roueff, Viti \& Williams}]{BELL2006}
Bell TA, Roueff E, Viti S, Williams DA. 2006.
\newblock \textit{\mnras} 371:1865--1872

\bibitem[{Bell, Viti \& Williams(2007)}]{BELL2007}
Bell TA, Viti S, Williams DA. 2007.
\newblock \textit{\mnras} 378:983--994

\bibitem[{Bennett et~al.(1994)Bennett, Fixsen, Hinshaw, Mather, Moseley
  et~al.}]{BENNETT1994}
Bennett CL, Fixsen DJ, Hinshaw G, Mather JC, Moseley SH, et~al. 1994.
\newblock \textit{\apj} 434:587--598

\bibitem[{Bertoldi \& McKee(1992)}]{BERTOLDI1992}
Bertoldi F, McKee CF. 1992.
\newblock \textit{\apj} 395:140--157

\bibitem[{{Blanc} et~al.(2013){Blanc}, {Schruba}, {Evans}, {Jogee}, {Bolatto}
  et~al.}]{BLANC2013}
{Blanc} GA, {Schruba} A, {Evans} II NJ, {Jogee} S, {Bolatto} A, et~al. 2013.
\newblock \textit{\apj} 764:117

\bibitem[{Blitz et~al.(1985)Blitz, Bloemen, Hermsen \& Bania}]{BLITZ1985}
Blitz L, Bloemen JBGM, Hermsen W, Bania TM. 1985.
\newblock \textit{Astronomy and Astrophysics (ISSN 0004-6361)} 143:267--273

\bibitem[{Blitz et~al.(2007)Blitz, Fukui, Kawamura, Leroy, Mizuno \&
  Rosolowsky}]{BLITZ2007}
Blitz L, Fukui Y, Kawamura A, Leroy A, Mizuno N, Rosolowsky E. 2007.
\newblock \textit{Protostars and Planets V} :81--96

\bibitem[{Blitz \& Shu(1980b)}]{BLITZ1980b}
Blitz L, Shu FH. 1980b.
\newblock \textit{\apj} 238:148--157

\bibitem[{Bloemen(1989)}]{BLOEMEN1989}
Bloemen H. 1989.
\newblock \textit{IN: Annual review of astronomy and astrophysics. Volume 27
  (A90-29983 12-90). Palo Alto} 27:469--516

\bibitem[{Bohlin, Savage \& Drake(1978)}]{BOHLIN1978}
Bohlin RC, Savage BD, Drake JF. 1978.
\newblock \textit{\apj} 224:132--142

\bibitem[{Bolatto, Jackson \& Ingalls(1999)}]{BOLATTO1999}
Bolatto AD, Jackson JM, Ingalls JG. 1999.
\newblock \textit{\apj} 513:275--286

\bibitem[{Bolatto et~al.(2003)Bolatto, Leroy, Israel \& Jackson}]{BOLATTO2003}
Bolatto AD, Leroy A, Israel FP, Jackson JM. 2003.
\newblock \textit{\apj} 595:167--178

\bibitem[{Bolatto et~al.(2011)Bolatto, Leroy, Jameson, Ostriker, Gordon
  et~al.}]{BOLATTO2011}
Bolatto AD, Leroy AK, Jameson K, Ostriker E, Gordon K, et~al. 2011.
\newblock \textit{\apj} 741:12

\bibitem[{Bolatto et~al.(2008)Bolatto, Leroy, Rosolowsky, Walter \&
  Blitz}]{BOLATTO2008}
Bolatto AD, Leroy AK, Rosolowsky E, Walter F, Blitz L. 2008.
\newblock \textit{\apj} 686:948--965

\bibitem[{Boselli, Lequeux \& Gavazzi(2002)}]{BOSELLI2002}
Boselli A, Lequeux J, Gavazzi G. 2002.
\newblock \textit{\aap} 384:33--47

\bibitem[{Bot et~al.(2004)Bot, Boulanger, Lagache, Cambr{\'e}sy \&
  Egret}]{BOT2004}
Bot C, Boulanger F, Lagache G, Cambr{\'e}sy L, Egret D. 2004.
\newblock \textit{\aap} 423:567--577

\bibitem[{Bot et~al.(2007)Bot, Boulanger, Rubio \& Rantakyro}]{BOT2007}
Bot C, Boulanger F, Rubio M, Rantakyro F. 2007.
\newblock \textit{\aap} 471:103--112

\bibitem[{Bot et~al.(2010)Bot, Rubio, Boulanger, Albrecht, Leroy
  et~al.}]{BOT2010}
Bot C, Rubio M, Boulanger F, Albrecht M, Leroy A, et~al. 2010.
\newblock \textit{\aap} 524:52

\bibitem[{Bothwell et~al.(2010)Bothwell, Chapman, Tacconi, Smail, Ivison
  et~al.}]{BOTHWELL2010}
Bothwell MS, Chapman SC, Tacconi L, Smail I, Ivison RJ, et~al. 2010.
\newblock \textit{\mnras} 405:219--233

\bibitem[{Bothwell et~al.(2012)Bothwell, Smail, Chapman, Genzel, Ivison
  et~al.}]{BOTHWELL2012}
Bothwell MS, Smail I, Chapman SC, Genzel R, Ivison RJ, et~al. 2012.
\newblock \textit{arXiv.org} 1205:1511

\bibitem[{Boulanger et~al.(1996)Boulanger, Abergel, Bernard, Burton, Desert
  et~al.}]{BOULANGER1996}
Boulanger F, Abergel A, Bernard JP, Burton WB, Desert FX, et~al. 1996.
\newblock \textit{\aap} 312:256--262

\bibitem[{Bradford et~al.(2003)Bradford, Nikola, Stacey, Bolatto, Jackson
  et~al.}]{BRADFORD2003}
Bradford CM, Nikola T, Stacey GJ, Bolatto AD, Jackson JM, et~al. 2003.
\newblock \textit{\apj} 586:891--901

\bibitem[{{Brand} \& {Wouterloot}(1995)}]{BRAND1995}
{Brand} J, {Wouterloot} JGA. 1995.
\newblock \textit{\aap} 303:851

\bibitem[{Braun et~al.(2009)Braun, Thilker, Walterbos \& Corbelli}]{BRAUN2009}
Braun R, Thilker DA, Walterbos RAM, Corbelli E. 2009.
\newblock \textit{\apj} 695:937--953

\bibitem[{Brown \& Vanden~Bout(1991)}]{BROWN1991}
Brown RL, Vanden~Bout PA. 1991.
\newblock \textit{\aj} 102:1956--1959

\bibitem[{Bryant \& Scoville(1996)}]{BRYANT1996}
Bryant PM, Scoville NZ. 1996.
\newblock \textit{\apj v.457} 457:678

\bibitem[{Bryant \& Scoville(1999)}]{BRYANT1999}
Bryant PM, Scoville NZ. 1999.
\newblock \textit{\aj} 117:2632--2655

\bibitem[{Burton et~al.(1975)Burton, Gordon, Bania \& Lockman}]{BURTON1975}
Burton WB, Gordon MA, Bania TM, Lockman FJ. 1975.
\newblock \textit{\apj} 202:30--49

\bibitem[{Cambr{\'e}sy, Jarrett \& Beichman(2005)}]{CAMBRESY2005}
Cambr{\'e}sy L, Jarrett TH, Beichman CA. 2005.
\newblock \textit{\aap} 435:131--139

\bibitem[{Carilli et~al.(2010)Carilli, Daddi, Riechers, Walter, Weiss
  et~al.}]{CARILLI2010}
Carilli CL, Daddi E, Riechers D, Walter F, Weiss A, et~al. 2010.
\newblock \textit{\apj} 714:1407--1417

\bibitem[{Chapman et~al.(2009)Chapman, Mundy, Lai \& Evans}]{CHAPMAN2009}
Chapman NL, Mundy LG, Lai SP, Evans NJI. 2009.
\newblock \textit{\apj} 690:496--511

\bibitem[{Cormier et~al.(2010)Cormier, Madden, Hony, Contursi, Poglitsch
  et~al.}]{CORMIER2010}
Cormier D, Madden SC, Hony S, Contursi A, Poglitsch A, et~al. 2010.
\newblock \textit{\aap} 518:L57

\bibitem[{Crawford et~al.(1985)Crawford, Genzel, Townes \&
  Watson}]{CRAWFORD1985}
Crawford MK, Genzel R, Townes CH, Watson DM. 1985.
\newblock \textit{\apj} 291:755--771

\bibitem[{Dabrowski(1984)}]{DABROWSKI1984}
Dabrowski I. 1984.
\newblock \textit{Canadian Journal of Physics (ISSN 0008-4204)} 62:1639--1664

\bibitem[{Daddi et~al.(2010b)Daddi, Bournaud, Walter, Dannerbauer, Carilli
  et~al.}]{DADDI2010b}
Daddi E, Bournaud F, Walter F, Dannerbauer H, Carilli CL, et~al. 2010b.
\newblock \textit{\apj} 713:686--707

\bibitem[{Daddi et~al.(2010)Daddi, Elbaz, Walter, Bournaud, Salmi
  et~al.}]{DADDI2010}
Daddi E, Elbaz D, Walter F, Bournaud F, Salmi F, et~al. 2010.
\newblock \textit{\apj Letters} 714:L118--L122

\bibitem[{Dame, Hartmann \& Thaddeus(2001)}]{DAME2001}
Dame TM, Hartmann D, Thaddeus P. 2001.
\newblock \textit{\apj} 547:792--813

\bibitem[{Dame et~al.(1987)Dame, Ungerechts, Cohen, de~Geus, Grenier
  et~al.}]{DAME1987}
Dame TM, Ungerechts H, Cohen RS, de~Geus EJ, Grenier IA, et~al. 1987.
\newblock \textit{\apj} 322:706--720

\bibitem[{Dannerbauer et~al.(2009)Dannerbauer, Daddi, Riechers, Walter, Carilli
  et~al.}]{DANNERBAUER2009}
Dannerbauer H, Daddi E, Riechers DA, Walter F, Carilli CL, et~al. 2009.
\newblock \textit{\apj Letters} 698:L178--L182

\bibitem[{de~Jong, Boland \& Dalgarno(1980)}]{DEJONG1980}
de~Jong T, Boland W, Dalgarno A. 1980.
\newblock \textit{\aap} 91:68--84

\bibitem[{Dickman(1978)}]{DICKMAN1978}
Dickman RL. 1978.
\newblock \textit{\apj Supplement Series} 37:407--427

\bibitem[{Dickman, Snell \& Schloerb(1986)}]{DICKMAN1986}
Dickman RL, Snell RL, Schloerb FP. 1986.
\newblock \textit{\apj} 309:326--330

\bibitem[{Dobashi et~al.(2008)Dobashi, Bernard, Hughes, Paradis, Reach \&
  Kawamura}]{DOBASHI2008}
Dobashi K, Bernard JP, Hughes A, Paradis D, Reach WT, Kawamura A. 2008.
\newblock \textit{\aap} 484:205--223

\bibitem[{Dobashi et~al.(2009)Dobashi, Bernard, Kawamura, Egusa, Hughes
  et~al.}]{DOBASHI2009}
Dobashi K, Bernard JP, Kawamura A, Egusa F, Hughes A, et~al. 2009.
\newblock \textit{\aj} 137:5099--5109

\bibitem[{Donovan~Meyer et~al.(2013)Donovan~Meyer, Koda, Momose \&
  Fukuhara}]{DONOVANMEYER2013}
Donovan~Meyer J, Koda J, Momose R, Fukuhara M. 2013.
\newblock \textit{\apj} submitted

\bibitem[{Donovan~Meyer et~al.(2012)Donovan~Meyer, Koda, Momose, Fukuhara,
  Mooney et~al.}]{DONOVANMEYER2012}
Donovan~Meyer J, Koda J, Momose R, Fukuhara M, Mooney T, et~al. 2012.
\newblock \textit{\apj} 744:42

\bibitem[{Downes \& Solomon(1998)}]{DOWNES1998}
Downes D, Solomon PM. 1998.
\newblock \textit{\apj} 507:615--654

\bibitem[{Downes, Solomon \& Radford(1993)}]{DOWNES1993}
Downes D, Solomon PM, Radford SJE. 1993.
\newblock \textit{\apj} 414:L13--L16

\bibitem[{Draine(1978)}]{DRAINE1978}
Draine BT. 1978.
\newblock \textit{\apj Supplement Series} 36:595--619

\bibitem[{Draine(2009)}]{DRAINE2009}
Draine BT. 2009.
\newblock \textit{Cosmic Dust - Near and Far ASP Conference Series} 414:453

\bibitem[{Draine et~al.(2007)Draine, Dale, Bendo, Gordon, Smith
  et~al.}]{DRAINE2007}
Draine BT, Dale DA, Bendo G, Gordon KD, Smith JDT, et~al. 2007.
\newblock \textit{\apj} 663:866--894

\bibitem[{Draine \& Li(2001)}]{DRAINE2001}
Draine BT, Li A. 2001.
\newblock \textit{\apj} 551:807--824

\bibitem[{Draine \& Li(2007b)}]{DRAINE2007b}
Draine BT, Li A. 2007b.
\newblock \textit{\apj} 657:810--837

\bibitem[{Dufour, Shields \& Talbot(1982)}]{DUFOUR1982}
Dufour RJ, Shields GA, Talbot RJJ. 1982.
\newblock \textit{\apj} 252:461--473

\bibitem[{Dwek(1998)}]{DWEK1998}
Dwek E. 1998.
\newblock \textit{\apj v.501} 501:643

\bibitem[{Dwek \& Cherchneff(2011)}]{DWEK2011}
Dwek E, Cherchneff I. 2011.
\newblock \textit{\apj} 727:63

\bibitem[{Elmegreen(2000)}]{ELMEGREEN2000}
Elmegreen BG. 2000.
\newblock \textit{\apj} 530:277--281

\bibitem[{Elmegreen, Morris \& Elmegreen(1980)}]{ELMEGREEN1980}
Elmegreen BG, Morris M, Elmegreen DM. 1980.
\newblock \textit{\apj} 240:455--463

\bibitem[{Evans et~al.(2009)Evans, Dunham, J{\o}rgensen, Enoch, Mer{\'\i}n
  et~al.}]{EVANS2009}
Evans NJI, Dunham MM, J{\o}rgensen JK, Enoch ML, Mer{\'\i}n B, et~al. 2009.
\newblock \textit{\apj Supplement} 181:321--350

\bibitem[{Falgarone et~al.(1994)Falgarone, Lis, Phillips, Pouquet, Porter \&
  Woodward}]{FALGARONE1994}
Falgarone E, Lis DC, Phillips TG, Pouquet A, Porter DH, Woodward PR. 1994.
\newblock \textit{\apj} 436:728--740

\bibitem[{Federman et~al.(1996)Federman, Rawlings, Taylor \&
  Williams}]{FEDERMAN1996}
Federman SR, Rawlings JMC, Taylor SD, Williams DA. 1996.
\newblock \textit{\mnras} 279:L41--L46

\bibitem[{Feldmann, Gnedin \& Kravtsov(2012)}]{FELDMANN2012}
Feldmann R, Gnedin NY, Kravtsov AV. 2012.
\newblock \textit{\apj} 747:124

\bibitem[{Fixsen, Bennett \& Mather(1999)}]{FIXSEN1999}
Fixsen DJ, Bennett CL, Mather JC. 1999.
\newblock \textit{\apj} 526:207--214

\bibitem[{Flagey et~al.(2009)Flagey, Noriega-Crespo, Boulanger, Carey, Brooke
  et~al.}]{FLAGEY2009}
Flagey N, Noriega-Crespo A, Boulanger F, Carey SJ, Brooke TY, et~al. 2009.
\newblock \textit{\apj} 701:1450--1463

\bibitem[{Frerking, Langer \& Wilson(1982)}]{FRERKING1982}
Frerking MA, Langer WD, Wilson RW. 1982.
\newblock \textit{\apj} 262:590--605

\bibitem[{Fukui \& Kawamura(2010)}]{FUKUI2010}
Fukui Y, Kawamura A. 2010.
\newblock \textit{\araa} 48:547--580

\bibitem[{Fukui et~al.(2008)Fukui, Kawamura, Minamidani, Mizuno, Kanai
  et~al.}]{FUKUI2008}
Fukui Y, Kawamura A, Minamidani T, Mizuno Y, Kanai Y, et~al. 2008.
\newblock \textit{\apj Supplement Series} 178:56--70

\bibitem[{Gabici, Aharonian \& Blasi(2007)}]{GABICI2007}
Gabici S, Aharonian FA, Blasi P. 2007.
\newblock \textit{Astrophysics and Space Science} 309:365--371

\bibitem[{Galliano et~al.(2011)Galliano, Hony, Bernard, Bot, Madden
  et~al.}]{GALLIANO2011}
Galliano F, Hony S, Bernard JP, Bot C, Madden SC, et~al. 2011.
\newblock \textit{\aap} 536:88

\bibitem[{Gao \& Solomon(2004)}]{GAO2004}
Gao Y, Solomon PM. 2004.
\newblock \textit{\apj} 606:271--290

\bibitem[{Garcia-Burillo, Combes \& Gerin(1993)}]{GARCIA-BURILLO1993}
Garcia-Burillo S, Combes F, Gerin M. 1993.
\newblock \textit{\aap} 274:148

\bibitem[{Genzel et~al.(2011)Genzel, Newman, Jones, F{\"o}rster~Schreiber,
  Shapiro et~al.}]{GENZEL2011}
Genzel R, Newman S, Jones T, F{\"o}rster~Schreiber NM, Shapiro K, et~al. 2011.
\newblock \textit{\apj} 733:101

\bibitem[{Genzel et~al.(2012)Genzel, Tacconi, Combes, Bolatto, Neri
  et~al.}]{GENZEL2012}
Genzel R, Tacconi LJ, Combes F, Bolatto A, Neri R, et~al. 2012.
\newblock \textit{\apj} 746:69

\bibitem[{Genzel et~al.(2010)Genzel, Tacconi, Gracia-Carpio, Sternberg, Cooper
  et~al.}]{GENZEL2010}
Genzel R, Tacconi LJ, Gracia-Carpio J, Sternberg A, Cooper MC, et~al. 2010.
\newblock \textit{\mnras} 407:2091--2108

\bibitem[{Glover et~al.(2010)Glover, Federrath, Mac~Low \&
  Klessen}]{GLOVER2010}
Glover SCO, Federrath C, Mac~Low MM, Klessen RS. 2010.
\newblock \textit{\mnras} 404:2--29

\bibitem[{Glover \& Mac~Low(2007)}]{GLOVER2007}
Glover SCO, Mac~Low MM. 2007.
\newblock \textit{\apj Supplement Series} 169:239--268

\bibitem[{Glover \& Mac~Low(2007b)}]{GLOVER2007b}
Glover SCO, Mac~Low MM. 2007b.
\newblock \textit{\apj} 659:1317--1337

\bibitem[{Glover \& Mac~Low(2011)}]{GLOVER2011}
Glover SCO, Mac~Low MM. 2011.
\newblock \textit{\mnras} 412:337--350

\bibitem[{Godard, Falgarone \& Pineau~des Forets(2009)}]{GODARD2009}
Godard B, Falgarone E, Pineau~des Forets G. 2009.
\newblock \textit{\aap} 495:847--867

\bibitem[{Goldreich \& Kwan(1974)}]{GOLDREICH1974}
Goldreich P, Kwan J. 1974.
\newblock \textit{\apj} 189:441--454

\bibitem[{Goldsmith et~al.(2008)Goldsmith, Heyer, Narayanan, Snell, Li \&
  Brunt}]{GOLDSMITH2008}
Goldsmith PF, Heyer M, Narayanan G, Snell R, Li D, Brunt C. 2008.
\newblock \textit{\apj} 680:428--445

\bibitem[{Gratier et~al.(2010)Gratier, Braine, Rodriguez-Fernandez, Israel,
  Schuster et~al.}]{GRATIER2010}
Gratier P, Braine J, Rodriguez-Fernandez NJ, Israel FP, Schuster KF, et~al.
  2010.
\newblock \textit{\aap} 512:68

\bibitem[{Grenier, Casandjian \& Terrier(2005)}]{GRENIER2005}
Grenier IA, Casandjian JM, Terrier R. 2005.
\newblock \textit{Science} 307:1292--1295

\bibitem[{Guelin et~al.(1995)Guelin, Zylka, Mezger, Haslam \&
  Kreysa}]{GUELIN1995}
Guelin M, Zylka R, Mezger PG, Haslam CGT, Kreysa E. 1995.
\newblock \textit{\aap} 298:L29

\bibitem[{Guelin et~al.(1993)Guelin, Zylka, Mezger, Haslam, Kreysa
  et~al.}]{GUELIN1993}
Guelin M, Zylka R, Mezger PG, Haslam CGT, Kreysa E, et~al. 1993.
\newblock \textit{Astronomy and Astrophysics (ISSN 0004-6361)} 279:L37--L40

\bibitem[{Habart et~al.(2010)Habart, Dartois, Abergel, Baluteau, Naylor
  et~al.}]{HABART2010}
Habart E, Dartois E, Abergel A, Baluteau JP, Naylor D, et~al. 2010.
\newblock \textit{\aap} 518:L116

\bibitem[{Harris et~al.(2010)Harris, Baker, Zonak, Sharon, Genzel
  et~al.}]{HARRIS2010}
Harris AI, Baker AJ, Zonak SG, Sharon CE, Genzel R, et~al. 2010.
\newblock \textit{\apj} 723:1139--1149

\bibitem[{Heiderman et~al.(2010)Heiderman, Evans, Allen, Huard \&
  Heyer}]{HEIDERMAN2010}
Heiderman A, Evans NJI, Allen LE, Huard T, Heyer M. 2010.
\newblock \textit{\apj} 723:1019--1037

\bibitem[{Heiles(1994)}]{HEILES1994}
Heiles C. 1994.
\newblock \textit{\apj} 436:720--727

\bibitem[{Heiles \& Troland(2003)}]{HEILES2003}
Heiles C, Troland TH. 2003.
\newblock \textit{\apj} 586:1067--1093

\bibitem[{Heyer et~al.(2009)Heyer, Krawczyk, Duval \& Jackson}]{HEYER2009}
Heyer M, Krawczyk C, Duval J, Jackson JM. 2009.
\newblock \textit{\apj} 699:1092--1103

\bibitem[{Heyer, Carpenter \& Snell(2001)}]{HEYER2001}
Heyer MH, Carpenter JM, Snell RL. 2001.
\newblock \textit{\apj} 551:852--866

\bibitem[{Hollenbach et~al.(2012)Hollenbach, Kaufman, Neufeld, Wolfire \&
  Goicoechea}]{HOLLENBACH2012}
Hollenbach D, Kaufman MJ, Neufeld D, Wolfire M, Goicoechea JR. 2012.
\newblock \textit{\apj} 754:105

\bibitem[{Hollenbach \& Tielens(1997)}]{HOLLENBACH1997}
Hollenbach DJ, Tielens AGGM. 1997.
\newblock \textit{\araa} 35:179--216

\bibitem[{Hollenbach \& Tielens(1999)}]{HOLLENBACH1999}
Hollenbach DJ, Tielens AGGM. 1999.
\newblock \textit{Reviews of Modern Physics} 71:173--230

\bibitem[{Hughes et~al.(2010)Hughes, Wong, Ott, Muller, Pineda
  et~al.}]{HUGHES2010}
Hughes A, Wong T, Ott J, Muller E, Pineda JL, et~al. 2010.
\newblock \textit{\mnras} 406:2065--2086

\bibitem[{Hunter et~al.(2001)Hunter, Kaufman, Hollenbach, Rubin, Malhotra
  et~al.}]{HUNTER2001}
Hunter DA, Kaufman M, Hollenbach DJ, Rubin RH, Malhotra S, et~al. 2001.
\newblock \textit{\apj} 553:121--145

\bibitem[{Hunter et~al.(1997)Hunter, Bertsch, Catelli, Dame, Digel
  et~al.}]{HUNTER1997}
Hunter SD, Bertsch DL, Catelli JR, Dame TM, Digel SW, et~al. 1997.
\newblock \textit{\apj v.481} 481:205

\bibitem[{Ingalls et~al.(2011)Ingalls, Bania, Boulanger, Draine, Falgarone \&
  Hily-Blant}]{INGALLS2011}
Ingalls JG, Bania TM, Boulanger F, Draine BT, Falgarone E, Hily-Blant P. 2011.
\newblock \textit{\apj} 743:174

\bibitem[{Iono et~al.(2007)Iono, Wilson, Takakuwa, Yun, Petitpas
  et~al.}]{IONO2007}
Iono D, Wilson CD, Takakuwa S, Yun MS, Petitpas GR, et~al. 2007.
\newblock \textit{\apj} 659:283--295

\bibitem[{Iono et~al.(2009)Iono, Wilson, Yun, Baker, Petitpas
  et~al.}]{IONO2009}
Iono D, Wilson CD, Yun MS, Baker AJ, Petitpas GR, et~al. 2009.
\newblock \textit{\apj} 695:1537--1549

\bibitem[{Israel(1997)}]{ISRAEL1997}
Israel FP. 1997.
\newblock \textit{\aap} 328:471--482

\bibitem[{Israel(1997b)}]{ISRAEL1997b}
Israel FP. 1997b.
\newblock \textit{\aap} 317:65--72

\bibitem[{Israel(2009)}]{ISRAEL2009}
Israel FP. 2009.
\newblock \textit{\aap} 506:689--702

\bibitem[{Israel(2009b)}]{ISRAEL2009b}
Israel FP. 2009b.
\newblock \textit{\aap} 493:525--538

\bibitem[{Israel \& Baas(2001)}]{ISRAEL2001}
Israel FP, Baas F. 2001.
\newblock \textit{\aap} 371:433--444

\bibitem[{Israel et~al.(1986)Israel, de~Graauw, van~de Stadt \&
  de~Vries}]{ISRAEL1986}
Israel FP, de~Graauw T, van~de Stadt H, de~Vries CP. 1986.
\newblock \textit{\apj} 303:186--197

\bibitem[{Israel et~al.(2003)Israel, Johansson, Rubio, Garay, de~Graauw
  et~al.}]{ISRAEL2003}
Israel FP, Johansson LEB, Rubio M, Garay G, de~Graauw T, et~al. 2003.
\newblock \textit{\aap} 406:817--828

\bibitem[{Israel \& Maloney(2011)}]{ISRAEL2011}
Israel FP, Maloney PR. 2011.
\newblock \textit{\aap} 531:19

\bibitem[{Israel et~al.(1996)Israel, Maloney, Geis, Herrmann, Madden
  et~al.}]{ISRAEL1996}
Israel FP, Maloney PR, Geis N, Herrmann F, Madden SC, et~al. 1996.
\newblock \textit{\apj} 465:738

\bibitem[{Israel, Tilanus \& Baas(2006)}]{ISRAEL2006}
Israel FP, Tilanus RPJ, Baas F. 2006.
\newblock \textit{\aap} 445:907--913

\bibitem[{Ivison et~al.(2011)Ivison, Papadopoulos, Smail, Greve, Thomson
  et~al.}]{IVISON2011}
Ivison RJ, Papadopoulos PP, Smail I, Greve TR, Thomson AP, et~al. 2011.
\newblock \textit{\mnras} 412:1913--1925

\bibitem[{Jackson et~al.(1995)Jackson, Paglione, Carlstrom \&
  Rieu}]{JACKSON1995}
Jackson JM, Paglione TAD, Carlstrom JE, Rieu NQ. 1995.
\newblock \textit{\apj} 438:695--701

\bibitem[{Kainulainen et~al.(2009)Kainulainen, Beuther, Henning \&
  Plume}]{KAINULAINEN2009}
Kainulainen J, Beuther H, Henning T, Plume R. 2009.
\newblock \textit{\aap} 508:L35--L38

\bibitem[{Kennicutt \& Evans(2012)}]{KENNICUTT2012}
Kennicutt RCJ, Evans NJI. 2012.
\newblock \textit{arXiv.org} 1204:3552

\bibitem[{Kutner \& Leung(1985)}]{KUTNER1985}
Kutner ML, Leung CM. 1985.
\newblock \textit{\apj} 291:188--201

\bibitem[{Kutner \& Ulich(1981)}]{KUTNER1981}
Kutner ML, Ulich BL. 1981.
\newblock \textit{\apj} 250:341--348

\bibitem[{Lada, Lombardi \& Alves(2010)}]{LADA2010}
Lada CJ, Lombardi M, Alves JF. 2010.
\newblock \textit{\apj} 724:687--693

\bibitem[{Lada \& Blitz(1988)}]{LADA1988}
Lada EA, Blitz L. 1988.
\newblock \textit{\apj} 326:L69--L73

\bibitem[{Langer et~al.(2010)Langer, Velusamy, Pineda, Goldsmith, Li \&
  Yorke}]{LANGER2010}
Langer WD, Velusamy T, Pineda JL, Goldsmith PF, Li D, Yorke HW. 2010.
\newblock \textit{\aap} 521:L17

\bibitem[{Larson(1981)}]{LARSON1981}
Larson RB. 1981.
\newblock \textit{\mnras} 194:809--826

\bibitem[{Lebrun et~al.(1983)Lebrun, Bennett, Bignami, Caraveo, Bloemen
  et~al.}]{LEBRUN1983}
Lebrun F, Bennett K, Bignami GF, Caraveo PA, Bloemen JBGM, et~al. 1983.
\newblock \textit{\apj} 274:231--236

\bibitem[{Lequeux et~al.(1994)Lequeux, Le~Bourlot, Pineau~des Forets, Roueff,
  Boulanger \& Rubio}]{LEQUEUX1994}
Lequeux J, Le~Bourlot J, Pineau~des Forets G, Roueff E, Boulanger F, Rubio M.
  1994.
\newblock \textit{\aap} 292:371--380

\bibitem[{Lequeux et~al.(1979)Lequeux, Peimbert, Rayo, Serrano \&
  Torres-Peimbert}]{LEQUEUX1979}
Lequeux J, Peimbert M, Rayo JF, Serrano A, Torres-Peimbert S. 1979.
\newblock \textit{\aap} 80:155--166

\bibitem[{Leroy et~al.(2007)Leroy, Bolatto, Stanimirovic, Mizuno, Israel \&
  Bot}]{LEROY2007}
Leroy A, Bolatto A, Stanimirovic S, Mizuno N, Israel F, Bot C. 2007.
\newblock \textit{\apj} 658:1027--1046

\bibitem[{Leroy et~al.(2006)Leroy, Bolatto, Walter \& Blitz}]{LEROY2006}
Leroy A, Bolatto A, Walter F, Blitz L. 2006.
\newblock \textit{\apj} 643:825--843

\bibitem[{Leroy et~al.(2009)Leroy, Bolatto, Bot, Engelbracht, Gordon
  et~al.}]{LEROY2009}
Leroy AK, Bolatto A, Bot C, Engelbracht CW, Gordon K, et~al. 2009.
\newblock \textit{\apj} 702:352--367

\bibitem[{Leroy et~al.(2011)Leroy, Bolatto, Gordon, Sandstrom, Gratier
  et~al.}]{LEROY2011}
Leroy AK, Bolatto A, Gordon K, Sandstrom K, Gratier P, et~al. 2011.
\newblock \textit{\apj} 737:12

\bibitem[{Leroy et~al.(2008)Leroy, Walter, Brinks, Bigiel, de~Blok
  et~al.}]{LEROY2008}
Leroy AK, Walter F, Brinks E, Bigiel F, de~Blok WJG, et~al. 2008.
\newblock \textit{\aj} 136:2782--2845

\bibitem[{Levrier et~al.(2012)Levrier, Le~Petit, Hennebelle, Lesaffre, Gerin \&
  Falgarone}]{LEVRIER2012}
Levrier F, Le~Petit F, Hennebelle P, Lesaffre P, Gerin M, Falgarone E. 2012.
\newblock \textit{\aap} 544:22

\bibitem[{Liszt(2011)}]{LISZT2011}
Liszt HS. 2011.
\newblock \textit{\aap} 527:45

\bibitem[{Liszt \& Pety(2012)}]{LISZT2012}
Liszt HS, Pety J. 2012.
\newblock \textit{\aap} 541:58

\bibitem[{Liszt, Pety \& Lucas(2010)}]{LISZT2010}
Liszt HS, Pety J, Lucas R. 2010.
\newblock \textit{\aap} 518:45

\bibitem[{Lombardi \& Alves(2001)}]{LOMBARDI2001}
Lombardi M, Alves J. 2001.
\newblock \textit{\aap} 377:1023--1034

\bibitem[{Lombardi, Alves \& Lada(2006)}]{LOMBARDI2006}
Lombardi M, Alves J, Lada CJ. 2006.
\newblock \textit{\aap} 454:781--796

\bibitem[{MacLaren, Richardson \& Wolfendale(1988)}]{MACLAREN1988}
MacLaren I, Richardson KM, Wolfendale AW. 1988.
\newblock \textit{\apj} 333:821--825

\bibitem[{Madden et~al.(1993)Madden, Geis, Genzel, Herrmann, Jackson
  et~al.}]{MADDEN1993}
Madden SC, Geis N, Genzel R, Herrmann F, Jackson J, et~al. 1993.
\newblock \textit{\apj} 407:579--587

\bibitem[{Madden et~al.(1997)Madden, Poglitsch, Geis, Stacey \&
  Townes}]{MADDEN1997}
Madden SC, Poglitsch A, Geis N, Stacey GJ, Townes CH. 1997.
\newblock \textit{\apj v.483} 483:200

\bibitem[{Magdis et~al.(2011)Magdis, Daddi, Elbaz, Sargent, Dickinson
  et~al.}]{MAGDIS2011}
Magdis GE, Daddi E, Elbaz D, Sargent M, Dickinson M, et~al. 2011.
\newblock \textit{\apj Letters} 740:L15

\bibitem[{Magnani et~al.(2003)Magnani, Chastain, Kim, Hartmann, Truong \&
  Thaddeus}]{MAGNANI2003}
Magnani L, Chastain RJ, Kim HC, Hartmann D, Truong AT, Thaddeus P. 2003.
\newblock \textit{\apj} 586:1111--1119

\bibitem[{Magnelli et~al.(2012)Magnelli, Saintonge, Lutz, Tacconi, Berta
  et~al.}]{MAGNELLI2012}
Magnelli B, Saintonge A, Lutz D, Tacconi LJ, Berta S, et~al. 2012.
\newblock \textit{arXiv.org} 1210:2760

\bibitem[{Maloney(1990)}]{MALONEY1990}
Maloney P. 1990.
\newblock \textit{\apj} 348:L9--L12

\bibitem[{Maloney \& Black(1988)}]{MALONEY1988}
Maloney P, Black JH. 1988.
\newblock \textit{\apj} 325:389--401

\bibitem[{Mannucci et~al.(2010)Mannucci, Cresci, Maiolino, Marconi \&
  Gnerucci}]{MANNUCCI2010}
Mannucci F, Cresci G, Maiolino R, Marconi A, Gnerucci A. 2010.
\newblock \textit{\mnras} 408:2115--2127

\bibitem[{Mannucci et~al.(2009)Mannucci, Cresci, Maiolino, Marconi, Pastorini
  et~al.}]{MANNUCCI2009}
Mannucci F, Cresci G, Maiolino R, Marconi A, Pastorini G, et~al. 2009.
\newblock \textit{\mnras} 398:1915--1931

\bibitem[{Mao et~al.(2010)Mao, Schulz, Henkel, Mauersberger, Muders \&
  {Dinh-V-Trung}}]{MAO2010}
Mao RQ, Schulz A, Henkel C, Mauersberger R, Muders D, {Dinh-V-Trung}. 2010.
\newblock \textit{\apj} 724:1336--1356

\bibitem[{Mauersberger et~al.(1999)Mauersberger, Henkel, Walsh \&
  Schulz}]{MAUERSBERGER1999}
Mauersberger R, Henkel C, Walsh W, Schulz A. 1999.
\newblock \textit{\aap} 341:256--263

\bibitem[{McKee(1999)}]{MCKEE1999}
McKee CF. 1999.
\newblock \textit{The Origin of Stars and Planetary Systems. Edited by Charles
  J. Lada and Nikolaos D. Kylafis. Kluwer Academic Publishers} :29

\bibitem[{McKee \& Ostriker(2007)}]{MCKEE2007}
McKee CF, Ostriker EC. 2007.
\newblock \textit{\araa} 45:565--687

\bibitem[{McKee \& Zweibel(1992)}]{MCKEE1992}
McKee CF, Zweibel EG. 1992.
\newblock \textit{\apj} 399:551--562

\bibitem[{Meier \& Turner(2001)}]{MEIER2001}
Meier DS, Turner JL. 2001.
\newblock \textit{\apj} 551:687--701

\bibitem[{{Meier} \& {Turner}(2004)}]{MEIER2004}
{Meier} DS, {Turner} JL. 2004.
\newblock \textit{\aj} 127:2069--2084

\bibitem[{M{\'e}ny et~al.(2007)M{\'e}ny, Gromov, Boudet, Bernard, Paradis \&
  Nayral}]{MENY2007}
M{\'e}ny C, Gromov V, Boudet N, Bernard JP, Paradis D, Nayral C. 2007.
\newblock \textit{\aap} 468:171--188

\bibitem[{Mirabel \& Sanders(1988)}]{MIRABEL1988}
Mirabel IF, Sanders DB. 1988.
\newblock \textit{\apj} 335:104--121

\bibitem[{Miville-Desch{\^e}nes et~al.(2010)Miville-Desch{\^e}nes, Martin,
  Abergel, Bernard, Boulanger et~al.}]{MIVILLE-DESCHENES2010}
Miville-Desch{\^e}nes MA, Martin PG, Abergel A, Bernard JP, Boulanger F, et~al.
  2010.
\newblock \textit{\aap} 518:L104

\bibitem[{Mizuno et~al.(2001)Mizuno, Rubio, Mizuno, Yamaguchi, Onishi \&
  Fukui}]{MIZUNO2001}
Mizuno N, Rubio M, Mizuno A, Yamaguchi R, Onishi T, Fukui Y. 2001.
\newblock \textit{\pasj} 53:L45--L49

\bibitem[{Moustakas et~al.(2010)Moustakas, Kennicutt, Tremonti, Dale, Smith \&
  Calzetti}]{MOUSTAKAS2010}
Moustakas J, Kennicutt RCJ, Tremonti CA, Dale DA, Smith JDT, Calzetti D. 2010.
\newblock \textit{VizieR On-line Data Catalog} 219:00233

\bibitem[{Mu{\~n}oz-Mateos et~al.(2009)Mu{\~n}oz-Mateos, Gil~de Paz, Boissier,
  Zamorano, Dale et~al.}]{MUNOZ-MATEOS2009}
Mu{\~n}oz-Mateos JC, Gil~de Paz A, Boissier S, Zamorano J, Dale DA, et~al.
  2009.
\newblock \textit{\apj} 701:1965--1991

\bibitem[{Murphy et~al.(2012)Murphy, Porter, Moskalenko, Helou \&
  Strong}]{MURPHY2012}
Murphy EJ, Porter TA, Moskalenko IV, Helou G, Strong AW. 2012.
\newblock \textit{\apj} 750:126

\bibitem[{Nakai \& Kuno(1995)}]{NAKAI1995}
Nakai N, Kuno N. 1995.
\newblock \textit{\pasj} 47:761--769

\bibitem[{Narayanan et~al.(2010)Narayanan, Hayward, Cox, Hernquist, Jonsson
  et~al.}]{NARAYANAN2010}
Narayanan D, Hayward CC, Cox TJ, Hernquist L, Jonsson P, et~al. 2010.
\newblock \textit{\mnras} 401:1613--1619

\bibitem[{Narayanan et~al.(2011)Narayanan, Krumholz, Ostriker \&
  Hernquist}]{NARAYANAN2011}
Narayanan D, Krumholz M, Ostriker EC, Hernquist L. 2011.
\newblock \textit{\mnras} 418:664--679

\bibitem[{Narayanan et~al.(2012)Narayanan, Krumholz, Ostriker \&
  Hernquist}]{NARAYANAN2012}
Narayanan D, Krumholz MR, Ostriker EC, Hernquist L. 2012.
\newblock \textit{\mnras} :2537

\bibitem[{Obreschkow \& Rawlings(2009)}]{OBRESCHKOW2009}
Obreschkow D, Rawlings S. 2009.
\newblock \textit{\mnras} 394:1857--1874

\bibitem[{Oka et~al.(1998)Oka, Hasegawa, Hayashi, Handa \& Sakamoto}]{OKA1998}
Oka T, Hasegawa T, Hayashi M, Handa T, Sakamoto S. 1998.
\newblock \textit{\apj v.493} 493:730

\bibitem[{Oka et~al.(2001)Oka, Hasegawa, Sato, Tsuboi, Miyazaki \&
  Sugimoto}]{OKA2001}
Oka T, Hasegawa T, Sato F, Tsuboi M, Miyazaki A, Sugimoto M. 2001.
\newblock \textit{\apj} 562:348--362

\bibitem[{Padoan, Jones \& Nordlund(1997)}]{PADOAN1997}
Padoan P, Jones BJT, Nordlund AP. 1997.
\newblock \textit{\apj v.474} 474:730

\bibitem[{Padovani, Galli \& Glassgold(2009)}]{PADOVANI2009}
Padovani M, Galli D, Glassgold AE. 2009.
\newblock \textit{\aap} 501:619--631

\bibitem[{Paglione, Jackson \& Ishizuki(1997)}]{PAGLIONE1997}
Paglione TAD, Jackson JM, Ishizuki S. 1997.
\newblock \textit{\apj v.484} 484:656

\bibitem[{Pak et~al.(1998)Pak, Jaffe, van Dishoeck, Johansson \&
  Booth}]{PAK1998}
Pak S, Jaffe DT, van Dishoeck EF, Johansson LEB, Booth RS. 1998.
\newblock \textit{\apj v.498} 498:735

\bibitem[{Papadopoulos \& Seaquist(1999)}]{PAPADOPOULOS1999}
Papadopoulos PP, Seaquist ER. 1999.
\newblock \textit{\apj} 516:114--126

\bibitem[{Papadopoulos et~al.(2012)Papadopoulos, van~der Werf, Xilouris, Isaak
  \& Gao}]{PAPADOPOULOS2012}
Papadopoulos PP, van~der Werf P, Xilouris E, Isaak KG, Gao Y. 2012.
\newblock \textit{\apj} 751:10

\bibitem[{Papadopoulos et~al.(2011)Papadopoulos, van~der Werf, Xilouris, Isaak,
  Gao \& Muehle}]{PAPADOPOULOS2011}
Papadopoulos PP, van~der Werf P, Xilouris EM, Isaak KG, Gao Y, Muehle S. 2011.
\newblock \textit{arXiv.org} 1109:4176

\bibitem[{Paradis et~al.(2012)Paradis, Dobashi, Shimoikura, Kawamura, Onishi
  et~al.}]{PARADIS2012}
Paradis D, Dobashi K, Shimoikura T, Kawamura A, Onishi T, et~al. 2012.
\newblock \textit{arXiv.org} 1205:3384

\bibitem[{Penzias(1975)}]{PENZIAS1975}
Penzias AA. 1975.
\newblock In \textit{In: Atomic and molecular physics and the interstellar
  matter; Proceedings of the Twenty-sixth Summer School of Theoretical
  Physics}. Bell Telephone Laboratories, Inc., Holmdel, N.J.

\bibitem[{Pilyugin \& Thuan(2005)}]{PILYUGIN2005}
Pilyugin LS, Thuan TX. 2005.
\newblock \textit{The Astrophysical Journal} 631:231--243

\bibitem[{Pineda, Caselli \& Goodman(2008)}]{PINEDA2008}
Pineda JE, Caselli P, Goodman AA. 2008.
\newblock \textit{\apj} 679:481--496

\bibitem[{Pineda et~al.(2010b)Pineda, Goldsmith, Chapman, Snell, Li
  et~al.}]{PINEDA2010b}
Pineda JL, Goldsmith PF, Chapman N, Snell RL, Li D, et~al. 2010b.
\newblock \textit{\apj} 721:686--708

\bibitem[{Pineda et~al.(2009)Pineda, Ott, Klein, Wong, Muller \&
  Hughes}]{PINEDA2009}
Pineda JL, Ott J, Klein U, Wong T, Muller E, Hughes A. 2009.
\newblock \textit{\apj} 703:736--751

\bibitem[{Pineda et~al.(2010)Pineda, Velusamy, Langer, Goldsmith, Li \&
  Yorke}]{PINEDA2010}
Pineda JL, Velusamy T, Langer WD, Goldsmith PF, Li D, Yorke HW. 2010.
\newblock \textit{\aap} 521:L19

\bibitem[{Planck~Collaboration et~al.(2011{\natexlab{a}})Planck~Collaboration,
  Ade, Aghanim, Arnaud, Ashdown et~al.}]{PLANCKCOLLABORATION2011XIX}
Planck~Collaboration XIX, Ade PAR, Aghanim N, Arnaud M, Ashdown M, et~al.
  2011{\natexlab{a}}.
\newblock \textit{\aap} 536:19

\bibitem[{Planck~Collaboration et~al.(2011{\natexlab{b}})Planck~Collaboration,
  Abergel, Ade, Aghanim, Arnaud et~al.}]{PLANCKCOLLABORATION2011XXI}
Planck~Collaboration XXI, Abergel A, Ade PAR, Aghanim N, Arnaud M, et~al.
  2011{\natexlab{b}}.
\newblock \textit{\aap} 536:21

\bibitem[{Planck~Collaboration et~al.(2011{\natexlab{c}})Planck~Collaboration,
  Abergel, Ade, Aghanim, Arnaud et~al.}]{PLANCKCOLLABORATION2011XXIV}
Planck~Collaboration XXIV, Abergel A, Ade PAR, Aghanim N, Arnaud M, et~al.
  2011{\natexlab{c}}.
\newblock \textit{\aap} 536:24

\bibitem[{Planck~Collaboration et~al.(2011{\natexlab{d}})Planck~Collaboration,
  Abergel, Ade, Aghanim, Arnaud et~al.}]{PLANCKCOLLABORATION2011XXV}
Planck~Collaboration XXV, Abergel A, Ade PAR, Aghanim N, Arnaud M, et~al.
  2011{\natexlab{d}}.
\newblock \textit{\aap} 536:25

\bibitem[{Poglitsch et~al.(1995)Poglitsch, Krabbe, Madden, Nikola, Geis
  et~al.}]{POGLITSCH1995}
Poglitsch A, Krabbe A, Madden SC, Nikola T, Geis N, et~al. 1995.
\newblock \textit{\apj v.454} 454:293

\bibitem[{Rachford et~al.(2009)Rachford, Snow, Destree, Ross, Ferlet
  et~al.}]{RACHFORD2009}
Rachford BL, Snow TP, Destree JD, Ross TL, Ferlet R, et~al. 2009.
\newblock \textit{\apj Supplement} 180:125--137

\bibitem[{Rand \& Kulkarni(1990)}]{RAND1990}
Rand RJ, Kulkarni SR. 1990.
\newblock \textit{\apj} 349:L43--L46

\bibitem[{Rand, Lord \& Higdon(1999)}]{RAND1999}
Rand RJ, Lord SD, Higdon JL. 1999.
\newblock \textit{\apj} 513:720--732

\bibitem[{Rangwala et~al.(2011)Rangwala, Maloney, Glenn, Wilson, Rykala
  et~al.}]{RANGWALA2011}
Rangwala N, Maloney PR, Glenn J, Wilson CD, Rykala A, et~al. 2011.
\newblock \textit{\apj} 743:94

\bibitem[{Rebolledo et~al.(2012)Rebolledo, Wong, Leroy, Koda \&
  Donovan~Meyer}]{REBOLLEDO2012}
Rebolledo D, Wong T, Leroy A, Koda J, Donovan~Meyer J. 2012.
\newblock \textit{arXiv.org} 1208:5499

\bibitem[{Rickard et~al.(1975)Rickard, Palmer, Morris, Zuckerman \&
  Turner}]{RICKARD1975}
Rickard LJ, Palmer P, Morris M, Zuckerman B, Turner BE. 1975.
\newblock \textit{\apj} 199:L75--L78

\bibitem[{Riechers et~al.(2011)Riechers, Carilli, Maddalena, Hodge, Harris
  et~al.}]{RIECHERS2011}
Riechers DA, Carilli CL, Maddalena RJ, Hodge J, Harris AI, et~al. 2011.
\newblock \textit{\apj Letters} 739:L32

\bibitem[{Riechers et~al.(2010)Riechers, Carilli, Walter \&
  Momjian}]{RIECHERS2010}
Riechers DA, Carilli CL, Walter F, Momjian E. 2010.
\newblock \textit{\apj Letters} 724:L153--L157

\bibitem[{Riechers et~al.(2011b)Riechers, Hodge, Walter, Carilli \&
  Bertoldi}]{RIECHERS2011b}
Riechers DA, Hodge J, Walter F, Carilli CL, Bertoldi F. 2011b.
\newblock \textit{\apj Letters} 739:L31

\bibitem[{Rieke \& Lebofsky(1985)}]{RIEKE1985}
Rieke GH, Lebofsky MJ. 1985.
\newblock \textit{\apj} 288:618--621

\bibitem[{R{\"o}llig et~al.(2006)R{\"o}llig, Ossenkopf, Jeyakumar, Stutzki \&
  Sternberg}]{ROLLIG2006}
R{\"o}llig M, Ossenkopf V, Jeyakumar S, Stutzki J, Sternberg A. 2006.
\newblock \textit{\aap} 451:917--924

\bibitem[{Roman-Duval et~al.(2010)Roman-Duval, Jackson, Heyer, Rathborne \&
  Simon}]{ROMAN-DUVAL2010}
Roman-Duval J, Jackson JM, Heyer M, Rathborne J, Simon R. 2010.
\newblock \textit{\apj} 723:492--507

\bibitem[{Rosolowsky(2007)}]{ROSOLOWSKY2007}
Rosolowsky E. 2007.
\newblock \textit{\apj} 654:240--251

\bibitem[{Rosolowsky \& Blitz(2005)}]{ROSOLOWSKY2005}
Rosolowsky E, Blitz L. 2005.
\newblock \textit{\apj} 623:826--845

\bibitem[{Rosolowsky et~al.(2003)Rosolowsky, Engargiola, Plambeck \&
  Blitz}]{ROSOLOWSKY2003}
Rosolowsky E, Engargiola G, Plambeck R, Blitz L. 2003.
\newblock \textit{\apj} 599:258--274

\bibitem[{Rosolowsky \& Leroy(2006)}]{ROSOLOWSKY2006}
Rosolowsky E, Leroy A. 2006.
\newblock \textit{The Publications of the Astronomical Society of the Pacific}
  118:590--610

\bibitem[{Rubio et~al.(2004)Rubio, Boulanger, Rantakyro \&
  Contursi}]{RUBIO2004}
Rubio M, Boulanger F, Rantakyro F, Contursi A. 2004.
\newblock \textit{\aap} 425:L1--L4

\bibitem[{Rubio, Lequeux \& Boulanger(1993)}]{RUBIO1993}
Rubio M, Lequeux J, Boulanger F. 1993.
\newblock \textit{\aap} 271:9

\bibitem[{Sanders \& Mirabel(1996)}]{SANDERS1996}
Sanders DB, Mirabel IF. 1996.
\newblock \textit{\araa} 34:749

\bibitem[{Sandstrom et~al.(2012)Sandstrom, Leroy, Walter, Bolatto, Croxall
  et~al.}]{SANDSTROM2012}
Sandstrom KM, Leroy AK, Walter F, Bolatto AD, Croxall KV, et~al. 2012.
\newblock \textit{arXiv.org} 1212:1208

\bibitem[{Savage et~al.(1977)Savage, Bohlin, Drake \& Budich}]{SAVAGE1977}
Savage BD, Bohlin RC, Drake JF, Budich W. 1977.
\newblock \textit{\apj} 216:291--307

\bibitem[{Schinnerer et~al.(2010)Schinnerer, Weiss, Aalto \&
  Scoville}]{SCHINNERER2010}
Schinnerer E, Weiss A, Aalto S, Scoville NZ. 2010.
\newblock \textit{\apj} 719:1588--1601

\bibitem[{Schlegel, Finkbeiner \& Davis(1998)}]{SCHLEGEL1998}
Schlegel DJ, Finkbeiner DP, Davis M. 1998.
\newblock \textit{\apj v.500} 500:525

\bibitem[{Schnee et~al.(2008)Schnee, Li, Goodman \& Sargent}]{SCHNEE2008}
Schnee S, Li J, Goodman AA, Sargent AI. 2008.
\newblock \textit{\apj} 684:1228--1239

\bibitem[{Schruba et~al.(2012)Schruba, Leroy, Walter, Bigiel, Brinks
  et~al.}]{SCHRUBA2012}
Schruba A, Leroy AK, Walter F, Bigiel F, Brinks E, et~al. 2012.
\newblock \textit{\aj} 143:138

\bibitem[{Scoville \& Good(1989)}]{SCOVILLE1989}
Scoville NZ, Good JC. 1989.
\newblock \textit{\apj} 339:149--162

\bibitem[{Scoville \& Hersh(1979)}]{SCOVILLE1979}
Scoville NZ, Hersh K. 1979.
\newblock \textit{\apj} 229:578--582

\bibitem[{Scoville \& Sanders(1987b)}]{SCOVILLE1987b}
Scoville NZ, Sanders DB. 1987b.
\newblock In \textit{IN: Interstellar processes; Proceedings of the Symposium}.
  California Institute of Technology, Pasadena

\bibitem[{Scoville \& Solomon(1975)}]{SCOVILLE1975}
Scoville NZ, Solomon PM. 1975.
\newblock \textit{\apj} 199:L105--L109

\bibitem[{Scoville, Yun \& Bryant(1997)}]{SCOVILLE1997}
Scoville NZ, Yun MS, Bryant PM. 1997.
\newblock \textit{\apj} 484:702

\bibitem[{Scoville et~al.(1987)Scoville, Yun, Sanders, Clemens \&
  Waller}]{SCOVILLE1987}
Scoville NZ, Yun MS, Sanders DB, Clemens DP, Waller WH. 1987.
\newblock \textit{\apj Supplement Series (ISSN 0067-0049)} 63:821--915

\bibitem[{Sheffer et~al.(2008)Sheffer, Rogers, Federman, Abel, Gredel
  et~al.}]{SHEFFER2008}
Sheffer Y, Rogers M, Federman SR, Abel NP, Gredel R, et~al. 2008.
\newblock \textit{\apj} 687:1075--1106

\bibitem[{Shetty et~al.(2011)Shetty, Glover, Dullemond \& Klessen}]{SHETTY2011}
Shetty R, Glover SC, Dullemond CP, Klessen RS. 2011.
\newblock \textit{\mnras} 412:1686--1700

\bibitem[{Shetty et~al.(2011b)Shetty, Glover, Dullemond, Ostriker, Harris \&
  Klessen}]{SHETTY2011b}
Shetty R, Glover SC, Dullemond CP, Ostriker EC, Harris AI, Klessen RS. 2011b.
\newblock \textit{\mnras} 415:3253--3274

\bibitem[{Shibai et~al.(1991)Shibai, Okuda, Nakagawa, Matsuhara, Maihara
  et~al.}]{SHIBAI1991}
Shibai H, Okuda H, Nakagawa T, Matsuhara H, Maihara T, et~al. 1991.
\newblock \textit{\apj} 374:522--532

\bibitem[{Sliwa et~al.(2012)Sliwa, Wilson, Petitpas, Armus, Juvela
  et~al.}]{SLIWA2012}
Sliwa K, Wilson CD, Petitpas GR, Armus L, Juvela M, et~al. 2012.
\newblock \textit{\apj} 753:46

\bibitem[{Smith et~al.(2012)Smith, Eales, Gomez, Roman-Duval, Fritz
  et~al.}]{SMITH2012}
Smith MWL, Eales SA, Gomez HL, Roman-Duval J, Fritz J, et~al. 2012.
\newblock \textit{\apj} 756:40

\bibitem[{Sodroski et~al.(1995)Sodroski, Odegard, Dwek, Hauser, Franz
  et~al.}]{SODROSKI1995}
Sodroski TJ, Odegard N, Dwek E, Hauser MG, Franz BA, et~al. 1995.
\newblock \textit{\apj} 452:262

\bibitem[{Sofia et~al.(2004)Sofia, Lauroesch, Meyer \& Cartledge}]{SOFIA2004}
Sofia UJ, Lauroesch JT, Meyer DM, Cartledge SIB. 2004.
\newblock \textit{\apj} 605:272--277

\bibitem[{Solomon \& de~Zafra(1975)}]{SOLOMON1975}
Solomon PM, de~Zafra R. 1975.
\newblock \textit{\apj} 199:L79--L83

\bibitem[{Solomon, Downes \& Radford(1992)}]{SOLOMON1992}
Solomon PM, Downes D, Radford SJE. 1992.
\newblock \textit{\apj} 398:L29--L32

\bibitem[{Solomon et~al.(1997)Solomon, Downes, Radford \&
  Barrett}]{SOLOMON1997}
Solomon PM, Downes D, Radford SJE, Barrett JW. 1997.
\newblock \textit{\apj} 478:144

\bibitem[{Solomon et~al.(1987)Solomon, Rivolo, Barrett \& Yahil}]{SOLOMON1987}
Solomon PM, Rivolo AR, Barrett J, Yahil A. 1987.
\newblock \textit{\apj} 319:730--741

\bibitem[{Solomon et~al.(1972)Solomon, Scoville, Penzias, Wilson \&
  Jefferts}]{SOLOMON1972}
Solomon PM, Scoville NZ, Penzias AA, Wilson RW, Jefferts KB. 1972.
\newblock \textit{\apj} 178:125--130

\bibitem[{Solomon \& Vanden~Bout(2005)}]{SOLOMON2005}
Solomon PM, Vanden~Bout PA. 2005.
\newblock \textit{\araa} 43:677--725

\bibitem[{Sonnentrucker et~al.(2007)Sonnentrucker, Welty, Thorburn \&
  York}]{SONNENTRUCKER2007}
Sonnentrucker P, Welty DE, Thorburn JA, York DG. 2007.
\newblock \textit{\apj Supplement Series} 168:58--99

\bibitem[{Stacey et~al.(1991)Stacey, Geis, Genzel, Lugten, Poglitsch
  et~al.}]{STACEY1991}
Stacey GJ, Geis N, Genzel R, Lugten JB, Poglitsch A, et~al. 1991.
\newblock \textit{\apj} 373:423--444

\bibitem[{Stanimirovic et~al.(2000)Stanimirovic, Staveley-Smith, van~der Hulst,
  Bontekoe, Kester \& Jones}]{STANIMIROVIC2000}
Stanimirovic S, Staveley-Smith L, van~der Hulst JM, Bontekoe TR, Kester DJM,
  Jones PA. 2000.
\newblock \textit{\mnras} 315:791--807

\bibitem[{Stecker et~al.(1975)Stecker, Solomon, Scoville \&
  Ryter}]{STECKER1975}
Stecker FW, Solomon PM, Scoville NZ, Ryter CE. 1975.
\newblock \textit{\apj} 201:90--97

\bibitem[{Strong \& Mattox(1996)}]{STRONG1996}
Strong AW, Mattox JR. 1996.
\newblock \textit{\aap} 308:L21--L24

\bibitem[{Strong et~al.(2004)Strong, Moskalenko, Reimer, Digel \&
  Diehl}]{STRONG2004}
Strong AW, Moskalenko IV, Reimer O, Digel S, Diehl R. 2004.
\newblock \textit{\aap} 422:L47--L50

\bibitem[{Swinbank et~al.(2010)Swinbank, Smail, Longmore, Harris, Baker
  et~al.}]{SWINBANK2010}
Swinbank AM, Smail I, Longmore S, Harris AI, Baker AJ, et~al. 2010.
\newblock \textit{Nature} 464:733--736

\bibitem[{Tacconi et~al.(2010)Tacconi, Genzel, Neri, Cox, Cooper
  et~al.}]{TACCONI2010}
Tacconi LJ, Genzel R, Neri R, Cox P, Cooper MC, et~al. 2010.
\newblock \textit{Nature} 463:781--784

\bibitem[{Tacconi et~al.(2008)Tacconi, Genzel, Smail, Neri, Chapman
  et~al.}]{TACCONI2008}
Tacconi LJ, Genzel R, Smail I, Neri R, Chapman SC, et~al. 2008.
\newblock \textit{\apj} 680:246--262

\bibitem[{Tacconi et~al.(2006)Tacconi, Neri, Chapman, Genzel, Smail
  et~al.}]{TACCONI2006}
Tacconi LJ, Neri R, Chapman SC, Genzel R, Smail I, et~al. 2006.
\newblock \textit{\apj} 640:228--240

\bibitem[{Tacconi \& Young(1987)}]{TACCONI1987}
Tacconi LJ, Young JS. 1987.
\newblock \textit{\apj} 322:681--687

\bibitem[{Taylor et~al.(1999)Taylor, Huttemeister, Klein \& Greve}]{TAYLOR1999}
Taylor CL, Huttemeister S, Klein U, Greve A. 1999.
\newblock \textit{\aap} 349:424--434

\bibitem[{Taylor, Kobulnicky \& Skillman(1998)}]{TAYLOR1998}
Taylor CL, Kobulnicky HA, Skillman ED. 1998.
\newblock \textit{\aj} 116:2746--2756

\bibitem[{Thompson, Quataert \& Murray(2005)}]{THOMPSON2005}
Thompson TA, Quataert E, Murray N. 2005.
\newblock \textit{\apj} 630:167--185

\bibitem[{Thronson(1988b)}]{THRONSON1988b}
Thronson HAJ. 1988b.
\newblock \textit{Galactic and Extragalactic Star Formation} 2:621

\bibitem[{Thronson et~al.(1988)Thronson, Greenhouse, Hunter, Telesco \&
  Harper}]{THRONSON1988}
Thronson HAJ, Greenhouse M, Hunter DA, Telesco CM, Harper DA. 1988.
\newblock \textit{\apj} 334:605--612

\bibitem[{Ueda et~al.(2012)Ueda, Iono, Petitpas, Yun, Ho et~al.}]{UEDA2012}
Ueda J, Iono D, Petitpas G, Yun MS, Ho PTP, et~al. 2012.
\newblock \textit{\apj} 745:65

\bibitem[{van Dishoeck \& Black(1986)}]{VANDISHOECK1986}
van Dishoeck EF, Black JH. 1986.
\newblock \textit{\apj Supplement Series (ISSN 0067-0049)} 62:109--145

\bibitem[{van Dishoeck \& Black(1988)}]{VANDISHOECK1988}
van Dishoeck EF, Black JH. 1988.
\newblock \textit{\apj} 334:771--802

\bibitem[{van Dishoeck, Jonkheid \& van Hemert(2006)}]{VANDISHOECK2006}
van Dishoeck EF, Jonkheid B, van Hemert MC. 2006.
\newblock \textit{in "Chemical Evolution of the Universe"} Faraday Discussions
  vol 133:231

\bibitem[{Velusamy et~al.(2012)Velusamy, Langer, Pineda \&
  Goldsmith}]{VELUSAMY2012}
Velusamy T, Langer WD, Pineda JL, Goldsmith PF. 2012.
\newblock \textit{\aap} 541:L10

\bibitem[{Velusamy et~al.(2010)Velusamy, Langer, Pineda, Goldsmith, Li \&
  Yorke}]{VELUSAMY2010}
Velusamy T, Langer WD, Pineda JL, Goldsmith PF, Li D, Yorke HW. 2010.
\newblock \textit{\aap} 521:L18

\bibitem[{Visser et~al.(2009)Visser, van Dishoeck, Doty \&
  Dullemond}]{VISSER2009}
Visser R, van Dishoeck EF, Doty SD, Dullemond CP. 2009.
\newblock \textit{\aap} 495:881--897

\bibitem[{Vogel, Boulanger \& Ball(1987)}]{VOGEL1987}
Vogel SN, Boulanger F, Ball R. 1987.
\newblock \textit{\apj} 321:L145--L149

\bibitem[{Vogel, Kulkarni \& Scoville(1988)}]{VOGEL1988}
Vogel SN, Kulkarni SR, Scoville NZ. 1988.
\newblock \textit{Nature (ISSN 0028-0836)} 334:402--406

\bibitem[{Walter et~al.(2001)Walter, Taylor, H{\"u}ttemeister, Scoville \&
  McIntyre}]{WALTER2001}
Walter F, Taylor CL, H{\"u}ttemeister S, Scoville N, McIntyre V. 2001.
\newblock \textit{\aj} 121:727--739

\bibitem[{Walter et~al.(2002)Walter, Weiss, Martin \& Scoville}]{WALTER2002}
Walter F, Weiss A, Martin C, Scoville N. 2002.
\newblock \textit{\aj} 123:225--237

\bibitem[{Ward et~al.(2003)Ward, Zmuidzinas, Harris \& Isaak}]{WARD2003}
Ward JS, Zmuidzinas J, Harris AI, Isaak KG. 2003.
\newblock \textit{\apj} 587:171--185

\bibitem[{Watanabe et~al.(2011)Watanabe, Sorai, Kuno \& Habe}]{WATANABE2011}
Watanabe Y, Sorai K, Kuno N, Habe A. 2011.
\newblock \textit{\mnras} 411:1409--1417

\bibitem[{Wei, Keto \& Ho(2012)}]{WEI2012}
Wei LH, Keto E, Ho LC. 2012.
\newblock \textit{\apj} 750:136

\bibitem[{Weiss et~al.(2007)Weiss, Downes, Walter \& Henkel}]{WEISS2007}
Weiss A, Downes D, Walter F, Henkel C. 2007.
\newblock \textit{From Z-Machines to ALMA: (Sub)Millimeter Spectroscopy of
  Galaxies ASP Conference Series} 375:25

\bibitem[{Wild et~al.(1992)Wild, Harris, Eckart, Genzel, Graf
  et~al.}]{WILD1992}
Wild W, Harris AI, Eckart A, Genzel R, Graf UU, et~al. 1992.
\newblock \textit{\aap} 265:447--464

\bibitem[{Wilson(1994)}]{WILSON1994}
Wilson CD. 1994.
\newblock \textit{\apj} 434:L11--L14

\bibitem[{Wilson(1995)}]{WILSON1995}
Wilson CD. 1995.
\newblock \textit{\apj Letters} 448:L97

\bibitem[{Wilson \& Scoville(1990)}]{WILSON1990}
Wilson CD, Scoville N. 1990.
\newblock \textit{\apj} 363:435--450

\bibitem[{Wilson et~al.(2003)Wilson, Scoville, Madden \&
  Charmandaris}]{WILSON2003}
Wilson CD, Scoville N, Madden SC, Charmandaris V. 2003.
\newblock \textit{\apj} 599:1049--1066

\bibitem[{Wilson et~al.(1988)Wilson, Scoville, Madore, Sanders \&
  Freedman}]{WILSON1988}
Wilson CD, Scoville N, Madore BF, Sanders DB, Freedman WL. 1988.
\newblock \textit{\apj} 333:611--615

\bibitem[{Wilson, Jefferts \& Penzias(1970)}]{WILSON1970}
Wilson RW, Jefferts KB, Penzias AA. 1970.
\newblock \textit{\apj} 161:L43

\bibitem[{Wilson(1999)}]{WILSON1999}
Wilson TL. 1999.
\newblock \textit{Reports on Progress in Physics} 62:143--185

\bibitem[{Wilson et~al.(1974)Wilson, Schwartz, Epstein, Johnson, Etcheverry
  et~al.}]{WILSON1974}
Wilson WJ, Schwartz PR, Epstein EE, Johnson WA, Etcheverry RD, et~al. 1974.
\newblock \textit{\apj} 191:357--374

\bibitem[{Wolfire, Hollenbach \& McKee(2010)}]{WOLFIRE2010}
Wolfire MG, Hollenbach D, McKee CF. 2010.
\newblock \textit{\apj} 716:1191--1207

\bibitem[{Wolfire, Hollenbach \& Tielens(1993)}]{WOLFIRE1993}
Wolfire MG, Hollenbach D, Tielens AGGM. 1993.
\newblock \textit{\apj} 402:195--215

\bibitem[{Wolfire et~al.(2008)Wolfire, Tielens, Hollenbach \&
  Kaufman}]{WOLFIRE2008}
Wolfire MG, Tielens AGGM, Hollenbach D, Kaufman MJ. 2008.
\newblock \textit{\apj} 680:384--397

\bibitem[{Wong et~al.(2011)Wong, Hughes, Ott, Muller, Pineda et~al.}]{WONG2011}
Wong T, Hughes A, Ott J, Muller E, Pineda JL, et~al. 2011.
\newblock \textit{\apj Supplement} 197:16

\bibitem[{Yang et~al.(2010)Yang, {Stancil, P. C.}, Balakrishnan \&
  Forrey}]{YANG2010}
Yang B, {Stancil, P. C.}, Balakrishnan N, Forrey RC. 2010.
\newblock \textit{\apj} 718:1062--1069

\bibitem[{Yao et~al.(2003)Yao, Seaquist, Kuno \& Dunne}]{YAO2003}
Yao L, Seaquist ER, Kuno N, Dunne L. 2003.
\newblock \textit{\apj} 588:771--791

\bibitem[{Young et~al.(1996)Young, Allen, Kenney, Lesser \& Rownd}]{YOUNG1996}
Young JS, Allen L, Kenney JDP, Lesser A, Rownd B. 1996.
\newblock \textit{\aj} 112:1903

\bibitem[{Young \& Knezek(1989)}]{YOUNG1989}
Young JS, Knezek PM. 1989.
\newblock \textit{\apj} 347:L55--L58

\bibitem[{Young \& Scoville(1982)}]{YOUNG1982}
Young JS, Scoville N. 1982.
\newblock \textit{\apj} 258:467--489

\bibitem[{Young \& Scoville(1991)}]{YOUNG1991}
Young JS, Scoville NZ. 1991.
\newblock \textit{\araa} 29:581--625

\bibitem[{Young et~al.(1995)Young, Xie, Tacconi, Knezek, Viscuso
  et~al.}]{YOUNG1995}
Young JS, Xie S, Tacconi L, Knezek P, Viscuso P, et~al. 1995.
\newblock \textit{\apj Supplement Series} 98:219

\bibitem[{Zhu et~al.(2009)Zhu, Papadopoulos, Xilouris, Kuno \&
  Lisenfeld}]{ZHU2009}
Zhu M, Papadopoulos PP, Xilouris EM, Kuno N, Lisenfeld U. 2009.
\newblock \textit{\apj} 706:941--959

\bibitem[{Zhu, Seaquist \& Kuno(2003)}]{ZHU2003}
Zhu M, Seaquist ER, Kuno N. 2003.
\newblock \textit{\apj} 588:243--263

\end{thebibliography}

\end{document}